\documentclass[12pt]{article}

\usepackage{amsmath,amssymb,amsfonts}

\usepackage{hyperref}
\usepackage{graphicx}
\usepackage{color}
\usepackage{array}
\usepackage{subfigure}
\usepackage{comment}
\usepackage{cite}

\newcommand{\w}{\wedge}
\newcommand{\beq}{\begin{equation}}
\newcommand{\eeq}{\end{equation}}
\newcommand{\ber}{\begin{eqnarray}}
\newcommand{\eer}{\end{eqnarray}}
\newcommand{\pt}{\partial}
\def\N{{\cal N}}
\def\Ic{{\cal I}}

\def\tr{{\mathrm{tr}}}
\def\nn{{\nonumber}}

\numberwithin{equation}{section}

\begin{document}

\thispagestyle{empty}

\title{\bf Pseudoscalar glueballs in the Klebanov-Strassler Theory}
\author{Dmitry Melnikov$^{a,b}$ and Corn\'elio Rodrigues Filho$^c$}

\maketitle

\vspace{-6cm}
\hfill{ITEP-TH-16/20}
\vspace{6cm}

\vspace{-30pt}
\begin{center}
\textit{\small $^a$  International Institute of Physics, Federal University of Rio Grande do Norte, \\ Campus Universit\'ario, Lagoa Nova, Natal-RN  59078-970, Brazil}\\ \vspace{6pt}
\textit{\small $^b$  Institute for Theoretical and Experimental Physics, \\ B.~Cheremushkinskaya 25, Moscow 117218, Russia}
\\ \vspace{6pt}
\textit{\small $^c$  Department of Theoretical and Experimental Physics, Federal University of Rio Grande do Norte, Campus Universit\'ario, Lagoa Nova, Natal-RN  59078-970, Brazil}
\\ \vspace{2cm}
\end{center}

\vspace{-2cm}

\begin{abstract}
In this paper we describe a pseudoscalar subsector of the Klebanov-Strassler 
model. This subsector completes the holographic reconstruction of the spectrum 
of the lowest-lying glueball states, which are singlet under the global symmetry 
group $SU(2)\times SU(2)$. We derive the linearized supergravity equations for 
the pseudoscalar fluctuations  and analyze their spectrum. The system of 
equations is shown to be compatible with six eigenmodes, as expected from 
supersymmetry. Our numerical analysis allows to reliably extract four of the 
corresponding towers. Their values match well the eigenvalues of the $0^{++}$ 
scalar states known from an earlier work. Assuming the masses of $0^{++}$ as a 
reference, we compare the lightest states of the holographic spectrum with 
lattice calculations in the quenched QCD at $N_c=3$ and $N_c=\infty$.

\end{abstract}

\pagebreak
\tableofcontents

\maketitle

\section{Introduction}
\label{sec:intro}

In~\cite{Klebanov:2000hb} Klebanov and Strassler derived a solution of the type IIB supergravity equations, which describes a holographic dual of a non-conformal ${\mathcal{N}}=1$ gauge theory. Contrary to initial expectations~\cite{Klebanov:1998hh,Gubser:1998fp,Klebanov:1999rd,Klebanov:2000nc} 
this theory did not quite provide a dual of ${\mathcal{N}}=1$ supersymmetric Yang-Mills theory (SYM) (the simplest extension of the bosonic gauge Yang-Mills theory) with large number of colors $N$. Instead it gave an interesting and novel example of the so-called ``cascading'' theories, with a less conventional RG flow~\cite{Strassler:2005qs}.

The difference of the Klebanov-Strassler (KS) cascading theories from the SYM in the IR limit is due to the spontaneous breaking of the baryonic $U(1)_B$ symmetry\cite{Gubser:2004qj,Gubser:2004tf,Benna:2006ib}. The spontaneous breaking is responsible for the presence of the Goldstone modes of $U(1)_B$ and for the corresponding tower of light states. Those light states mix with the light states of the SYM sector of the theory.

The KS theory is constructed as a solution of type IIB supergravity equations on 
$AdS^5\times T^{1,1}$, where $T^{1,1}$ is the space $S^3\times S^2$ with a 
special choice of the metric compatible with ${\mathcal{N}}=1$ supersymmetry. 
$T^{1,1}$ space, and hence  the dual gauge theory, has an $SU(2)\times SU(2)$ 
global symmetry. The spectrum of such a theory should be organized in its 
irreducible representations. The structure of the mutliplets of the conformal 
version  of the KS model on $T^{1,1}$ (known as Klebanov-Witten 
theory\cite{Klebanov:1998hh}) was analyzed in~\cite{Ceresole:1999ht}. Meanwhile, 
in the KS theory the conformal symmetry is broken by the flux of the 3-form 
field $F_3$. The supergravity metric does not have isometries of the $AdS_5$: it 
is a ``warped'' metric with a non-conformal logarithmic running of the $AdS_5$ 
radius, as well as different scaling of $S^3$ and $S^2$ in $T^{1,1}$. Hence, in 
the KS theory superconformal multiplets of~\cite{Ceresole:1999ht} break into 
multiplets of ${\mathcal{N}}=1$ supersymmetry. This implies, in particular, that 
 massive bosonic states of this theory appear in pairs. 

Fields of the pure SYM sector of the theory are singlets under the global $SU(2)\times SU(2)$. One might be interested in the structure of this particular sector as of a cousin of the simplest $\N=1$ SYM theory, and even as a more distant cousing of the pure glue (quenched) Yang-Mills. The spectrum of glueballs in the latter theory has been computed on a lattice~\cite{Morningstar:1999rf,Chen:2005mg} and one may wonder if a meaningful comparison with lattice results can be made.

The first observation is that, from the string theory point of view, the classical gravity solution that we use here is only a valid approximation for a very small string tension, which practically means that all states with spin higher or equal than two (with the exception of $2^{++}$) cannot be seen in the gravity description. Therefore one should focus only on the low-spin states, whose mass has the lowest order in the string tension.

The second observation is that classical gravity is only valid in the limit of large number of colors $N_c$. This is less of an obstacle because the glueball masses are expected to weakly depend on $N_c$:
\beq
m(N_c) \ \simeq \ m(\infty) + \frac{c}{N_c^2}\,.
\eeq
Moreover, calculation for different values of $N_c$ are available on the 
lattice~\cite{Teper:1998kw,Lucini:2004my,Lucini:2010nv,Holligan:2019lma,
Bennett:2020hqd,Athenodorou:2020ani}, from which the $N_c\to\infty$ limit can 
be extrapolated. Lattice calculations also show that if fermions are introduced, 
unquenching the theory, the values of the glueball masses will only vary by a 
little~\cite{Gregory:2012hu}, so that the pure glue values can be viewed as 
reasonable approximations for the masses in QCD and experiments. One can then 
expect that the supersymmetric extension and even additional matter fields in 
the KS theory would not spoil the structure of the light spectrum.

The goal of this paper is to complete the study of the SYM sector of the KS 
theory. Its structure is in general understood from a series of previous 
works~\cite{Berg:2005pd,Dymarsky:2006hn,Berg:2006xy,Dymarsky:2007zs,Benna:2007mb
,Dymarsky:2008wd,Gordeli:2009nw}. See also~\cite{Gordeli:2013jea} for the 
summary of this sector, 
\cite{Krasnitz:2000ir,Caceres:2000qe,Bianchi:2003ug,Amador:2004pz} for some 
earlier works and 
\cite{Caceres:2005yx,Elander:2009bm,Bianchi:2010cy,Pufu:2010ie,Elander:2014ola,
Elander:2017cle,Elander:2017hyr} for extensions beyond the singlet regime, and 
beyond the KS theory. Almost all glueball states from the singlet sector were 
explicitly 
constructed as perturbations of the KS background. The masses of the lowest 
states in the conformal towers were computed. The only missing subsector in the 
previous analysis is that of the $0^{-+}$ states, although it was explained 
in~\cite{Gordeli:2009nw} that this sector should consist of six modes degenerate 
with six of the seven of $0^{++}$ scalars described 
in~\cite{Berg:2005pd,Berg:2006xy}.

Here we will explicitly construct the pseudoscalar modes. In the holographic approach the spectrum is derived from linearized perturbations over a background (vacuum) solution. We will explain the structure of the $J^{PC}=0^{-+}$ perturbations in the KS background and present the corresponding linearized supergravity equations.

The perturbations are described by a complicated set of six coupled second order differential equations. However, it is more practical to write them with a help of an auxiliary field, which makes the equations more compact at the expense of introducing an additional first, or second order equation.

We find no obvious way to decouple the equations. Before analysing their spectrum numerically, we make a few consistency checks. First, the full number of equations obtained in the analysis of the perturbations is eight. We check that one of the eight equations can be derived from the remaining seven in a non-trivial way. 

In deriving the equations we worked in explicitly gauge invariant setting. Gauge (diffeomorphism) invariance provides an additional check of the consistency of the derived equations. Finally, after fixing the gauge the system indeed can be reduced to six second order equations, matching the expected number of the $0^{-+}$ modes. 

We expect that the spectrum of the pseudoscalar modes matches the spectrum of $0^{++}$ calculated in~\cite{Berg:2006xy}\footnote{The spectrum of~\cite{Berg:2006xy} was independently checked in~\cite{Elander:2014ola}.}. In our numerical simulations we were able to observe what appears as four towers of eigenvalues that indeed match well the masses of $0^{++}$. However, our current numerical method does not resolve for the remaining two towers. A different approach with seven unconstrained modes produces two more towers, that do not match the spectrum in~\cite{Berg:2006xy}. The nature of the latter two towers is not completely clear from our analysis and we expect that the subsequent studies will either confirm or discard them.

We show that a meaningful comparison of the holographic spectrum with lattice calculations can indeed be made. We can compare the lightest states in the spectrum with $0^{++}$ and $0^{-+}$ of the $SU(3)$ theory~\cite{Chen:2005mg} and with its the $SU(\infty)$ extrapolation~\cite{Lucini:2004my}. The states in the $C$-odd sector studied in~\cite{Benna:2007mb,Dymarsky:2008wd} can be compared with the $SU(3)$ values~\cite{Chen:2005mg}. Some of them can also be compared with $SU(\infty)$~\cite{Lucini:2010nv}.

The comparative plot is shown on figure~\ref{fig:lattice}. In general the 
holographic calculation reproduces quite well the structure of the lattice 
spectrum, which confirms the expected weak dependence of the glueball spectrum 
on the details of the theory, such as $N_c$ and additional matter. Since the KS 
theory is supersymmetric, one sees more states than there are in the pure glue 
theory, so the holographic calculation may be expected to provide a reasonable 
estimate for the structure of the spectrum of a supersymmetric Yang-Mills 
theory. Further details of the comparison will appear in a separate 
work~\cite{inprogress}.

The remainder of this paper is organized as follows. In section~\ref{sec:methods} we summarize the necessary background material. Section~\ref{sec:KS} contains a brief review of the Klebanov-Strassler solution and section~\ref{sec:glueballs} comments on the dual theory. In section~\ref{sec:PC} we review the quantum numbers of particles from both the field theory and supergravity points of view. We also discuss the expected operator content in the pseudoscalar sector by comparing with the structure of superconformal multiplets on $AdS_5\times T^{1,1}$ from
the analysis of~\cite{Ceresole:1999ht}.  Section~\ref{sec:pseudoscalars} contains the main analytical results of our work. Following the earlier analysis of quantum numbers, in section~\ref{sec:pflt} we specify the general ansatz for pseudoscalar fluctuations. In section~\ref{sec:eqs} we present the corresponding linearized equations. We analyze their asymptotic behavior in section~\ref{sec:asymptotic}. Finally, in section~\ref{sec:gaugesym} we check the consistency of the derived equations by checking their gauge invariance and analysing the number of independent modes. In section~\ref{sec:numerics} we explain the results of the numerical analysis of the spectrum. Concluding remarks are made in section~\ref{sec:conclude}, where we also present the results of our comparison with the lattice data. The paper also contains two appendices~\ref{sec:UV} and~\ref{sec:IRasympt}, in which we describe the asymptotic solutions.

\section{Glueballs from a holographic model}
\label{sec:methods}

Holographic approach~\cite{Maldacena:1997re,Gubser:1998bc,Witten:1998qj} provides a powerful tool to analyze a few explicitly known interacting gauge theories in the regime of extremely strong coupling, for reviews see\cite{Aharony:1999ti,Nastase:2015wjb,Erdmenger:2018xqz,DeWolfe:2018dkl,Harlow:2018fse}. In particular, the spectrum of light states in a theory can be extracted from classical gravity equations. In this section we describe a specific gravity system found by Klebanov and Strassler~\cite{Klebanov:2000hb}, based on earlier developments in~\cite{Klebanov:1998hh,Gubser:1998fp,Klebanov:1999rd,Klebanov:2000nc}, dual to a $\N=1$ supersymmetric gauge theory with large number of colors at strong coupling. 

\subsection{Brief review of the Klebanov-Strassler theory}
\label{sec:KS}

The Klebanov-Strassler (KS) model~\cite{Klebanov:2000hb} is based on a solution of the equations of motion of type IIB supergravity~\cite{Schwarz:1983qr}. The bosonic sector of this theory reduces to the Einstein equation
\begin{multline}
    R_{MN} \ = \ \frac{1}{2}\pt_M\Phi\pt_N\Phi+\frac{1}{2}e^{2\Phi}\pt_M C\pt_N
C+\frac{1}{96}g_s^2\tilde{F}_{MPQRS}\tilde{F}^{PQRS}_N+\\
+\frac{g_s}{4}\bigg(e^{-\Phi}H_{MNP}H^{PQ}_N+e^{\Phi}\tilde{F}_{MPQ}\tilde{F}^{PQ}_N\bigg)-
\\ -\frac{g_s}{48}G_{MN}\bigg(e^{-\Phi}H_{PQR}H^{PQR}+e^{\Phi}\tilde{F}_{PQR}\tilde{F}^{PQR}\bigg)\,,\label{RMN}
\end{multline}
here written in the Einstein frame, and equations for the matter fields
\ber
d(e^{2\Phi}*dC)&=&-g_se^{\Phi}H_3\w *\tilde{F}_3\label{ceom}\,,\\
d(e^{\Phi}*\tilde{F}_3)&=&g_s F_5\w H_3\,,\label{f3eom}\\
d*(e^{-\Phi}H_3-C e^{\Phi}\tilde{F}_3)&=&-g_sF_5\w F_3\label{h3eom}\,,\\
*\tilde{F}_5&=&\tilde{F}_5\,.\label{dualf}
\eer
The following notations are commonly used:
\ber
F_3=dC_2,\ H_3=dB_2,\ F_5=dC_4, \ \tilde{F}_3=F_3-C H_3,\ \tilde{F}_5=F_5+B_2\w F_3\,.
\eer
Here $\Phi$, $B_2\equiv B_{MN}$ and $G_{MN}$ are the NS-NS sector fields (respectively, the dilaton, the antisymmetric rank two tensor and the metric, whose Ricci tensor is denoted $R_{MN}$) of the type IIB theory in ten dimensions $M,N=0,1,\ldots,9$. $C\equiv C_0$, $C_2$ and $C_4$ are the R-R fields (the scalar and antisymmetric tensors of rank 2 and 4). For compactness the equations are written in the differential form notations. In particular, $\ast$ denote the Hodge dual. Additionally, there is a Bianchi identity for the 5-form field $\tilde{F}_5$,
\beq\label{bif5}
d\tilde{F}_5=H_3\w F_3.
\eeq

The KS solution of the above equations starts from the metric
\beq\label{mks}
ds_{KS}^2=h^{-1/2}(\tau)\eta_{\mu\nu}dx^\mu dx^\nu + h^{1/2}(\tau)ds_{6}^{2}\,,
\eeq
where $h(\tau)$ is called the warp factor and $ds_6^2$ is the metric of the deformed conifold, a six dimensional cone with the base $S^3\times S^2$~\cite{Candelas:1989js},
\begin{multline}
ds_{6}^2 \  = \ \frac{\epsilon^{4/3} K(\tau)}{2}\Big[\frac{1}{3K(\tau)^3}(d\tau^2+(g^5)^2)+\cosh^2\Big({\frac{\tau}{2}}\Big)[(g^3)^2+(g^4)^2]+ \\
+\sinh^2\Big({\frac{\tau}{2}}\Big)[(g^2)^2+(g^1)^2]\Big]\,, \label{dc}
\end{multline}
with
\beq
K(\tau)=\frac{(\sinh(2\tau)-2\tau)^{1/3}}{2^{1/3}\sinh\tau}.
\eeq
Here $\tau$ is the radial coordinate on the conifold, measuring the coordinate distance away from its tip $\tau=0$. $\epsilon$ is the deformation parameter, controlling the curvature of the conifold at the tip.

The base of the conifold can be parameterized by angular coordinates. Equation~(\ref{dc}) uses the following basis of 1-forms on the base
\ber
\nonumber
g^1&=&\frac{e^1-e^3}{\sqrt{2}}, \ g^2=\frac{e^2-e^4}{\sqrt{2}},\\
\nonumber
g^3&=&\frac{e^1+e^3}{\sqrt{2}}, \ g^4=\frac{e^2+e^4}{\sqrt{2}},\\
g^5&=&e^5. \label{gb}
\eer
where
\ber
\nonumber
e^1&\equiv& -\sin\theta_1 d\phi_1,\  e^2\equiv d\theta_1,\\
\nonumber
e^3&\equiv& \cos\psi\sin\theta_2d\phi_2-\sin\psi d\theta_2,\\
\nonumber
e^4&\equiv& \sin\psi\sin\theta_2d\phi_2+\cos\psi d\theta_2,\\
\nonumber
e^5&\equiv& d\psi + \cos\theta_1 d\phi_1+\cos\theta_2 d\phi_2. \label{eb}
\eer

The solution also contains a non-trivial $F_3$, the field strength of $C_2$, which can be written as 
\beq\label{f3k}
F_{3}=\frac{M\alpha'}{2}\{g^5\w g^3 \w g^4 +d[F(\tau)(g^1\w g^3+g^2 \w g^4] \},
\eeq
and a non-trivial $H_3$, conveniently defined through
\ber
B_{2}&=&\frac{g_s M \alpha'}{2}[f(\tau)g^1\w g^2+k(\tau)g^3\w g^4],\\
\nonumber
H_{3}&=&dB_2=\frac{g_s M\alpha'}{2}\Big[d\tau(f'(\tau)g^1\w g^2+\\&&+k'(\tau)g^3\w g^4)+\frac{1}{2}(k(\tau)-f(\tau))g^5\w(g^1\w g^3+g^2\w g^4)\Big]. \label{h3k}
\eer
In the above equations $g_s$ is the string coupling constant, $\alpha'=M_{\rm Pl}^{-2}$ is the string scale parameter and $M$ is an integer explained below.

Finally, there is a self-dual 5-form, which is decomposed as 
\beq\label{f5k}
\tilde{F}_5=(1+\ast)\mathcal{F}_5 \,,
\eeq
where
\beq
\mathcal{F}_5=\frac{g_s M^2(\alpha')^2}{4}l(\tau)g^1\w g^2 \w g^3\w g^4\w g^5\,,
\eeq
with
\beq
l=f(\tau)(1-F(\tau))+k(\tau)F(\tau).
\eeq

The above ansatz for the differential form is written in terms of functions $F(\tau)$, $f(\tau)$,  $k(\tau)$, $l(\tau)$ and $h(\tau)$. The explicit form of these functions is given by~\cite{Klebanov:2000hb}
\ber
F(\tau)&=&\frac{\sinh\tau-\tau}{2\sinh\tau},\label{Ftau}\\
f(\tau)&=&\frac{\tau \coth\tau-1}{2\sinh\tau}(\cosh\tau-1),\label{ftau}\\
k(\tau)&=&\frac{\tau\coth\tau-1}{2\sinh\tau}(\cosh\tau-1),\label{ktau}\\
l(\tau)&=&\frac{\tau\coth\tau-1}{4\sinh^2\tau}(\sinh2\tau-2\tau),
\eer
while the warp factor is found from an integral
\beq
\label{warp}
h(\tau)=(g_s M \alpha')^2 2^{2/3}\epsilon^{-8/3}I(\tau)\,,
\eeq
with
\beq\label{eI}
I(\tau)\equiv\int_\tau^\infty dx \frac{x\coth x -1}{\sinh^2 x}(\sinh 2x-2x)^{1/3}.
\eeq
The asymptotic behavior of this integral is
\ber
I(\tau)& \rightarrow & 3\cdot2^{-1/3}\bigg(\tau-\frac{1}{4}\bigg)e^{-{4\tau}/3}\,, \qquad \tau \rightarrow\infty\,, \\
I(\tau) & \rightarrow & I_0+\mathcal{O}(\tau^2)\,, \qquad \tau \rightarrow 0\,,
\eer
with $I_0\approx0.71805$.


\subsection{Field theory interpretation and glueball states}
\label{sec:glueballs}

Let us also briefly discuss the field theory interpretation of the dual geometry and explain how the spectrum of light particles of the gauge theory can be extracted from it. First, the metric explicitly breaks the conformal group $SO(4,2)$ of isometries of $AdS_5$ and hence corresponds to a non-conformal theory. It is only compatible with $\mathcal{N}=1$ supersymmetry. This geometry can be obtained from $N$ D3-branes and $M$ $D5$ branes, $N\gg M$, such that four of the spacetime dimensions of the D5 coincide with those of the D3, and the remaining two are wrapped around the $S^2\in T^{1,1}$.  Consequently, the dual theory is a SYM theory with $SU(M+N)\times SU(N)$ gauge group and global symmetry $SU(2)\times SU(2)\times U(1)_B$. 

The gauge theory is coupled to two chiral superfields,  $A_1$, $A_2$,  in the representation ${(M+N,\overline{N})}$ and two anti-chiral superfields, $B_1$, $B_2$, in the representation $({\overline{M+N},{N}})$. The chiral and antichiral superfields transform as doublets of the respective $SU(2)$ factor, while  $U(1)_B$ is the baryon symmetry that acts as $A_i\rightarrow e^{i\alpha}A_i$ and $B_i\rightarrow e^{-i\alpha}B_i$. The theory has a superpotential of the form
\beq
\mathcal{W}=\lambda\epsilon^{ik}\epsilon^{jl}\, \tr{A_i B_k A_j B_l}\,.
\eeq

The gauge couplings of the two factors of the gauge group flow in opposite 
directions and the theory exhibits a ``cascade'' of Seiberg dualities whenever 
one of the couplings becomes infinitely strong: the spectrum of the theory 
changes and the direction of the flow flips. At the IR end of the cascade the 
theory becomes a strongly coupled $SU(M)$ SYM with light excitations 
corresponding to bound states of gluons and gluinos (the glueballs). The theory 
also possesses a non-zero condensate spontaneously breaking the baryon symmetry, 
which comes with an associated tower of light mesons and hybrid states of mesons 
and glue~\cite{Gubser:2004tf,Gubser:2004qj,Benna:2006ib}. Below we will refer to 
all the above as glueballs.

The particle spectrum can be found from the poles of two-point correlation functions. In the AdS/CFT dictionary~\cite{Gubser:1998bc,Witten:1998qj}, the two-point functions can be computed from the solutions of the equations of type IIB supergravity, linearized over the background solution. See~\cite{Csaki:1998qr,Brower:2000rp} for early examples of the glueball spectrum calculations. Our purpose here will be to select the perturbations of the type IIB fields corresponding to pseudoscalar modes, singlet under the global symmetry. The particle spectrum is commonly classified by the $J^{PC}$ quantum numbers, where $J$ is the particle spin, $P$ its parity and $C$ the charge conjugation. Hence we will be interested in $0^{-+}$ states.

In the next section we will discuss how the $J^{PC}$ quantum numbers can be determined for the supergravity fluctuations. We will also summarize our expectations about the dimensions of the dual gauge theory operators.

\section{Symmetries and quantum numbers of glueballs}
\label{sec:PC}

\subsection{Quantum numbers}
\label{sec:qnumbers}

In this section we will review how the quantum numbers of the glueball states are determined in the holographic approach. We refer the reader to~\cite{Csaki:1998qr,Brower:2000rp} for some original literature.

The matter sector of the Klebanov-Strassler theory has $SU(2)\times SU(2)\times U(1)_B$ continuous global symmetry. The particle states should be classified by its irreducible representations. For example, under the $SU(2)\times SU(2)$, the states are classified by a pair of half-integer numbers $(j_1,j_2)$. They also carry charge under $U(1)_B$ baryon symmetry.

The glueball states of the Yang-Mills sector of the theory are singlets with respect to $SU(2)\times SU(2)$, so we will be interested in all states with $j_1=j_2=0$. They also carry no baryon number. One should keep in mind, however, that this sector mixes with ``hybrid'' glueballs, containing $A$ and $B$ fields, charged under $U(1)_B$. Due to the presence of these fields there is also a large non-singlet sector, classified by the $A$ and $B$ composition of the hybrids. Invariant combinations of $A$ and $B$ also contribute to the singlet sector.

The axial symmetry $U(1)_A$ and the $U(1)_R$ symmetry of the SUSY algebra, are anomalous in the KS theory~\cite{Klebanov:1998hh}. The $U(1)_R$ is broken down to a $\mathbb{Z}_{2M}$ subgroup~\cite{Klebanov:2002gr}. The vacuum further breaks the remaining symmetry spontaneously down to $\mathbb{Z}_2$.\footnote{Note that $U(1)_B$ is also spontaneously broken in the vacuum by the expectation values of baryon operators~\cite{Gubser:2004qj}.} Nevertheless, because it is only broken by the anomaly, $U(1)_R$ remains a convenient symmetry to classify the supermultiplet structure of the states. Since the superpotential must have R-charge $2$, and supercoordinates transform as $\vartheta\to {\rm e}^{i\alpha}\vartheta$ under the R-symmetry, $A$ and $B$ have charge $1/2$, while the charge of the components is defined in such a way that the transformation of the superfield is homogeneous.

Let us now discuss the discrete symmetries. Realization of parity in the Klebanov-Strassler theory is straightforward. It reflects the spatial coordinates of the Minkowski factor,
\beq
{\mathcal P}:\qquad \vec{x} \ \to \ - \vec{x}\,.
\eeq
Charge conjugation involves complex conjugation of the fields and respectively, of their representations. Since fields $A$ and $B$ belong to $(N+M,\overline{N})$ and $(\overline{N+M},N)$ representations of the gauge group, respectively, one can combine the charge conjugation with an exchange of $A$ and $B$. Note that the superpotential is odd under this transformation, $\mathcal{W}\to - \mathcal{W}$. If, together with the exchange of the fields, the supercoordinates are rotated, $\vartheta\to i\vartheta$, the combined transformation will be a symmetry of the action:
\beq
{\mathcal I}: \qquad 
\begin{array}{c}
     A  \\
     B  \\
     \vartheta
\end{array}
\ \to \ 
\begin{array}{c}
     \sqrt{i}\,\overline{B}  \\
     \sqrt{-i}\,\overline{A} \\
     i\bar{\vartheta}
\end{array}\,.
\eeq
Following~\cite{Gubser:2004tf} we call this ${\mathcal I}$-symmetry. It can also be understood from the embedding to the ${\mathcal N}=4$ theory, where $A$ and $\bar{B}$ are combined into the fundamental multiplet of the $SU(4)$ R-symmetry. $\Ic$-symmetry is an unbroken $\mathbb{Z}_2$ subgroup of that symmetry, mixing the factors of the continuous $SU(2)\times SU(2)$. We note that on the pure gauge sector,  $\Ic$-symmetry acts simply as a charge conjugation. So, for the purpose of this paper, $C$ will be the eigenvalue of the $\Ic$ operation.

We now come to the discussion of the realization of the above symmetries in the type IIB SUGRA. First of all, the continuous $SU(2)\times SU(2)$ is the isometry of the two $S^2$ factors of conifold metric~(\ref{dc}). Since we are interested in the singlet sector, we need to define a basis of differential forms that is invariant under the isometries.

\begin{itemize}
    \item One forms. The only $SU(2)\times SU(2)$-invariant one-form on $T^{1,1}$ is $g^5$~(\ref{gb}), so the full basis is provided by
    \beq
    \label{1forms}
    \left\{dx^\mu,d\tau,g^5\right\}.
    \eeq
    \item Antisymmetric two-forms. There are four invariant two-forms on $T^{1,1}$ and the full basis is provided by 
    \beq
    \label{2forms}
    \left\{g^1\wedge g^2,g^3\wedge g^4,g^1\wedge g^3+g^2\wedge g^4,dg^5,\cdots\right\},
    \eeq
    where dots stand for external products of invariant one-forms~(\ref{1forms}).
    
    \item Symmetric two-forms are needed to construct metric fluctuations. The corresponding basis is provided by
    \beq
    \label{S2forms}
    \left\{(g^1)^2+(g^2)^2,(g^3)^2+(g^4)^2,g^1\cdot g^4-g^2\cdot g^3,\cdots\right\},
    \eeq
    where dots denote terms obtained from internal products of invariant one-forms~(\ref{1forms}).
    \item All higher rank invariant antisymmetric forms can be obtained by evaluating the exterior products of the forms listed above.
\end{itemize}

The $U(1)_R$ symmetry acts by shifts of the coordinate $\psi$ on $T^{1,1}$, $\psi\to \psi+\zeta$. The metric and $F_3$ form have an explicit dependence on $\psi$, which means that this symmetry is broken by the KS background. The dependence is compatible with the anomaly~\cite{Klebanov:2002gr}. In particular, since $\psi$ is a double cover of a circle, there is a remaining $\mathbb{Z}_2$ symmetry $\psi\to\psi +2\pi$.

The $U(1)_B$ symmetry is not realized as an action on $T^{1,1}$. As $U(1)_R$, it is not an isometry either, because it is spontaneously broken. Consequently, it generates a one-parametric family of deformations away from the KS background~\cite{Gubser:2004qj}. This family is called the baryonic branch~\cite{Butti:2004pk}.

The parity $P$ in gravity theory is realized as inversion of the sign of the purely spatial coordinates $\vec{x}$, but also as an action on the internal coordinates (which is a remanence of the higher-dimensional parity of ten-dimensional string theory). This implies that some gravity fields also transform. We assume that parity acts on the angular coordinates as
\beq
\phi_i\ \to\ \phi_i+\pi\,, \qquad \theta_i\ \to \ \pi - \theta_i\,, \qquad \psi \ \to \ 2\pi - \psi\,, \qquad i\ = \ 1,2\,.
\eeq
Comparing this with the background solution, parity is a conserved quantity if $B_2$ ($H_3$) and $C_4$ ($F_5$) have negative ``intrinsic'' parity. Besides, the ``axion'' $C$ is a pseudoscalar.

$\Ic$ symmetry is an internal symmetry of the gauge theory. It acts only on the $T^{1,1}$ part of the geometry exchanging two $S^2$ spheres within $T^{1,1}$. In terms of the coordinates, it swaps
\beq
\theta_1\ \leftrightarrow\  \theta_2\,, \qquad \phi_1\ \leftrightarrow\  \phi_2\,.
\eeq
Besides, the $\Ic$ symmetry flips the sign of the $F_3$ and $H_3$ forms.

It is useful to classify the invariant differential forms according to their $P$ and $\Ic$ transformations. We summarize the charges of the forms on $T^{1,1}$ in table~\ref{tab:charges}.

\begin{table}[htb]
    \centering
    \begin{tabular}{c|c|c|c}
        Form & $P$ & $C$ & ${\cal{R}}$  \\
        \hline $g^5$ & - & + & 0 \\
        $(g^1)^2+(g^2)^2+(g^3)^2+(g^4)^2$ & +  & + & 0\\
        $(g^1)^2+(g^2)^2-(g^3)^2-(g^4)^2$ & + & + &  $\pm2$\\
        $g^1\cdot g^4- g^2\cdot g^3$ & - & + & $\pm2$\\
        $g^1\wedge g^2+g^3\wedge g^4$ & - & - & $0$\\
        $g^1\wedge g^2-g^3\wedge g^4$ & - & - &$\pm2$ \\
        $g^1\wedge g^3+ g^2\wedge g^4$ & + & - & $\pm2$
    \end{tabular}
    \caption{Parity, $\Ic$ and ${\cal R}$ charges of the differential forms.}
    \label{tab:charges}
\end{table}


\subsection{Dual operators}
\label{sec:modes}

Knowing the parity and $\Ic$ transformations of the forms, it is straightforward to construct an $SU(2)\times SU(2)$ invariant ansatz for the $0^{-+}$ modes. We will do this in section~\ref{sec:pflt}. Before that we can anticipate the spectrum of operators, which will appear in the $0^{-+}$ sector, by looking at the superconformal structure of the modes on $AdS^5\times T^{1,1}$ studied in~\cite{Ceresole:1999ht}.

It is useful to establish the charges of the forms under the R-symmetry. Since only $\psi$ coordinate is transformed, the charge depends on the degree of the trigonometric function, which appears in the form. It is easy to check, that the linear combinations of forms shown in table~\ref{tab:charges} can be assigned either zero R-charge, or ${\cal R}=\pm 2$.

On should be looking for operators of rational dimension, in particular those, corresponding to $\tr \lambda\lambda$ and $\tr F_{\mu\nu}F^{\mu\nu}$ and $\tr F_{\mu\nu}\tilde{F}^{\mu\nu}$ that combine into short multiplets of supersymmetry. In the superconformal theory the dimensions of the operators in the short multiplets are protected. Indeed, the analysis of the conformal dimensions of the $0^{++}$ modes studied in~\cite{Berg:2006xy} showed that all of them have integer dimensions~(table~\ref{tab:dimsApreda}) and the spectrum contains $\Delta=3$ and $\Delta=4$ modes. This analysis was originally done in~\cite{Apreda:2003gc}.

\begin{table}[htb]
\begin{center}
\begin{tabular}{l|ccccccc}
\hline
Mode & $y$ & $N_2$ & $s$ & $\Phi$ & $f$ & $N_1$ & $q$  \\ \hline
$AdS_5$ mass, $m_5^2$ & $-3$ & $-3$ & $0$ & $0$ & $12$ & $21$ & $32$  \\
Dimension, $\Delta$ & $3$ & $3$ & $4$ & $4$ & $6$ & $7$ & $8$    \\ 
\hline
\end{tabular}
\end{center}
\caption{Spectrum of $SU(2)\times SU(2)$ singlet $0^{++}$ scalar operators in the KS theory~\cite{Apreda:2003gc}.}
\label{tab:dimsApreda}
\end{table}

From the analysis of the spectrum of Kaluza-Klein modes~\cite{Kim:1985ez,Ceresole:1999ht,Apreda:2003gc} one knows that the $\Delta=4$ modes come from the fluctuations of $B_2$ or $C_2$ proportional to the $S^2$ volume form $\omega_2=g^1\wedge g^2 + g^3\wedge g^4$. Besides, fluctuations of the dilaton and the axion are $\Delta=4$. Since dilaton $\Phi$ and axion $C$, as well as $B_2$ and $C_2$ have opposite parity, we expect to have two pseudoscalar modes of dimension $\Delta=4$. These modes have ${\cal R}=0$.

One pair of dimension three modes comes from the fluctuations of the metric, proportional to  $(g^1)^2+(g^2)^2-(g^3)^2-(g^4)^2$ and $g^1\cdot g^4 - g^2\cdot g^3$. The first one is scalar, and the second - pseudoscalar. Another pair of $\Delta=3$ modes comes from the fluctuations of the 3-form potentials proportional to $g^1\wedge g^3 + g^2\wedge g^4$ and $g^1\wedge g^2 - g^3\wedge g^4$. Again, one of these forms is parity-even and the other parity-odd. Besides these modes have ${\cal R}=\pm 2$. Therefore, we expect two $\Delta=3$ modes in the pseudoscalar sector.

One can expect the $\Delta=3$ modes to pair up with $\Delta=4$ modes and form chiral multiplets of ${\cal N}=1$ supersymmetry. This is compatible with the structure of short superconformal multiplets of~\cite{Ceresole:1999ht}. In particular, shortened Vector Multiplets III and IV of~\cite{Ceresole:1999ht} contain $\Delta=3$ and $\Delta=4$ operators. The two multiplets differ by the sign of the R-charge of the lowest component, ${\cal R}=-2$ for type III and ${\cal R}=2$ for type IV.

From the analysis of~\cite{Apreda:2003gc} one also observes modes with dimension $\Delta=7$, $\Delta=8$ and $\Delta=6$. The first one comes from the second possible combination of the fluctuations of the 3-form potentials, proportional to $g^1\wedge g^3 + g^2\wedge g^4$ and $g^1\wedge g^2 - g^3\wedge g^4$, while the other two modes come from linear combinations of fluctuations of traces of the metric on $AdS^5$ and on $T^{1,1}$. Out of singlet superconformal multiplets of~\cite{Ceresole:1999ht} only Vector Multiplet II can accommodate scalars with such a high dimension. This multiplet is not short, but nevertheless has a rational dimension. It corresponds to an unconstrained vector multiplet $V$ of ${\cal N}=1$ symmetry, which accommodates four spin zero fields. Without conformal symmetry, this multiplet decomposes into on-shell massive vector mutliplet and two massive chiral multiplets.  The vector component of the vector multiplet of dimension $\Delta=7$ was found in~\cite{Gordeli:2009nw}. To complete the Vector Multiplet II, one is missing a $\Delta=7$ scalar and the longitudinal part of the vector mode.

Below we will explicitly identify the pseudoscalar modes and reproduce their spectrum from the linearized equations. Further details of the multiplet structure and the explicit form of dual operators can be found in~\cite{Ceresole:1999ht,Apreda:2003gc,Gordeli:2013jea}.


\section{Singlet Pseudoscalars of the KS theory}
\label{sec:pseudoscalars}


\subsection{Ansatz for the modes}
\label{sec:pflt}

In this section we construct the ansatz for the $SU(2)\times SU(2)$ singlet $0^{-+}$ modes in the KS theory. From the $P$ and $\Ic$ transformation properties the most general form of the ansatz is
\ber
\delta (ds^2_{T^{1,1}})&=&B(g^1\cdot g^4-g^2\cdot g^3),\label{dg13}\\
\delta (ds^2_5) &=& (*_4{d a}+A d\tau ) \wedge g^5,\label{daA}\\
\delta C&=&C, \label{C}\\
\delta C_{2}&=&C_{2}^{-}(g^{1}\wedge g^{2}-g^{3}\wedge
g^{4})+C_{2}^{+}(g^{1}\wedge g^{2}+g^{3}\wedge g^{4}),\label{cpcm}\\
\delta B_{2}&=&B_{2} \left( g^{1}\wedge g^{3}+g^{2}\wedge
g^{4}\right)\label{b2},
\eer
\begin{multline}
\delta F_{5} \ = \ \frac{lG^{55}}{2}\left[\left(\partial_{\mu}(a +\phi_{1})dx^{\mu}+( A+\phi _{2}) d\tau\right) \wedge g^{1}\wedge g^{2}\wedge g^{3}\wedge
g^{4} - \right. 
\\
 - \sqrt{-G}\left( G^{11}G^{33}\right)^{2}\left(h^{1/2}  \ast_4d( a -\phi _{1}) \wedge d\tau  +
  (A -\phi _{2}) 
 G^{55}d^{4}x\right)\wedge g^{5} -
\\
\left. - h^{1/2}\sqrt{-G} 
G^{11}G^{33}\left( G^{55}\right)^{2}*_4d\phi
_{3} \wedge dg^{5}+\partial_{\mu}\phi _{3}dx^{\mu}\wedge d\tau
\wedge g^{5}\wedge dg^{5}\right].
\label{df5}
\end{multline}
Here we have considered fluctuations of the metric, R-R scalar, R-R 2-form, NS-NS 2-form potential and R-R 5-form, respectively.
$*_4$ is the Hodge star operator in 4-dimensions with $d*_4d P=\square_4 P=-m^2 P$. $G^{11}$, $G^{33}$ and $G^{55}$ are the inverse coefficients of the $(g^{11})^2$, $(g^{33})^2$ and the $(g^{55})^2$ terms in metric~(\ref{dc}) and $h$ is the warp factor~(\ref{warp}). The fluctuation of $F_5$ looks so complicated because it is constructed to satisfy the self-duality condition.

The ansatz is constructed in terms of ten modes: $B$, $B_{2}$, $C_{2}^{+}$, $C_{2}^{-}$, $C$, $\phi _{1},$ $\phi _{2},$ $\phi _{3}$, $a$ and $A$. However, we expect that there are only six physical modes, whose mass spectrum should match six of the seven towers of scalars in~\cite{Berg:2005pd,Berg:2006xy}. In particular, Bianchi identity~\eqref{bif5} allows to solve explicitly for two modes,
\begin{equation*}
\phi _{1}=a-\frac{\left( h^{1/2}\sqrt{-G}\left( G^{55}\right)
^{2}G^{33}G^{11}\phi _{3}\right) ^{\prime }}{h^{1/2}\sqrt{-G}\left(
G^{33}G^{11}\right) ^{2}}\qquad \text{and} \qquad \phi _{2}=A+\frac{h^{1/2}G^{55}}{%
G^{33}G^{11}}\square _{4}\phi _{3}.
\end{equation*}%
So modes $\phi_1$ and $\phi_2$ can be dropped in favor of $\phi_3$, $A$ and $a$.

Below we will write the linearized equations of type IIB supergravity over the KS background for the remaining eight fluctuations and show that the equations indeed describe six independent modes.


\subsection{Linearized Equations}
\label{sec:eqs}

We will spare the reader the details of the derivation of the linearized equations and simply present the result. We will discuss the consistency of the derived system in section~\ref{sec:gaugesym}. Thus, the only independent equations generated by fluctuations~\eqref{dg13}-\eqref{df5} are the following:
\begin{scriptsize}
 \begin{multline}
  \left( \frac{27IK^{4}\sinh ^{2}\tau }{32}\left( \frac{I^{\prime
}K^{6}\sinh ^{2}\tau }{I^{3/2}}\phi _{3}\right) ^{\prime }\right) ^{\prime }+%
\frac{3I^{\prime }K^{4}\sinh ^{2}\tau \phi _{3}}{4I^{1/2}}\left( \frac{%
9IK^{4}\sinh ^{2}\tau }{8}\tilde{m}^{2}-1\right)-\\
-\left( \frac{3I^{\prime}K^{4}\sinh ^{2}\tau }{4I^{1/2}}a\right) ^{\prime }+
\frac{3I^{\prime }K^{4}\sinh ^{2}\tau A}{4I^{1/2}}+\frac{2^{1/3}I^{\prime
}B_{2}}{K}+\left( \frac{2^{1/3}I^{\prime }}{K}\right) ^{\prime
}C_{2}^{-}
+\left( \frac{2^{1/3}I^{\prime }\cosh \tau }{K}\right) ^{\prime}C_{2}^{+}=0;   
\label{eqf3}
 \end{multline}
\begin{multline}
\left( \frac{\cosh ^{2}\tau +1}{I\sinh ^{2}\tau }C_{2}^{-\prime }\right)
^{\prime }-\frac{C_{2}^{-}}{I}+\frac{\left( \cosh ^{2}\tau +1\right) \tilde{m%
}^{2}C_{2}^{-}}{K^{2}\sinh ^{2}\tau }+\left( \frac{2\cosh \tau }{I\sinh
^{2}\tau }C_{2}^{+\prime }\right) ^{\prime }+\frac{2\cosh \tau \tilde{m}%
^{2}C_{2}^{+}}{K^{2}\sinh ^{2}\tau }-\\
-\left( \frac{2^{1/3}I^{\prime }B}{%
2I^{3/2}K^{2}\sinh ^{2}\tau }\right) ^{\prime }+  \frac{K^{2}B}{2^{1/3}I^{3/2}}-\frac{\tau }{2\sinh \tau }\left( \frac{C}{I}%
\right) ^{\prime }+\frac{I^{\prime }B_{2}}{I^{2}}+\frac{2^{1/3}3}{8}\left( \frac{K^{2}}{I^{3/2}}\left( \frac{%
I^{\prime }}{K}\right) ^{\prime }A\right) ^{\prime }+\\ +\frac{2^{1/3}3I^{\prime }KA%
}{8I^{3/2}}+\left( \frac{I^{\prime }%
}{K}\right) ^{\prime }\frac{3\tilde{m}^{2}a}{2^{2/3}4I^{1/2}}+  \frac{27I^{\prime }K^{6}\sinh ^{2}\tau \tilde{m}^{2}\phi _{3}}{%
2^{2/3}32I^{3/2}}\left( \frac{I^{\prime }}{K}\right) ^{\prime }=0;
\label{eqC2p}
\end{multline}
\begin{multline}
\label{eqC2m}
\left( \frac{\cosh ^{2}\tau +1}{I\sinh ^{2}\tau }C_{2}^{+\prime }\right)
^{\prime }+\frac{\left( \cosh ^{2}\tau +1\right) \tilde{m}^{2}C_{2}^{+}}{%
K^{2}\sinh ^{2}\tau }+\left( \frac{2\cosh \tau C_{2}^{-\prime }}{I\sinh
^{2}\tau }\right) ^{\prime }+\frac{2\cosh \tau \tilde{m}^{2}C_{2}^{-}}{%
K^{2}\sinh ^{2}\tau }
-\left( \frac{C}{2I}\right) ^{\prime }+\\+\frac{2^{1/3}3}{8}\left( \frac{K^{2}%
}{I^{3/2}}\left( \frac{I^{\prime }}{K}\cosh \tau \right) ^{\prime }A\right)
^{\prime }+\left( \frac{I^{\prime }}{K}\cosh \tau \right) ^{\prime }\frac{3%
\tilde{m}^{2}a}{2^{2/3}4I^{1/2}}-\left( \frac{2^{1/3}I^{\prime }\cosh \tau B}{%
2I^{3/2}K^{2}\sinh ^{2}\tau }\right) ^{\prime }+\\ +\frac{27I^{\prime }K^{6}\sinh ^{2}\tau 
\tilde{m}^{2}\phi _{3}}{2^{2/3}32I^{3/2}}\left( \frac{I^{\prime }}{K}\cosh
\tau \right) ^{\prime }=0;
\end{multline}
\begin{eqnarray}
&&B_{2}^{\prime \prime }-\frac{I^{\prime }B_{2}^{^{\prime }}}{I}-\frac{%
\left( \cosh ^{2}\tau +1\right) B_{2}}{\sinh ^{2}\tau }+\frac{\tilde{m}%
^{2}IB_{2}}{K^{2}}+\frac{I\sqrt{K^{3}\sinh \tau }}{2^{1/3}}\left( \sqrt{%
\frac{K}{I^{3}\sinh \tau }}B\right) ^{\prime }+ \frac{3I^{1/2}I^{\prime }\tilde{m}^{2}a}{2^{8/3}K}+ \notag \\\nonumber
&&+\frac{3I^{\prime }I\sinh
^{2}\tau }{2^{8/3}K}\left( \frac{K^{2}}{I^{3/2}\sinh ^{2}\tau }A\right)
^{\prime }+\frac{2^{2/3}I^{\prime
}I}{4K}\left( \frac{C}{I}\right) ^{\prime }+\frac{I^{\prime }C_{2}^{-}}{I}+\\\label{eqb2}&&
+
\frac{2^{1/3}27I^{\prime 2}K^{5}\sinh ^{2}\tau \tilde{m}^{2}\phi _{3}}{%
64I^{1/2}}=0.
\end{eqnarray}%
\begin{eqnarray}
&&C^{\prime \prime }+\frac{2\left( K\sinh \tau \right) ^{\prime }C^{\prime }%
}{K\sinh \tau }+\frac{\tilde{m}^{2}IC}{K^{2}}+\left( \frac{I^{\prime \prime }%
}{I}+2\frac{I^{\prime }}{I}\frac{\left( K\sinh \tau \right) ^{\prime }}{%
K\sinh \tau }\right) C-\frac{2^{5/3}I^{\prime }B}{I^{3/2}K\sinh ^{2}\tau }-%
\frac{3K^{6}A}{2I^{1/2}}+\frac{2^{4/3}\tau C_{2}^{-\prime }}{IK^{2}\sinh
^{3}\tau }+  \notag \\
\label{eqc}
&&+\frac{2I^{\prime }C_{2}^{-}}{IK^{3}\sinh ^{3}\tau }+\frac{%
2^{4/3}C_{2}^{+\prime }}{IK^{2}\sinh ^{2}\tau }-\frac{2}{IK^{2}\sinh
^{2}\tau }\left( \frac{I^{\prime }}{K}B_{2}\right) ^{\prime }=0;
\end{eqnarray}
\begin{eqnarray}
\nonumber
&&B^{\prime \prime }-\frac{I^{\prime }B^{\prime }}{I}+\frac{\tilde{m}^{2}IB}{%
K^{2}}+\frac{4B}{9K^{6}\sinh ^{2}\tau }+\frac{3B}{4}\left( \frac{I^{\prime }%
}{I}\right) ^{2}+\frac{K^{\prime 2}B}{K^{2}}+\frac{\left( K^{2}\coth \tau
\right) ^{\prime }B}{K^{2}}+\frac{2^{4/3}I^{\prime }\cosh \tau C_{2}^{+\prime }}{I^{1/2}K^{2}\sinh
^{2}\tau }+\\
\nonumber
&&+\left( \frac{I^{\prime \prime }}{2I}-\frac{%
2^{1/3}K^{4}}{I}+2\frac{I^{\prime }}{I}\frac{\left( K\sinh \tau \right)
^{\prime }}{K\sinh \tau }\right) B  + \frac{3IK\tilde{m}^{2}a}{2}+\frac{3I^{5/2}}{2K}\left( \frac{K^{4}}{I^{5/2}%
}A\right) ^{\prime }-\frac{2^{4/3}I^{\prime }K}{I^{1/2}}C-\\&&-2^{5/3}\sqrt{\frac{%
K}{I\sinh \tau }}\left( \sqrt{K^{3}\sinh \tau }B_{2}\right) ^{\prime }+\frac{%
2^{5/3}K^{2}C_{2}^{-}}{I^{1/2}}+\frac{2^{4/3}I^{\prime }C_{2}^{-\prime }}{%
I^{1/2}K^{2}\sinh ^{2}\tau }=0
\end{eqnarray}
\begin{eqnarray}
\nonumber
&&a^{\prime \prime }+\frac{2\left( K^{2}\sinh \tau \right) ^{\prime }a}{%
K^{2}\sinh \tau }-\frac{8a}{9K^{6}\sinh ^{2}\tau }+\left( \frac{I^{\prime
\prime }}{2I}+\frac{I^{\prime }}{I}\frac{\left( K^{2}\sinh \tau \right)
^{\prime }}{K^{2}\sinh \tau }+\frac{1}{4}\left( \frac{I^{\prime }}{I}\right)
^{2}\right) a-\frac{4B}{3K^{3}\sinh ^{2}\tau }-\\
\nonumber
&&-\frac{\left(
I^{1/2}K^{2}\sinh ^{2}\tau A\right) ^{\prime }}{I^{1/2}K^{2}\sinh ^{2}\tau }-\left( \frac{I^{\prime }}{K}\right) ^{\prime }\frac{2^{7/3}C_{2}^{-}}{%
3I^{1/2}K^{4}\sinh ^{2}\tau }-\left( \frac{I^{\prime }}{K}\cosh \tau \right)
^{\prime }\frac{2^{7/3}C_{2}^{+}}{3I^{1/2}K^{4}\sinh ^{2}\tau }-\\
&&-\frac{%
2^{7/3}I^{\prime }B_{2}}{3I^{1/2}K^{5}\sinh ^{2}\tau }-\frac{9I^{\prime }}{%
8I^{1/2}}\left( \frac{I^{\prime }K^{6}\sinh ^{2}\tau }{I^{3/2}}\phi
_{3}\right) ^{\prime }=0,  \label{eqa}
\end{eqnarray}
\begin{eqnarray}
\nonumber
&&-I\tilde{m}^{2}A+K^{2}\left( -\frac{I^{\prime \prime }}{I}+\frac{2}{\sinh
^{2}\tau }-2\frac{I^{\prime }}{I}\frac{\left( K\sinh \tau \right) ^{\prime }%
}{K\sinh \tau }\right) A+2I\left( \frac{K^{\prime }}{K}-\frac{1}{4}\frac{%
I^{\prime }}{I}\right) \tilde{m}^{2}a+I\tilde{m}^{2}a^{\prime }+\\
\nonumber
&&+\frac{4}{%
3I^{3/2}\sinh ^{2}\tau }\left( \frac{I^{3/2}B}{K}\right) ^{\prime }+ 
\frac{2^{5/3}K^{4}C}{3I^{1/2}}-\frac{2^{7/3}I^{\prime }C_{2}^{-}}{%
3I^{1/2}K^{3}\sinh ^{2}\tau }+\frac{2^{7/3}C_{2}^{-\prime }}{%
3I^{1/2}K^{2}\sinh ^{2}\tau }\left( \frac{I^{\prime }}{K}\right) ^{\prime }+\\
\nonumber
&&+
\frac{2^{7/3}C_{2}^{+\prime }}{3I^{1/2}K^{2}\sinh ^{2}\tau }\left( \frac{%
I^{\prime }}{K}\cosh \tau \right) ^{\prime }+\frac{2^{7/3}}{3I^{1/2}K^{2}\sinh ^{4}\tau }\left( \frac{I^{\prime }\sinh
^{2}\tau }{K}B_{2}\right) ^{\prime }-\\&&-\frac{9I^{\prime 2}K^{6}\sinh ^{2}\tau 
\tilde{m}^{2}\phi _{3}}{8I}=0.  \label{eqA}
\end{eqnarray}
\end{scriptsize}

Here, we Fourier transformed the modes, substituting $e^{ik\cdot x}$ for the spacetime coordinates with $k^2=m_4^2$. We also renormalized the mass eigenvalue,
\beq
\tilde{m}^2 \ = \ \frac{ 2^{5/3}}{3\epsilon^{4/3}} m_4^2\,.
\eeq
For the remaining parameters we used the convention $\alpha^{\prime}=g_s=1$ and $M=2$. As a result, we arrive at a system of eight coupled ODE's (seven second order and one first order) for eight unknown functions.

Equation~(\ref{eqf3}) comes from Bianchi identity~\eqref{bif5}, equations~(\ref{eqC2p}) and~(\ref{eqC2m}) come from the $F_{3}$ EoM~\eqref{f3eom}, equation~(\ref{eqb2}) is the  $H_{3}$ EoM~\eqref{h3eom}, equation~(\ref{eqc}) is the $C$ EoM~\eqref{ceom} and the remaining three equations come from the Einstein equations~\eqref{RMN}. One can show that not all of the eight equations are independent. For example, one can take $\tau$ derivative of equation~(\ref{eqA}) and find a linear combination of the remaining equations, equal to the result of the differentiation.

This system of equations can be simplified if one uses gauge invariance of the gravity equations~(see more details in section~\ref{sec:gaugesym}). One convenient gauge choice is to set $a=0$. Since the eight equations are linearly dependent, one can drop equation~(\ref{eqa}). Equation~(\ref{eqA}) is only algebraic equation for $A$. It can be solved, so that one remains with six second order ODE's for six unknown functions. 

 However, the choice $a=0$ is not convenient for the numerical analysis, because the coefficient of $A$ in equation~(\ref{eqA}) vanishes for finite $\tau$. The gauge that we will use for numerics below consists of choosing $A=0$ and keeping~\eqref{eqA} as an additional constraint. This constraint eliminates one of the seven modes leaving only six on shell.


\subsection{Asymptotic behavior of the solutions}
\label{sec:asymptotic}

Before solving the system numerically we need to carefully analyze the 
asymptotic behavior of the solutions. In particular, we need to identify 
fourteen linearly independent solutions of the second order equations, determine 
which of those are regular, and eliminate those that do not satisfy the first 
order constraint.\footnote{We have recently learned about a powerful generic 
method for approaching the asymptotic analysis introduced 
in~\cite{Elander:2010wd} and applied to similar systems 
in~\cite{Elander:2012yh,Elander:2017hyr,Elander:2017cle}.} The asymptotic 
analysis will also allow us to extract the dimensions of the dual gauge theory 
operators and compare them to the analysis made in section~\ref{sec:modes}. 

\subsubsection{UV}
We start from the analysis in the UV limit, $\tau\to\infty$. The coefficients are not analytic functions in this limit, so that equations can be organized as an expansion in 
\beq
\label{UVexpansion}
P(\tau)e^{\lambda_0\tau}\left(1 + O(e^{-2\tau/3})\right)\,,
\eeq
where $\lambda_0$ is the leading UV exponent, which will determine the dimension of the dual operator. $P(\tau)$ is some analytic function of $1/\tau$. We will construct the asymptotic solution only in the leading exponential order, determining a few first expansion coefficients of $P(\tau)$. 

The detailed results of the leading exponential expansion of the UV solutions can be found in appendix~\ref{sec:UV}. The asymptotic form of the seven second order equations can be read from formulas~(\ref{UVeq1})-(\ref{UVeq7}). The solutions to those equations are summarized in section~\ref{sec:UV-sing} (for dominant modes) and section~\ref{sec:UVreg} (for regular modes). In section~\ref{sec:UVcomp} we check the compatibility of the regular solutions with the first order constraint~(\ref{UVeq8}), which is the UV asymptotic form of equation~(\ref{eqA}). Let us give a simplified version of the analysis here.

One first observes that the modes roughly separate into two groups: one containing $\phi_3$, $C_2^+$, $C$ and $a$, and another with $C_2^-$, $B_2$ and $B$. (Specifically, one observes that for most of the modes, the modes within a group contribute the same order to the equations.) One can further decouple the pairs $\phi_3$ and $a$, $C_2^+$ and $C$, $C_2^-$ and $B_2$ and the single mode $B$. This decoupling is only approximate, as in most cases the modes of the decoupled equations induce the other modes, but it does correctly determine the leading exponents of the modes. In this way one is left with the following system
\ber
\label{phire}
{\phi_3}''+\frac{2{\phi_3}'}{3}-\frac{8\phi_3}{9}-\frac{8{a}'}{9} & = & 0\,,\\
\label{c2mre}
{C_2^-}''+\frac{4{C_2}'}{3}-C_2^--\frac{4B_2}{3}&=&0\,,\\
\label{c2pre}
{C_2^+}''+\frac{4{C_2^+}'}{3}-\frac{C'}{2}-\frac{2C}{3}&=&0\,,\\
\label{b2re}
{B_2}''+\frac{4{B_2}'}{3}-B_2-\frac{4C_2^-}{3}&=&0\,,\\
\label{cre}
C''+\frac{4C'}{3}&=&0\,,\\
\label{bre}
B''+\frac{4B'}{3}+\frac{B}{3}&=&0\,,\\
\label{are}
a''+\frac{2a'}{3} -2\phi_3' -\frac{4{\phi_3}}{3} & = & 0.
\eer
This decoupled system can be solved analytically:
\ber
\label{UVas1}
\phi_3&=& \alpha_1 e^{-2 \tau} + \alpha_2 e^{4 \tau/3} + \alpha_{14}e^{-2\tau/3}\,,\\
C_2^-&=& \alpha_4 e^{-{7 \tau}/{3}}+\alpha_5 e^{-\tau}+ \alpha_{13}e^{-\tau/3} + \alpha_6 e^\tau\,,\\
C_2^+&=&\alpha_7 e^{-{4 \tau}/{3}}+\alpha_8 \tau + \alpha_9\,,\\
B_2&=&\alpha_4 e^{-{7 \tau}/{3}}- \alpha_5 e^{-\tau} - \alpha_{13}e^{-\tau/3} + \alpha_6 e^\tau\,,\\
C&=& \alpha_{10} e^{-{4 \tau}/{3}}+2\alpha_{8} \,,\\
B&=&\alpha_{11} e^{-{\tau}/{3}}+\alpha_{12} e^{-\tau}\,,\\
a & = & -\alpha_1 e^{-2\tau} + \frac{3}{2} \alpha_2 e^{4 \tau/3} + \alpha_3 + \frac{3}{2}\alpha_{14}e^{-2\tau/3} .
\label{UVas7}
\eer
As expected there are fourteen independent modes. However, not all of them satisfy the constraint imposed by equation~(\ref{eqA}). In view of the above splitting of the modes one can write the asymptotic form of equation~(\ref{eqA}) as four simple constraints:
\ber
\label{favin}
a'-2\phi_3 & = & 0\,,\\
\label{b2c2mvin}
{B_2}'+{B_2}-{C_2^-}'-{C_2^-}&=&0\,,\\
\label{cc2pvin}
C-2 {C_2^+}'&=&0\,, \\
3B'-5B & = & 0\,.
\eer
Out of modes in equations~(\ref{UVas1})-(\ref{UVas7}) only $\alpha_1$, $\alpha_2$, $\alpha_3$, $\alpha_4$, $\alpha_5$, $\alpha_6$, $\alpha_8$ and $\alpha_9$ satisfy the above constraints. Besides, the following linear combination also of the two remaining modes also satisfies the constraints:
\ber
\alpha_7 & = & - \frac{3}{8}\alpha_{10}\,,
\eer
It turns out that more modes satisfy the constraints imposed by equation~(\ref{eqA}), but to see that, one has to go beyond the naive analysis given here. What will be important for the numerical analysis is that the appropriate generalization of modes
\beq
\alpha_1\,, \quad \alpha_4\,, \quad \alpha_5\,, \quad \alpha_7= -\frac{3}{8}\alpha_{10}\,, \quad \alpha_{10} = \frac{27}{16} \sqrt{\frac{3}{2}}m^2 \alpha_{14}\,, \quad \alpha_{12} = \frac{\alpha_5}{\sqrt{6\tau}}\,. 
\eeq
satisfy all the asymptotic equations.\footnote{The fact that the last relation is $\tau$-dependent simply means that the expansion of $B$ and $B_2$, $C_2^-$, that is  $P(\tau)$ in equation~(\ref{UVexpansion}), starts from different powers of $\tau$. This is not captured by the leading order analysis here, but can be seen explicitly from the full expansion in equation~(\ref{Rmode1}).} These are the six regular solutions, corresponding to the physical modes, as we confirm using equation~(\ref{UVaction}) in appendix~\ref{sec:UVcomp}.

Similarly, the following appear to be physical singular modes
\beq
\label{singmodes}
\alpha_2\,, \quad \alpha_3\,, \quad \alpha_6\,, \quad \alpha_8\,, \quad \alpha_{9}  \quad \alpha_{11} = 2\sqrt{\frac{2}{3}}{\sqrt{\tau}}\alpha_{13}\,. 
\eeq
From the exponents associated with these modes we can obtain the spectrum of the dual operators. For this one changes the radial variable according to 
\beq
\tau \ = \  -3\log {z}\,.
\eeq

After an appropriate field redefinition one can cast equations~(\ref{phire})-(\ref{are}) for every singular mode~(\ref{singmodes}) in the form
\beq
z^2f'' - 3zf' - m_5^2f \ = \ 0
\eeq
and use the standard formula,
\beq
\Delta \ =\ 2 + \sqrt{4+m_5^2}\,,
\eeq
to compute the dimensions. We summarize our dimensions in table~\ref{tab:dims}.

\begin{table}[htb]
\begin{center}
\begin{tabular}{l|cccccc}
\hline
Mode & $\alpha_2$ & $\alpha_3$ & $\alpha_6$ & $\alpha_{8}$ & $\alpha_9$ & $\alpha_{11}$  \\ \hline
$AdS_5$ mass, $m_5^2$ & $5$ & $-3$ & $21$ & $0$ & $0$ & $-3$  \\ 
Dimension, $\Delta$ & $5$ & $3$ & $7$ & $4$ & $4$ & $3$    \\ 
\hline
\end{tabular}
\end{center}
\caption{Five-dimensional masses of the pseudoscalar modes and the corresponding scaling dimensions of the dual operators.}
\label{tab:dims}
\end{table}

We see that the dimensions of the operators, in general, match our expectations. The only subtlety is the dimension $\Delta=5$ mode coming from $\phi_3$. By construction, this mode should correspond to the longitudinal part of the vector appearing in the Vector Multiplet II in the classification of~\cite{Ceresole:1999ht}. Consequently, a more natural dimension for it would be $\Delta=7$. However, one observes that the $\phi_3$ mode is unusual. It enters in the equations multiplied by the factor of $\Box_4$. Consequently, what couples to the operator of dimension 5 is $\Box_4\phi_3$, while $\phi_3$ itself should couple to an operator of dimension 7. The remaining modes complete the Vector multiplets II, III and IV in~\cite{Ceresole:1999ht}.

\subsubsection{IR}

The analysis for small $\tau$ is technically simpler as the solutions are analytic there. We look for the solutions as a power series expansion
\beq
\label{IRexpansion}
\beta_i\tau^{\gamma_i}\left(1+c_i\tau^2+c_j\tau^4+O(\tau^6)\right).
\eeq
More generally, the coefficients of the expansion can contain $\log\tau$ terms. We will see that all the relevant solutions (those that also satisfy constraint~(\ref{eqA})) will not contain logarithmic terms.

As in the large $\tau$ case, here we will give a simplified version of the analysis of the asymptotic solutions, while the full expansions are collected in appendix~\ref{sec:IRasympt}. For example, equations~(\ref{f3eqIR})-(\ref{AeqIR}) in appendix~\ref{sec:IRasympt} correspond to equations~(\ref{eqf3})-(\ref{eqA}) with the leading asymptotic of the coefficients for $\tau\to 0$. In order to find the set of $\gamma_i$, the exponents in the IR expansion of the modes, as in equation~(\ref{IRexpansion}) we consider the following simplified system: 
\ber
\tau^2\phi_3'' + 8\tau \phi_3' +10\phi_3 & = & 0\,, \label{eqf3IRs}\\
\tau^2{C_2^-}''- 2\tau{C_2^+}' - 2C_2^- - \frac{\tau}{6^{2/3}I_0^{1/2}}B' + \frac{3}{6^{2/3}I_0^{1/2}}B & = & 0 \,,\\
\tau^2{C_2^+}''- 2\tau{C_2^-}' + 2C_2^- + \frac{\tau}{6^{2/3}I_0^{1/2}}B' - \frac{3}{6^{2/3}I_0^{1/2}}B& = & 0 \,,\\
\tau^2 B_2'' - 2B_2 & = & 0 \,,\\
\tau C'' + 2C' & = & 0 \,,\\
3^{2/3}I_0^{1/2}\tau B'' - 2^{4/3}{C_2^+}' - 2^{4/3}{C_2^-}' & = & 0 \,,\\
\tau^2a''+ 2\tau a' - 2a & = & 0 \,. \label{eqaIRs}
\eer
The solution to this simplified system consists of the following modes:
\ber
\phi_3 & = & \beta_1\tau^{-5} + \beta_2\tau^{-2}\,, \label{Bmode12}\\
C_2^- & = & \beta_3\tau^{-2} + \beta_4\tau + \frac{3}{5}\beta_6\tau^3 + \frac{\beta_{11}}{2I_0} + \frac{2^{2/3}\beta_{12}}{27I_0}\tau\left(1+3\log\tau\right)\,, \\
C_2^+ & = & \beta_5 - \beta_3\tau^{-2} - \beta_4\tau  + \frac{2}{5}\beta_6\tau^3 - \frac{\beta_{11}}{2I_0} - \frac{2^{2/3}\beta_{12}}{27I_0}\tau\left(1+3\log\tau\right) \,, \\
B_2 & = & \beta_7\tau^{-1} + \beta_8\tau^2\,, \\
C & = & \beta_9\tau^{-1} + \beta_{10}\,,  \\
B & = & 6^{1/3}\beta_{11}+ 3^{1/3}\beta_{12}\tau+\beta_6\tau^3\,, \\
a & = & \beta_{13}\tau^{-2} + \beta_{14}\tau\,. \label{Bmode1314}
\eer
This solution demonstrates the leading modes present in the system. The complete solution of the asymptotic equations is constructed in appendix~\ref{sec:IRasympt}, where the modes are in general, linear combinations of the above modes.

We also have to identify, which of the fourteen modes are regular. This is defined with respect to the action functional for the modes, which appears in equation~(\ref{IRaction}) in section~\ref{sec:IRaction} of the appendix. Our analysis shows that all the modes $\beta_i$ with even $i$ are regular, besides all of them, except $\beta_{12}$ satisfy equation~(\ref{eqA}).


\subsection{Are the equations correct?}
\label{sec:gaugesym}

Before moving on to the numerics we would like to discuss the consistency of our analysis. Equations~(\ref{eqf3})-(\ref{eqA}) look fairly complicated and a reasonable question is how one can know that they are correct. 

Some non-trivial checks are provided by considerations that the eight equations are not all independent and by supersymmetry must describe only six physical towers of states. This is indeed true. Equation~(\ref{eqA}) is algebraic so that mode $A$ can be eliminated. Some of the remaining modes can be eliminated by a gauge choice, for example $a=0$, and the remaining seven equations can be shown to be linearly dependent. The analysis of the asymptotic solutions in the appendix also confirms this: we find that out of seven independent modes of the second order equations~(\ref{eqf3})-(\ref{eqa}) only six satisfy equation~(\ref{eqA}), which lead to six physical modes.

Gauge invariance provides another non-trivial constraint on the equations.\footnote{We thank A.~Dymarsky for suggesting this and checking some of our equations.} Since we did not fix the gauge, we can check, whether system~(\ref{eqf3})-(\ref{eqA}) is gauge (diffeomorphism) invariant. Infinitesimal diffeomorphisms are generated by Lie
derivatives, which for an arbitrary rank tensor read 
\begin{equation*}
\mathcal{L}_{\xi }T_{i_{1}\cdots i_{k}}=\xi ^{j}\partial _{j}T_{i_{1}\cdots
i_{k}}+\sum_{k=1}T_{j\cdots i_{k}}\partial _{i_{i}}\xi ^{j}+\cdots
+T_{i_{1}\cdots j}\partial _{i_{k}}\xi ^{j},
\end{equation*}%
where $\xi $ is a vector, which for the KS theory we can write as $\vec{\xi}=\tilde{a}%
\left( x,\tau \right) \hat{e}_{\psi }$. In terms of the modes one finds that
\begin{equation}
\begin{array}{lllll}
\delta C  = 0, && \delta a  = 2G_{55}\tilde{a},  && \delta B=-\frac{1}{2}\varepsilon ^{4/3}h^{1/2}K%
\tilde{a}, \\
&&&&\\
\delta C_{2}^{+}=\frac{%
\tilde{a}}{2},  && \delta A=2G_{55}\tilde{a}^{\prime },  &&  \delta C_{2}^{-}=\frac{1}{2}\left( 2F-1\right) \tilde{a}, \\
&&&&\\
\delta\phi
_{3}=0,  && \delta B_{2} =\frac{\left( k-f\right) }{2}\tilde{a}.  && 
\end{array}
\end{equation}%
One can check that the above transformations indeed generate a symmetry of system~\eqref{eqf3}-\eqref{eqA}.

Another check is provided by the analysis of the dimensions of the asymptotic behavior of the fields and by the dimensions of the dual operators. We find a set of integer dimensions, summarized in table~\ref{tab:dims}, which match our expectations.

Finally, the spectrum of the equations should provide one more consistency check of the equations. It is expected to match the eigenvalues of~\cite{Berg:2006xy} without the tower of~\cite{Gordeli:2009nw}. We will explain the results of our numerical analysis in section~\ref{sec:numerics}. Studying the spectrum of this particular system of seven coupled equations turns out to be a difficult exercise, and our success is only partial so far. In our numerical studies we are able to see four towers of eigenvalues found in~\cite{Berg:2005pd} and~\cite{Berg:2006xy}, whose numerical values match quite well our spectrum. As for the two remaining towers, a direct method, which we used, was not able to detect them. Instead, if we add another degree of freedom by removing the constraint, we see two additional towers that do not match the eigenvalues of~\cite{Berg:2006xy}, so their status is unclear. In the meantime, we see indications that our direct method might not have sufficient resolution, which might explain the absence of the remaining modes.


\section{Numerical analysis and spectrum}
\label{sec:numerics}

In this section we describe the results of our numerical analysis of the system of equations~(\ref{eqf3})-(\ref{eqA}). 

The basis of our analysis is the midpoint method, also used in~\cite{Berg:2006xy}, for the analysis of the $0^{++}$ glueballs. This technique is a generalization of the shooting method. In the regular shooting method, one replaces the boundary problem by the initial value problem and solves equations either going from IR (small $\tau$) to UV (large $\tau$) or vice versa. The midpoint method involves shooting from both the UV and IR ends and gluing the solutions at some intermediate point $\tau_{mid}$. Consequently, for $n$ equations for $n$ fields, one sets $2n$ regular initial conditions in the UV (for the fields and derivatives) and $2n$ regular initial conditions in the IR. The equations are solved until $\tau_{mid}$, where one builds a square matrix of dimension $2n\times2n$,
\beq\label{deta}
\gamma=
\left(
\begin{array}{cc}
a_{IR} & a_{UV}\\
\partial a_{IR} & \partial a_{UV}
\end{array}\right)
_{\tau =  \tau_{mid}},
\eeq
where $a_i$ is the set of fields, $i=1,\ldots,n$, and $\partial a_i$ is the set of derivatives. The eigenvalues of the system of equations are values of $\tilde{m}^2$, for which matrix $\gamma$ has a zero eigenvalue. In particular, the determinant of $\gamma$, as a function of $\tilde{m}^2$, will change the sign at the eigenvalue point.

In this work we will be solving seven equations for six initial conditions in the UV and the IR, satisfying a first order constraint. One possibility in this case is to remove one of the fields and its derivative from the matrix $\gamma$, to form a $12\times 12$ matrix. Alternatively, one can calculate a rectangular $\gamma$, $12\times 14$, keeping all the fields, and compute the singular value decomposition. At the eigenvalue point, the smallest singular value should have a zero. This can be observed as a cusp in singular value as a function of $m^2$. 

Yet another possibility that we will explore is to solve the system of seven equations with seven unconstrained regular initial conditions and look for zeroes of the determinant of a $14\times 14$ matrix. In this case one can expect an additional tower of eigenstates, which could be separated and removed when the spectrum is compared with that of~\cite{Berg:2006xy}.

We first discuss the results obtained by the calculation of singular values of the $12\times 14$ matrix. This approach, based on the analysis of the six physical modes showed the best convergence with the spectrum of $0^{++}$ for smaller mass values. From now on we renormalize mass parameter $\tilde{m}^2$ to the conventions of~\cite{Berg:2006xy}. 
\beq
{m}^2\ = \ 0.9409\tilde{m}^2\,.
\eeq

\begin{figure}
    \centering
    \includegraphics[width=0.9\linewidth]{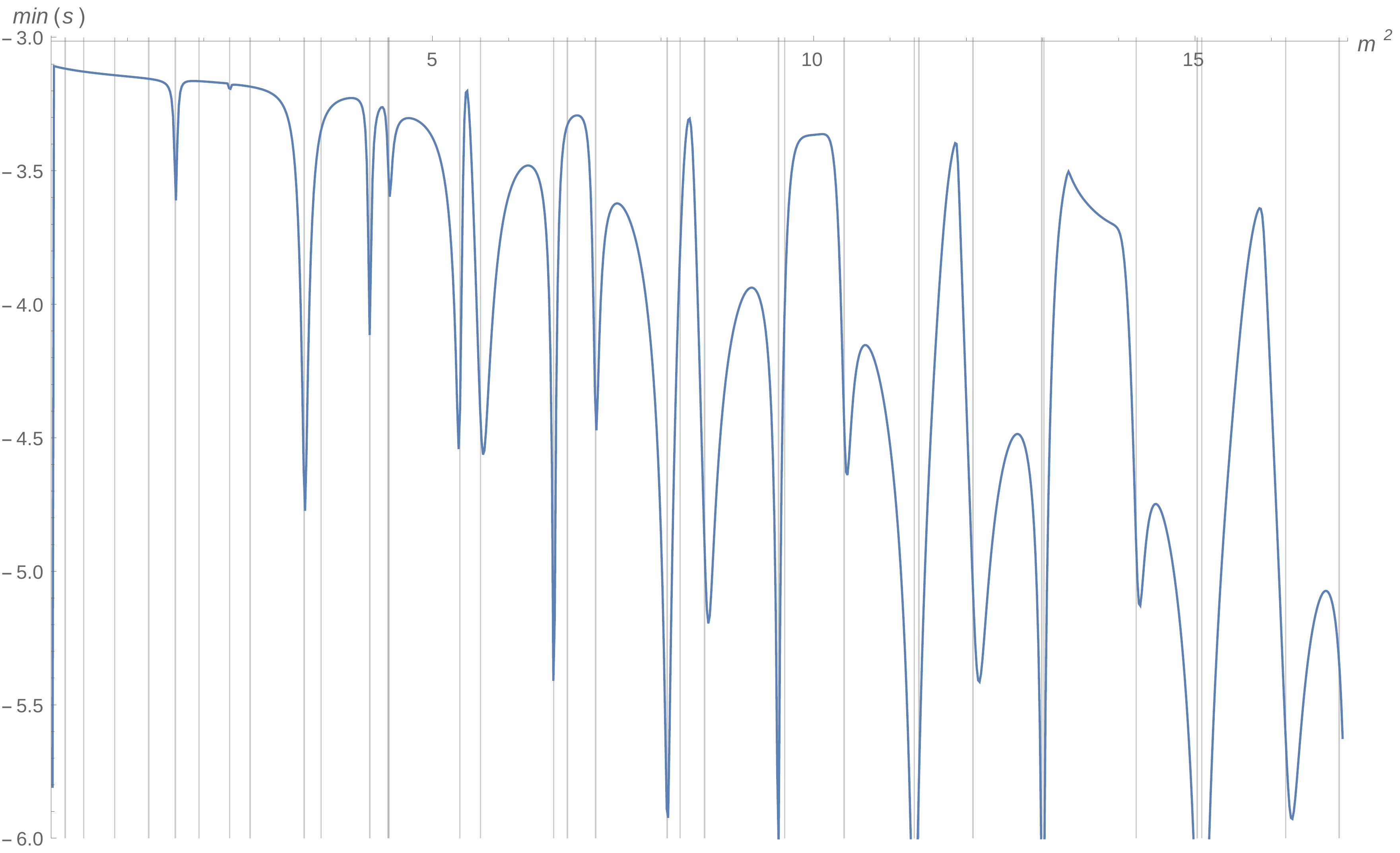}
    \caption{The spectrum of the lowest states ($m^2<17$) from four $0^{-+}$ towers seen from the the smallest singular value of the $12\times 14$ Wronskian matrix~(\ref{deta}). Vertical grid indicates the spectrum of the $0^{++}$ states of~\cite{Berg:2006xy}.}
    \label{fig:singvalspec}
\end{figure}

The results of the calculation can be found in table~\ref{tab:sing}, where the values are compared with the values of~\cite{Berg:2006xy}. We see that the found masses match rather well the values in~\cite{Berg:2006xy}, however, they do not reproduce the whole spectrum. See figure~\ref{fig:singvalspec}, for example. 

\begin{table}[htb]
\begin{center}
\resizebox{\textwidth}{!}{%
\begin{tabular}{|l|l||l|l|l|l||l|l|l|l|l|l|l|l|}
\hline 
\bf{$n$}                      & \bf{0}  & \bf{1}    & \bf{2}    & \bf{3}    & \bf{4}    & \bf{5}   & \bf{6}     & \bf{7}     & \bf{8} \\\hline 
${m}^2$  & 0.543*  & 1.635 & 2.34 & 3.33  & 4.18  & 4.44  & 5.345  & 5.67  & 6.59 \\\hline
${m}_{BHM}^2$  & 0.428  & 1.63  & 2.34  & 3.32  & 4.18  & 4.43 & 5.36  & 5.63      & 6.59 \\\hline\hline
\bf{$n$}   &       & \bf{9}    & \bf{10}    & \bf{11}    & \bf{12}    & \bf{13}    & \bf{14} & \bf{15}       & \bf{16}    \\ \hline
${m}^2$   &  &  7.15 & 8.08  &  8.63 & 9.535 & 10.44 & 11.32 & 12.17  & 13.01 \\\hline
${m}_{BHM}^2$   &     & 7.14   & 8.08     & 8.57& 9.54   &    10.40 &  11.32 & 12.09   & 13.02  \\ \hline\hline
\bf{$n$}    &     & \bf{17}    & \bf{18}      & \bf{19}      & \bf{20}      & \bf{21}     & \bf{22} & \bf{23}     & \bf{24}   \\ \hline
${m}^2$  &    & 14.28  & 15.09  & 16.27  &  17.02  & 18.70 & 19.39  &  20.93  &  21.56 \\\hline
${m}_{BHM}^2$  &   & 14.23   & 15.09 & 16.19     & 17.03 &  18.61  &   19.40 & 20.79  & 21.58 \\ \hline\hline
$n$ &    & \bf{25}    & \bf{26}      & \bf{27}      & \bf{28}      & \bf{29} & \bf{30} & \bf{31} & \bf{32}   \\ \hline
${m}^2$ &  & 23.64 & 24.23 & 25.98 & 26.58  & 28.96  & 29.59  & 31.58 & 32.15 \\\hline
${m}_{BHM}^2$ &    & 23.53 & 24.24 & 25.94    & 26.67  & 28.95     & 29.62  & 31.57 & 32.30    \\ \hline\hline
$n$     &    & \bf{33}      & \bf{34} & \bf{35}& \bf{36} & \bf{37}& \bf{38}& \bf{39} & \bf{40}      \\ \hline
${m}^2$  &  & 34.87 &  35.43   &  37.70  &  38.48  &  41.11  & 42.01  &  44.27   &  45.16 \\\hline
${m}_{BHM}^2$ &   & 34.82  &    35.54  & 37.65      & 38.47  & 41.15  &  42.01  & 44.22  & 45.19   \\ \hline
\end{tabular}%
}
\end{center}
\caption{The $m^2$ values obtained by the singular value method and the corresponding values of the $0^{++}$ masses calculated in~\cite{Berg:2006xy}. The results were obtained using the Mathematica Explicit Runge Kutta method fixed step $\delta\tau \leq 5\times 10^{-4}$, $\tau_{UV} = 20$, $\tau_{mid}=\tau_{IR} = 0.1$, for $1\leq m^2\leq 23$; $\tau_{mid}=2$, for $23\leq m^2\leq 38$; $\tau_{mid}=5$, for $38\leq m^2\leq 46$. The value $m^2=0.543$ does not appear as a cusp, but rather as a broad minimum, for some values of $\tau_{mid}$.}
\label{tab:sing}
\end{table}

By analyzing the periodic pattern of the eigenvalues on figure~\ref{fig:singvalspec} and similar plots, we conclude that we observe four of the six expected towers. The singular value shows cusp-like dips around some of the values of~\cite{Berg:2006xy}. For for $m^2\leq 23$ the best convergence so far was achieved by selecting $\tau_{mid}=\tau_{IR}$, that is effectively using the shooting method. However, for $m^2>23$ the method starts failing on some of the eigenvalues. The sharp cusps of some of the towers are replaced by smoother minima and eventually disappear. This explains why in principle, the remaining two towers are not seen by our calculation. For the four towers the problem can partially solved by moving $\tau_{mid}$ to higher values. In particular, to see the eigenvalues $38\leq m^2\leq 46$, we used $\tau_{mid}=5$. The general experience tells that whenever we have a sharp dip, the eigenvalue matches very well with the table of~\cite{Berg:2006xy}. Smoother shallow minima probably indicate the problem with the numerics. Their positions typically do not match the values in~\cite{Berg:2006xy} so well.

\begin{figure}
    \centering
    \includegraphics[width=0.7\linewidth]{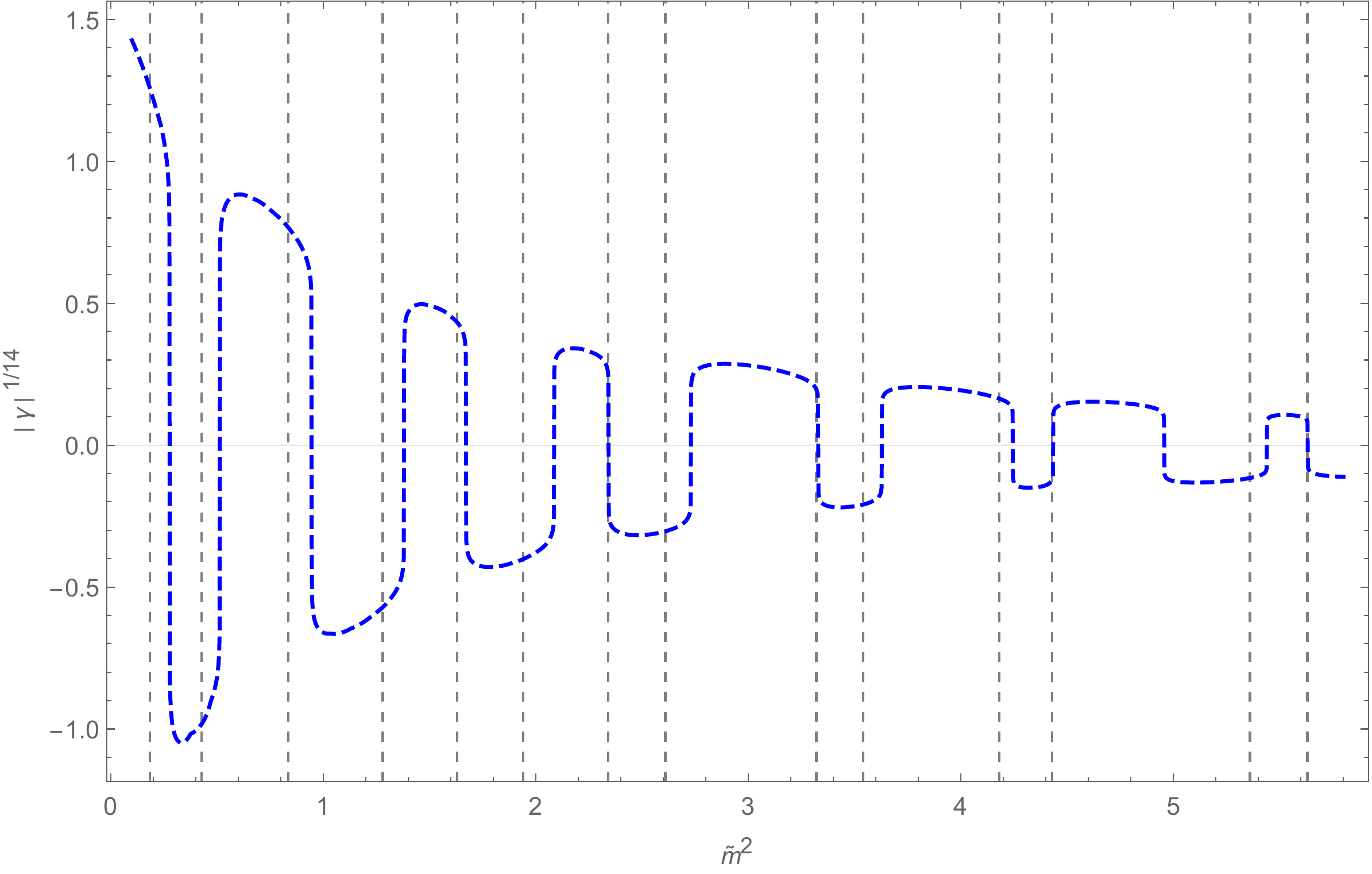}
    \caption{Behavior of the determinant of~(\ref{deta}) around the lowest eigenvalues. The midpoint was performed at $\tau_{mid}=1$ with $\tau_{IR}=0.1$, $\tau=20$ and step $\Delta\tau=10^{-2}$.}
    \label{fig:det}
\end{figure}

In an attempt to see the remaining modes, we also tried to solve the system, 
including the non-physical mode, that does not satisfy constraint~(\ref{eqA}). 
In this case we computed the determinant of the $14\times 14$ matrix, and there 
is no need to through away any information, as would be the case of a $12\times 
12$ determinant. In this approach we used the $\tau_{IR}=0.1$, $\tau_{UR}=20$, 
$\tau_{mid}=1$, $\delta\tau=10^{-2}$, although the calculation worked equally 
well for other choices of the parameters, and for a smaller value of the step 
$\delta\tau=10^{-3}$. The results for the first 65 eigenvalues are summarized in 
table~\ref{tab:mid7}, where they are compared with the values 
of~\cite{Berg:2006xy}. The behavior of the lower values is demonstrated in 
figure~\ref{fig:det}.

\begin{table}[htb]
\begin{center}
\resizebox{\textwidth}{!}{%
\begin{tabular}{|l|l|l|l|l|l|l|l|l|l|l|l|l|l|l|l|l|}
\hline 
\bf{$n$}                 & \bf{1}      & \bf{2}      & \bf{3}     & \bf{4}    & \bf{5}    & \bf{6}    & \bf{7}    & \bf{8}    & \bf{9} &\bf{10} &\bf{11} \\\hline 
$m^2$ & {\bf 0.273}    & {\bf 0.513} & {\bf 0.946} & {\bf 1.38} & 1.67  & {\bf 2.09} & 2.34  & {\bf 2.73} & 3.33  & {\bf 3.63}  & 4.24\\\hline
${m}_{BHM}^2$ & 0.185  & 0.428  & 0.835 & 1.28 & 1.63 & 1.94 & 2.34 & 2.61 & 3.32  & 3.54  & 4.18 \\\hline\hline
\bf{$n$}                 & \bf{12}     & \bf{13}     & \bf{14}     & \bf{15}    & \bf{16}    & \bf{17}    & \bf{18}    & \bf{19}    & \bf{20}  &\bf{21} &\bf{22}    \\ \hline
${m}^2$ & 4.43    & {\bf 4.96}   & 5.44  & 5.63  &  {\bf 6.25} &   6.63  &    \bf{6.96} & 7.61  & 8.09     &\bf{8.43} & 8.93 \\\hline
${m}_{BHM}^2$ & 4.43  & 4.43  & 5.36  & 5.63  & 5.63    & 6.59    & 6.77   & 7.14   & 8.08    & 8.25 & 8.57  \\ \hline\hline
\bf{$n$}                  & \bf{23}      & \bf{24}      & \bf{25}      & \bf{26}     & \bf{27}  & \bf{28}     & \bf{29}  & \bf{30}    & \bf{31}      & \bf{32}  & \bf{33}      \\ \hline
${m}^2$   & 9.56 &\bf{9.93} & 10.86  & 11.33 & \bf{11.74}  &  12.51 &  13.01 & \bf{13.63} & 14.55  &  15.09 & \bf{15.64}  \\\hline
${m}_{BHM}^2$ & 9.54   &   9.62& 10.40  & 11.32 & 11.38  & 12.09  & 12.99  & 13.02   & 14.23 & 15.03  & 15.09  \\ \hline\hline
$n$          & \bf{34} & \bf{35} & \bf{36}  & \bf{37}    & \bf{38}      & \bf{39} & \bf{40}& \bf{41}& \bf{42}& \bf{43}& \bf{44}  \\ \hline
${m}^2$    & 16.40     & 17.0 & \bf{17.68} & 18.84    & 19.38 & \bf{20.06}  & 20.95   & 21.55  &\bf{22.47}   &23.62  & 24.22 \\\hline
${m}_{BHM}^2$   & 16.19  & 16.89  & 17.03 &  18.61  &  19.22 & 19.40   & 20.79  & 21.58 &22.10   & 23.53 & 23.95      \\ \hline\hline
$n$     
 & \bf{45}  & \bf{46}& \bf{47}& \bf{48}& \bf{49} & \bf{50}  & \bf{51} & \bf{52}    & \bf{53}      & \bf{54} & \bf{55}    \\ \hline
${m}^2$  & \bf{25.02}  & 25.97     & 26.62 & \bf{27.62}  & 28.95  & 29.59 & \bf{30.55} & 31.56  & 32.26 & \bf{33.48}&34.83     \\\hline
${m}_{BHM}^2$& 24.24 & 25.94& 26.32  & 26.67  & 28.95  & 29.25  & 29.62  & 31.57 &31.93 & 32.30  & 34.82  \\ \hline
$n$ & \bf{56}& \bf{57}& \bf{58}& \bf{59} & \bf{60}&\bf{61}& \bf{62} & \bf{63}& \bf{64}&\bf{65}& \bf{66}  \\ \hline
${m}^2$     &   35.51 &\bf{36.58} & 37.68   & 38.42  &\bf{39.72}  & 41.22   & 41.97&\bf{43.19} & 44.33    & 45.15&---\\\hline
${m}_{BHM}^2$  & 35.21 & 35.54  & 37.65 & 38.17      & 38.47  & 41.15 & 41.79 &  42.01  & 44.22 & 45.01 & 45.19   \\ \hline

\end{tabular}%
}
\end{center}
\caption{The values of $m^2$ computed from the $14\times 14$ matrix which includes an unphysical boundary condition. Apart from the four towers that match those, reported in table~\ref{tab:sing} the spectrum shows two additional towers (shown in bold) which do not match the eigenvalues in~\cite{Berg:2006xy}. The calculation is done for $\delta\tau=10^{-2}$, $\tau_{IR}=0.1$, $\tau_{UV}=20$ and $\tau_{mid}=1$.}
\label{tab:mid7}
\end{table}

\begin{figure}
    \centering
    \includegraphics[width=0.7\linewidth]{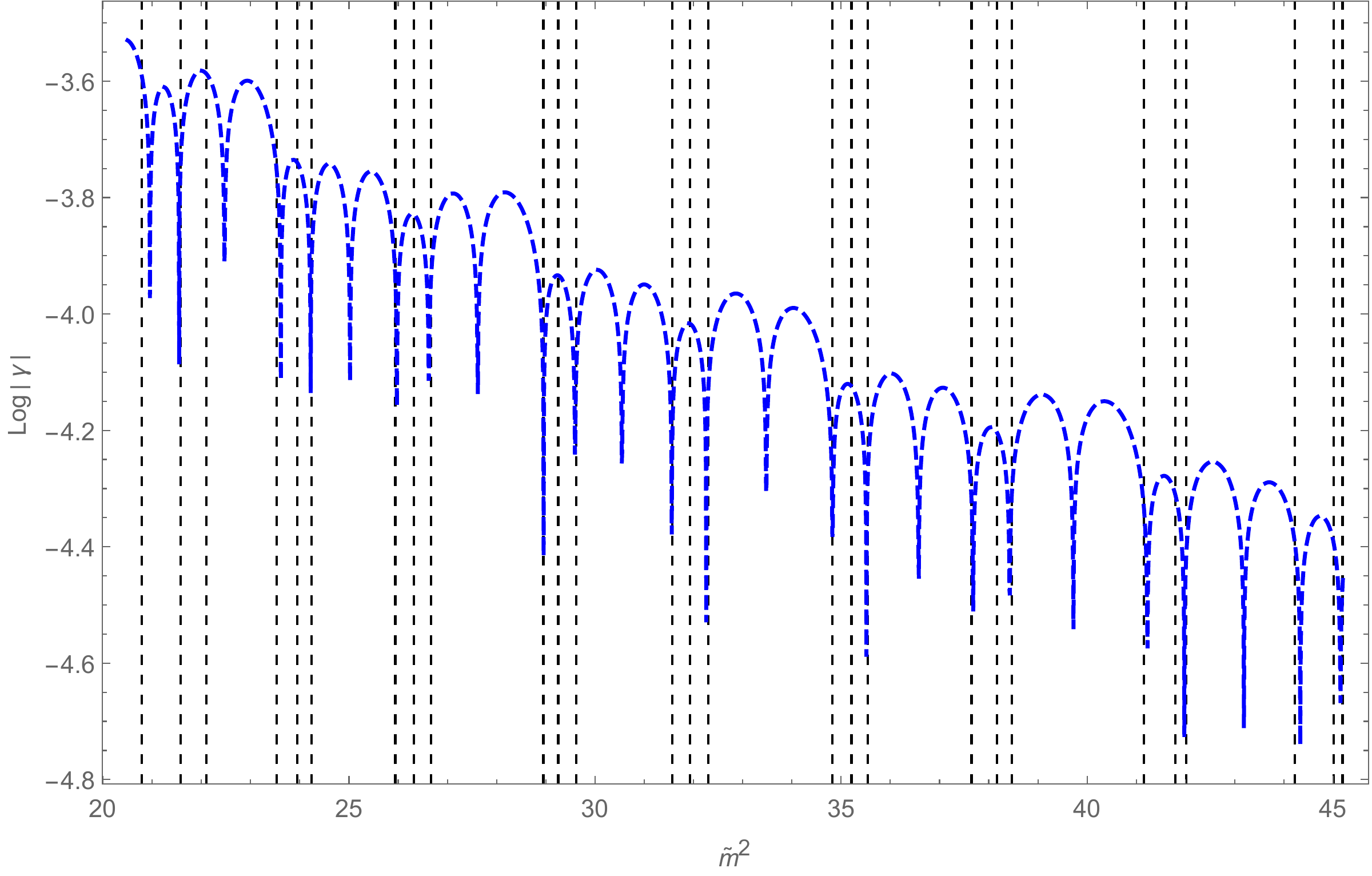}
    \caption{Behavior of the logarithm of the $14\times 14$ determinant of~(\ref{deta}) for higher masses. The calculation was performed at $\tau_{mid}=1$ with $\tau_{IR}=0.1$, $\tau=20$ and step $\delta\tau=10^{-2}$.}
    \label{fig:heavy}
\end{figure}

In the $14\times 14$ case, we are able to see more eigenvalues and the periodicity patterns indicates to the presence of six towers. Out of the six towers one can distinguish the four towers discovered in the singular value approach. The two methods give compatible sets of values. However, the two new towers (their eigenvalues appear in bold in table~\ref{tab:mid7}) are not the towers seen in~\cite{Berg:2006xy}. This is clearly seen on figure~\ref{fig:heavy}, where a set of zeroes of the determinant insert in the gaps of the heavier part of the $0^{++}$ spectrum.

Comparison of the two methods allows to separate the spectrum of~\cite{Berg:2006xy} into groups of four and two towers, in accordance with the separation of values in table~\ref{tab:mid7}, although we have to assume that the singular value methods captures all the eigenvalues of the four towers. This is apparently true for higher masses, but can be more subtle for the lightest ones. For example, for $\tau_{mid}=5$, in the singular value method, we see a broad minimum around $m^2=0.543$, which could be an onset of a lighter mode in the four towers, or a mode from a different tower.

\section{Conclusions and discussion}
\label{sec:conclude}

In this work we have analysed the $0^{-+}$ pseudoscalar $SU(2)\times SU(2)$ singlet fluctuations of the Klebanov-Strassler theory. We constructed the ansatz for the perturbations of the bosonic type IIB supergravity fields over the KS background, with $0^{-+}$ quantum numbers, and derived the linearized equations. 

The linearized equations that we have found can be reduced to a system of six coupled second order equations, so they describe six independent towers of $0^{-+}$ states, however, this system has a more compact presentation in terms of seven second order equations, with a first order constraint. We analyzed the asymptotic behavior of the compact system and recovered asymptotic expansions of the solutions at the two ends of the KS geometry (UV for $\tau\to\infty$ and IR for $\tau\to 0$). This allowed to extract the spectrum of the $0^{-+}$ operators in the dual gauge theory and complete their classification in terms of the multiplets of a ${\cal N}=1$ superconformal theory dual to strings on $AdS_5\times T^{1,1}$~\cite{Ceresole:1999ht}.

We discussed different self-consistency checks of our complicated system, which included a non-trivial linear dependence of eight differential equations, gauge invariance and consistency of the operator spectrum. 

As an ultimate consistency check we have been analyzing the numerical mass spectrum of the six $0^{-+}$ modes produced by the linearized equations and comparing them with the known spectrum of $0^{++}$ modes~\cite{Berg:2006xy}. We were able to show that our equations reproduce four of the six towers of the $0^{++}$ modes with a reasonable accuracy. However, our main method of the numerical analysis was not able to resolve for the remaining two towers. We expect that a more powerful numerical approach should be able to solve this problem. We left this for a future work~\cite{inprogress}.

As an alternative method we have been analyzing the spectrum of seven equations without the constraint. In this case we were able to capture six independent towers of eigenvalues: four compatible with the previous analysis and with the $0^{++}$ spectrum, and two other towers quite different from the remaining eigenvalues of~\cite{Berg:2006xy}. We have to interpret the presence of a new pair of modes. One of those modes could be an unphysical mode, which does not satisfy constraint~(\ref{eqA}). But there are two modes, so there is a possibility that the spectrum of our equations is different from that of~\cite{Berg:2006xy}. So far, our numerical analysis does not allow to definitely conclude this. One problem is that we have not seen all the seven modes. Moreover, analyzing the different initial conditions, we observe that even if one starts from an unphysical initial condition at one end, the system will evolve into the physical subspace, satisfying the constraint, in the other end. So, the solution is projected onto a smaller subspace in the evolution. This is demonstrated by the plots on figures~\ref{fig:uvtoir} and~\ref{fig:irtouv} shown in appendices~\ref{sec:UV} and~\ref{sec:IRasympt}.

\begin{figure}
    \centering
    \includegraphics[width=0.47\linewidth]{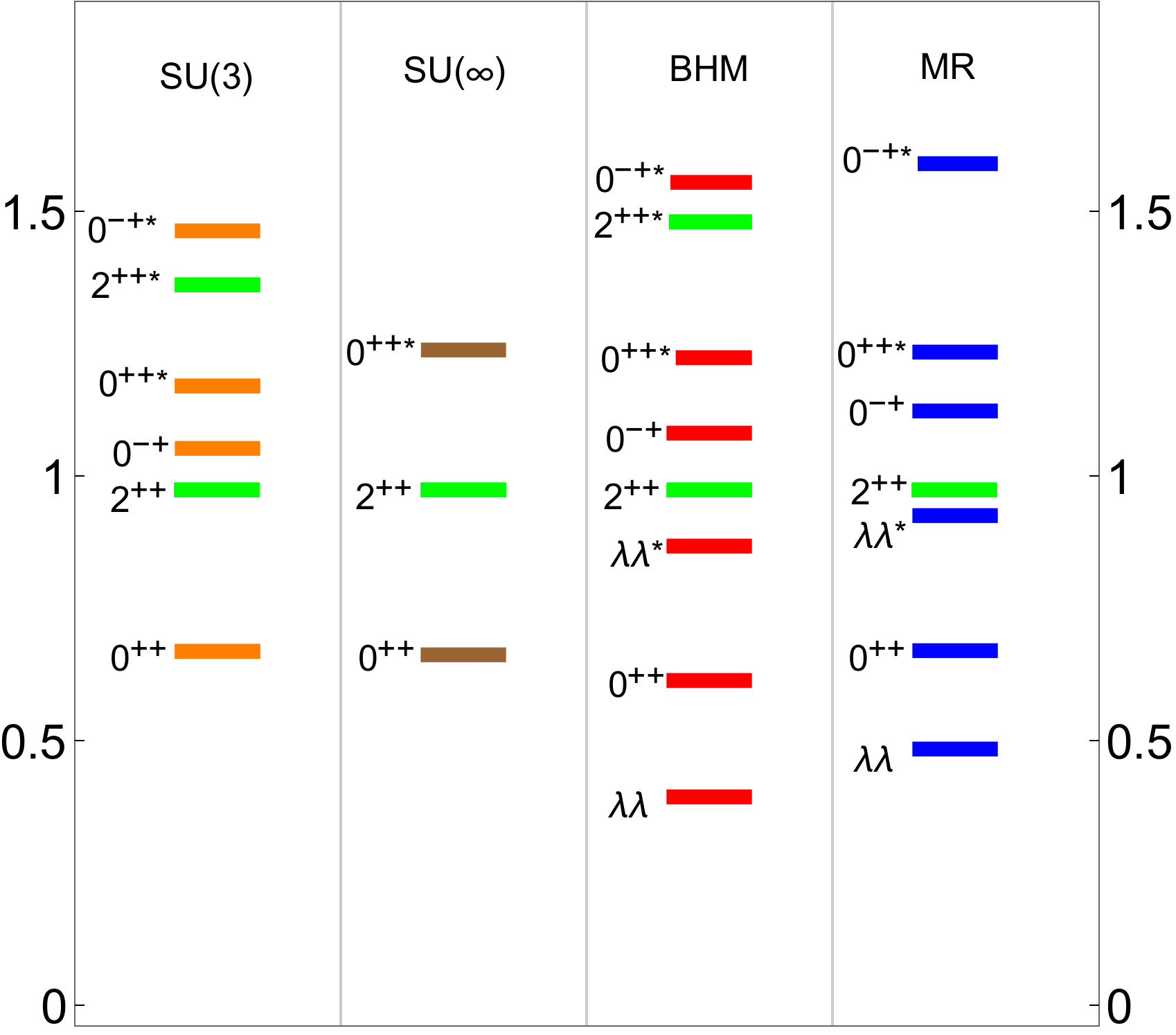}
    \hfill
    \includegraphics[width=0.47\linewidth]{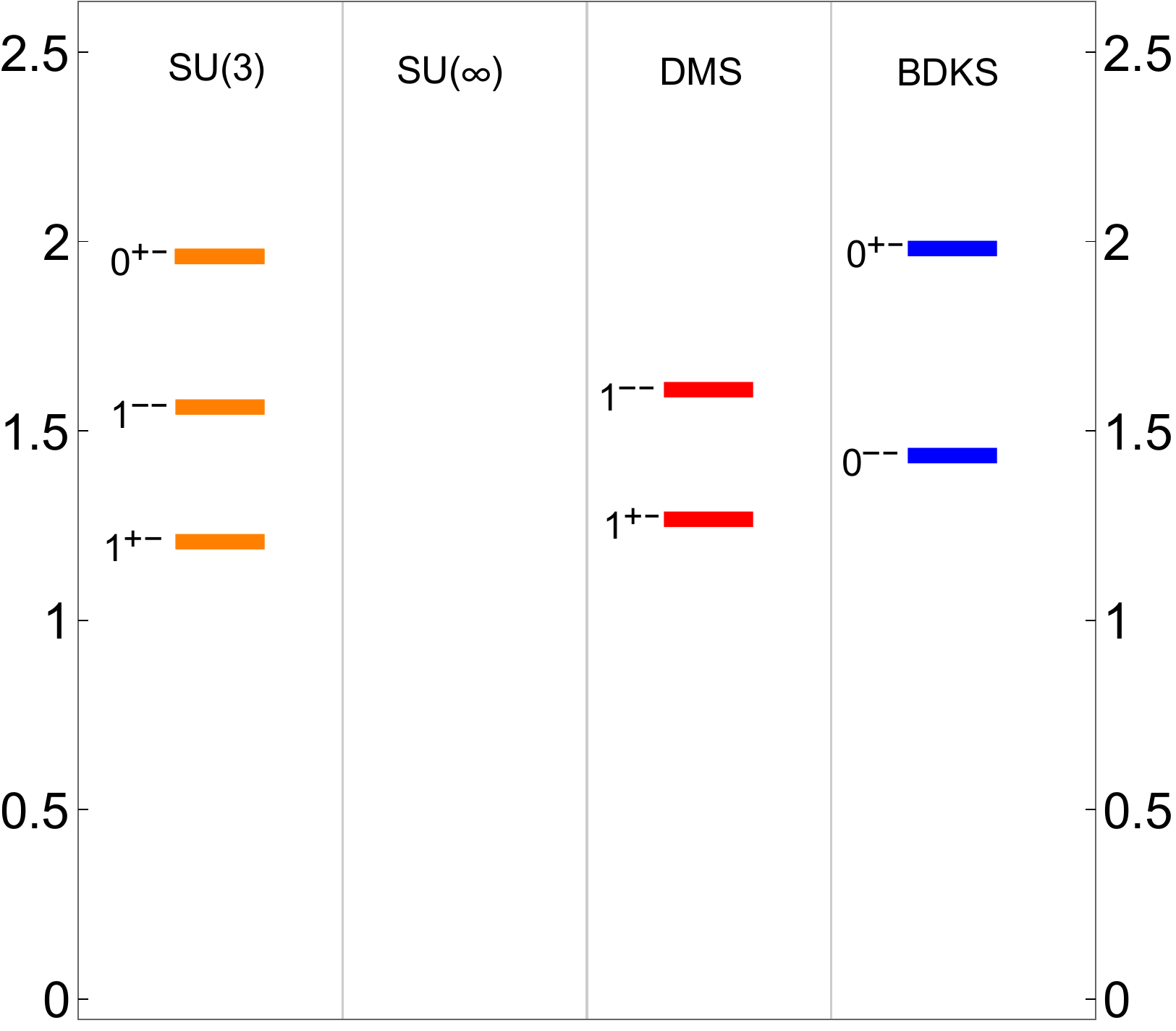}
    \caption{Comparison of the spectrum of the lightest glueballs on the lattice 
with the holographic calculation in the Klebanov-Strassler theory. We only 
compare states of low spin, which can be obtained from the holographic analysis 
($0^{PC}$, $1^{PC}$ and $2^{++}$). All masses are normalized to the mass of the 
$2^{++}$ state. Lattice data are taken from~\cite{Athenodorou:2020ani} in the 
positive $C$ sector and from~\cite{Chen:2005mg} in the opposite one. The column 
BHM refers to the result of~\cite{Berg:2006xy}, MR compares the results of the 
present work, DMS and BDKS are the results of~\cite{Dymarsky:2008wd} 
and~\cite{Benna:2007mb}
respectively.}
    \label{fig:lattice}
\end{figure}

It is interesting to compare the holographic prediction of the spectrum with the 
spectrum of the pure glue $SU(3)$ theory calculated on the 
lattice~\cite{Morningstar:1999rf,Chen:2005mg,Athenodorou:2020ani} and its 
extrapolation to $SU(\infty)$~\cite{Teper:1998kw,Lucini:2004my,Lucini:2010nv}. 
One can observe a reasonable match of the pattern of all of the six lowest 
glueballs of the lattice spectrum, which can in principle be captured by the 
classical gravity approximation, including $0^{++}$, $0^{-+}$, $2^{++}$, 
$1^{+-}$, $1^{--}$ and $0^{+-}$. Our comparison is shown on 
figure~\ref{fig:lattice}.

To compare, we normalized all the glueball masses to the value of the $2^{++}$ 
state mass. In the $C=-1$ sector, the match is quite good (right panel of 
figure~\ref{fig:lattice}), as previously discussed 
in~\cite{Dymarsky:2008wd,Gordeli:2013jea}. There is somewhat less 
lattice data in this sector, so we only compare with the $SU(3)$ results. See 
however~\cite{Lucini:2010nv,Holligan:2019lma}.

In the $C=+1$ sector, the lightest scalar on the lattice, corresponding to the $\tr F_{\mu\nu}F^{\mu\nu}$ operator, matches the second lightest state in the $0^{++}$ spectrum of~\cite{Berg:2006xy}. One has to remember that the KS theory is supersymmetric, so the lightest expected $0^{++}$ state in the supersymmetric theory is dual to the gluino bilinear $\tr\lambda\lambda$, not present in the pure glue theory. The same is true for the $0^{--}$ state that appears in the $C=-1$ sector, and some heavier states that we do not show here. Moreover, from our analysis, we find it plausible, that the state that roughly matches the first excited state $0^{\ast++}$ on the lattice is the first excited state of the lowest bosonic holographic $0^{++}$. In other words, the matching can be extended to the excited states as well. There is also a holographic state that matches the lightest pseudoscalar on the lattice. This state is dual to the operator ${\rm Tr}F_{\mu\nu}\tilde{F}^{\mu\nu}$.

We can also speculate on the possibility that the two towers of pseudoscalar states seen in our analysis, which do not match the spectrum $0^{++}$ computed in~\cite{Berg:2006xy}, are, in fact, the correct values. The lightest states of this spectrum are shown in the last column of the left panel of figure~\ref{fig:lattice}. The values would give a comparable match for the lightest states in the pure glue $SU(\infty)$ sector. Besides the $14\times 14$ spectrum turns out to be more fittable with quadratic fits in comparison to the spectrum of~\cite{Berg:2006xy}. This in particular supports the expectation of the third and the fifth states being the first excited states, of $\tr\lambda\lambda$ and $\tr F_{\mu\nu}F^{\mu\nu}$ respectively. In the following work, with a better numerical resolution, we expect to validate or discard our conjectures.

As a possible alternative way of testing our equations one can consider the 
method of~\cite{Bianchi:2003ug,Berg:2005pd} of deriving the truncated 
five-dimensional sigma model action for the pseudoscalar fields. A similar work 
was done in~\cite{Elander:2018aub} for the dual of the Romans 
supergravity~\cite{Romans:1985tw}. It would also be interesting to explicitly 
derive the pseudoscalar equations from the scalar ones, either using the 
supersymmetric quantum mechanics approach~\cite{Dymarsky:2007zs}, or the full 
ten-dimensional supersymmetry transformations.

\paragraph{Acknowledgements} We are grateful to Marcus Benna, Ivan Gordeli and
Sasha Solovyov for collaboration on the early stages of this project, and to 
Igor Klebanov for insightful discussions and comments. We are especially 
grateful to Anatoly Dymarsky for both collaboration and discussions. We would 
like to thank Daniel Elander and Maurizio Piai for the correspondence on the 
relevant methods and recent lattice results. The work of 
DM and CR was supported by the Brazilian ministry of education (MEC). The work 
of DM was also partially supported by the Universal grant \#433935/2018-9 of the 
Brazilian national agency CNPq. DM would also like to thank FAPESP grant 
2016/01343-7 for funding the visit to ICTP-SAIFR in August 2019, where part of 
this work was done.

\begin{appendix}

\section{UV asymptotics}
\label{sec:UV}

In this section we analyze the asymptotic behavior of the solutions to the system of linearized equations~(\ref{eqf3})-(\ref{eqA}) for large $\tau$ (UV regime). We will construct the solutions as expansions in $1/\tau$, working in the leading exponential approximation, as in equation~(\ref{UVexpansion}). In such an approximation equations~(\ref{eqf3})-(\ref{eqa}) take the following asymptotic form.

\begin{multline}
\label{UVeq1}
    \frac{\sqrt[6]{2} \sqrt{3} (\tau -1) }{\sqrt{4 \tau -1}} a' \ + \ \frac{\sqrt[6]{2} \sqrt{3} (2 \tau +1) }{(4 \tau -1)^{3/2}}a \ - \  2\ 2^{2/3} e^{-\tau } (\tau -1) B_2+2\ 2^{2/3} e^{-\tau } (\tau -2) C_2^- \ - \ 2^{2/3} C_2^+ \\ - \ \frac{3 \sqrt{3} \left(4 \tau ^2-5 \tau +10\right) }{2\ 2^{5/6} (4 \tau -1)^{3/2}}\phi_3' \ + \ \frac{\sqrt{3} \left(64 \tau ^3-120 \tau ^2+12 \tau +125\right) }{2\ 2^{5/6} (4 \tau -1)^{5/2}}\phi_3 \ - \ \frac{9 \sqrt{3} (\tau -1) }{4\ 2^{5/6} \sqrt{4 \tau -1}}\phi_3'' \ = \ 0\,,
\end{multline}{}

\begin{multline}
\label{UVeq2}
    \frac{\sqrt{3} m^2 e^{-\frac{\tau }{3}} (\tau -2) }{\sqrt[6]{2} \sqrt{4 \tau -1}} a \ + \ \frac{32\ 2^{5/6} e^{-\frac{2 \tau }{3}} (\tau -1) }{3 \sqrt{3} (4 \tau -1)^{3/2}}B' \ + \ \frac{8\ 2^{5/6} e^{\frac{4 \tau }{3}} }{3 \sqrt{3} (4 \tau -1)^{3/2}}B \ -\ \frac{64 \sqrt[3]{2} e^{\frac{4 \tau }{3}} (\tau -1) }{9 (1-4 \tau )^2}B_2 \\ +\ \frac{4 \sqrt[3]{2} e^{\tau /3} \tau  }{3-12 \tau }C' \ - \ \frac{64 \sqrt[3]{2} e^{\tau /3} (\tau -1) \tau  }{9 (1-4 \tau )^2}C \ +\ \frac{4 \sqrt[3]{2} e^{\frac{4 \tau }{3}} }{12 \tau -3}{C_2^-}'' \ +\ \frac{64 \sqrt[3]{2} e^{\frac{4 \tau }{3}} (\tau -1) }{9 (1-4 \tau )^2}{C_2^-}' \ +\ \frac{4 \sqrt[3]{2} e^{\frac{4 \tau }{3}} }{3-12 \tau }C_2^- \\ + \ \frac{16 \sqrt[3]{2} e^{\tau /3} }{12 \tau -3}{C_2^+}'' \ + \ \frac{16 \sqrt[3]{2} e^{\tau /3} (4 \tau -13) }{9 (1-4 \tau )^2}{C_2^+}' \ +\ 2 \sqrt[3]{2} m^2 e^{-\frac{\tau }{3}} C_2^+ \\ + \ \frac{3\ 2^{5/6} \sqrt{3} m^2 e^{-\frac{\tau }{3}} (\tau -2) (\tau -1) }{(4 \tau -1)^{3/2}}\phi_3 \ = \ 0\,,
\end{multline}{}

\begin{multline}
\label{UVeq3}
    -\frac{\sqrt{3} m^2 e^{\frac{2 \tau }{3}} }{2 \sqrt[6]{2} \sqrt{4 \tau -1}}a \ + \ \frac{16\ 2^{5/6} e^{\tau /3} (\tau -1) }{3 \sqrt{3} (4 \tau -1)^{3/2}}B' \ +\ \frac{16\ 2^{5/6} e^{\tau /3} \left(4 \tau ^2-11 \tau +16\right) }{9 \sqrt{3} (4 \tau -1)^{5/2}}B \ +\ \frac{2 \sqrt[3]{2} e^{\frac{4 \tau }{3}} }{3-12 \tau }C' \\ - \ \frac{32 \sqrt[3]{2} e^{\frac{4 \tau }{3}} (\tau -1) }{9 (1-4 \tau )^2}C \ +\ \frac{16 \sqrt[3]{2} e^{\tau /3} }{12 \tau -3}{C_2^-}'' \ + \ \frac{16 \sqrt[3]{2} e^{\tau /3} (4 \tau -13) }{9 (1-4 \tau )^2}{C_2^-}' \ +\ 2 \sqrt[3]{2} m^2 e^{-\frac{\tau }{3}} C_2^- \\ + \ \frac{4 \sqrt[3]{2} e^{\frac{4 \tau }{3}} }{12 \tau -3}{C_2^+}'' \ +\ \frac{64 \sqrt[3]{2} e^{\frac{4 \tau }{3}} (\tau -1) }{9 (1-4 \tau )^2}{C_2^+}' \ +\ \frac{m^2 e^{\frac{2 \tau }{3}} }{2^{2/3}}C_2^+ \ + \ \frac{3 \sqrt{3} m^2 e^{\frac{2 \tau }{3}} (\tau -1) }{\sqrt[6]{2} (4 \tau -1)^{3/2}}\phi_3 \ = \ 0\,,
\end{multline}{}

\begin{multline}
\label{UVeq4}
-\frac{3}{4} m^2 e^{-\frac{5 \tau }{3}} (\tau -1) \sqrt{6 \tau -\frac{3}{2}} a \ +\ 2 \sqrt{\frac{2}{3}} \sqrt{\frac{1}{(4 \tau -1)}} B'  + \ \frac{4}{3} \sqrt{\frac{2}{3}} \sqrt{\frac{1}{(4 \tau -1)^3}} (8 \tau -11) B \\ +\ B_2'' \ +\ \frac{16 (\tau -1) }{12 \tau -3}B_2' \ -\ B_2\ -\ e^{-\tau } (\tau -1) C'+\frac{16 e^{-\tau } (\tau -1)^2 }{3 (1-4 \tau )}C \ +\ \frac{16 (\tau -1) }{3-12 \tau }C_2^- \\ +\ \frac{9 \sqrt{\frac{3}{2}} m^2 e^{-\frac{5 \tau }{3}} (\tau -1)^2 }{2 \sqrt{4 \tau -1}}\phi_3 \ = \ 0\,,    
\end{multline}{}

\begin{multline}
\label{UVeq5}
    \frac{256 \sqrt{\frac{2}{3}} e^{-\tau } (\tau -1) }{3 (4 \tau -1)^{3/2}}B \ +\ \frac{64 e^{-\tau } (\tau -1) }{3 (4 \tau -1)}B_2' \ +\ \frac{64 e^{-\tau } (\tau -2) }{3 (1-4 \tau )}B_2 \ - \ \frac{32 }{3-12 \tau }{C_2^+}' \\ +\ C''\ + \ \frac{4 }{3}C' \ +\ \frac{16 }{3-12 \tau }C \ - \ \frac{64 e^{-\tau } \tau  }{3-12 \tau }{C_2^-}' \ +\ \frac{64 e^{-\tau } (\tau -1) }{3 (1-4 \tau )}C_2^-  \ = \ 0\,,
\end{multline}{}

\begin{multline}
\label{UVeq6}
    \frac{9}{8} m^2 e^{-\frac{5 \tau }{3}} (4 \tau -1) a\ +\ B''\ +\ \frac{16 (\tau -1) }{12 \tau -3}B' \ +\ \frac{\left(16 \tau ^2-168 \tau +77\right) }{3 (1-4 \tau )^2}B \\ -\ 8 \sqrt{\frac{2}{3}} \sqrt{\frac{1}{4 \tau -1}} B_2'\ -\ 32 \sqrt{\frac{2}{3}} e^{-2 \tau } (\tau -1) \sqrt{\frac{1}{4 \tau -1}} B_2\ + \ \frac{16 e^{-\tau } (\tau -1) }{\sqrt{6 \tau -\frac{3}{2}}}C \\ -\ \frac{32 e^{-2 \tau } (\tau -1) }{\sqrt{6 \tau -\frac{3}{2}}}{C_2^-}' \ +\ \frac{8 }{\sqrt{6 \tau -\frac{3}{2}}}C_2^- \ -\ \frac{16 e^{-\tau } (\tau -1) }{\sqrt{6 \tau -\frac{3}{2}}}{C_2^+}' \ = \ 0\,,
\end{multline}{}

\begin{multline}
\label{UVeq7}
    a''\ +\ \frac{2 }{3}a' \ -\ \frac{16 (5 \tau -2) }{3 (1-4 \tau )^2}a \ -\ \frac{8}{3} e^{-\tau } B\ +\ \frac{32 e^{-\tau } (\tau -1) }{3 \sqrt{6 \tau -\frac{3}{2}}}B_2 \ -\ \frac{32 e^{-\tau } (\tau -2) }{3 \sqrt{6 \tau -\frac{3}{2}}}C_2^- \\ +\ \frac{16}{3 \sqrt{6 \tau -\frac{3}{2}}} C_2^+ \ +\ \frac{32 \left(8 \tau ^3-24 \tau ^2+33 \tau -17\right) }{3 (1-4 \tau )^3}\phi_3\ -\ \frac{32 (\tau -1)^2 }{(1-4 \tau )^2}\phi_3' \ =\ 0\,.
\end{multline}{}

In section~\ref{sec:asymptotic} we presented a simplified version of this system, see equations~(\ref{phire})-(\ref{are}), which was solved analytically and the modes were classified according to equations~(\ref{UVas1})-(\ref{UVas7}). Here we will construct solutions to the full system.

Below, in section~\ref{sec:UV-sing} we will construct the singular modes of the seven above equations, while in section~\ref{sec:UVreg} we will do the same for the regular modes. The modes will be constructed in the leading exponential and up to the next-to-next-to-leading order (NNLO). In section~\ref{sec:UVcomp} we justify the classification of the modes as singular and regular and analyze their compatibility with the constraint imposed by equation~(\ref{eqA}).

\subsection{Singular UV solutions}
\label{sec:UV-sing}

The UV asymptotic equations~(\ref{UVeq1})-(\ref{UVeq7}) possess seven singular and seven regular modes. One can choose the following basis of the singular modes, which roughly follows the classification~(\ref{UVas1})-(\ref{UVas7}):

\begin{itemize}
    \item $\alpha_2$ mode
\begin{eqnarray}
\phi_3 & = & \frac{2\alpha_2}{3}\frac{1}{\sqrt{\tau}}\left(1+\frac{7}{8\tau}+\frac{111}{128\tau^2}+O(\tau^{-3})\right)e^{4\tau/3}  \,,\nn\\
C_2^+ & = & 0 \,, \nn\\
C_2^- & = & \frac{1215\alpha_2}{656}\sqrt{\frac{3}{2}}m^2\tau\left(1+O(\tau^{-5})\right)e^{-\tau/3}
\,, \nn\\
B_2 & = & \frac{27\alpha_2}{16}\sqrt{\frac{3}{2}}{m^2{\tau}}\left(1+\frac{31}{41\tau}+O(\tau^{-5})\right)e^{-\tau/3} \,, \nn\\
C & = & 0 \,, \nn\\
B & = & \frac{1161\alpha_2}{164}m^2\tau^{3/2}\left(1-\frac{177}{344\tau}+\frac{225}{5504\tau^2}+O(\tau^{-3})\right)e^{-\tau/3} \,, \nn\\
a & = &  {\alpha_2}\frac{1}{\sqrt{\tau}}\left(1+\frac{1}{8\tau}+\frac{3}{128\tau^2}+O(\tau^{-3})\right)e^{4\tau/3} \,,
\end{eqnarray}
    
    \item $\alpha_3$ mode
\begin{eqnarray}
\phi_3 & = & \frac{\alpha_3}{\sqrt{\tau}}\left(-1- \frac{7}{8\tau}-\frac{111}{128\tau^2}+O(\tau^{-3})\right)  \,,\nn\\
C_2^+ & = & \frac{27\alpha_3}{16}\sqrt{\frac{3}{2}}m^2\tau\left(-1-\frac{17}{4\tau}+O(\tau^{-5})\right)e^{-2\tau/3} \,, \nn\\
C_2^- & = &  \frac{27\alpha_3}{16}\sqrt{\frac{3}{2}}m^2\tau^2\left(-1-\frac{7}{2\tau}-\frac{21}{16\tau^2}+O(\tau^{-5})\right)e^{-5\tau/3}\,, \nn\\
B_2 & = & \frac{27\alpha_3}{16} \sqrt{\frac{3}{2}}m^2\tau^2 \left(1-\frac{1}{2\tau}+\frac{109}{16\tau^2}+O(\tau^{-5})\right)e^{-5\tau/3}  \,, \nn\\
C & = &  \frac{27\alpha_3}{4}\sqrt{\frac{3}{2}}m^2\left(1+O(\tau^{-5})\right)e^{-2\tau/3} \,, \nn\\
B & = & \frac{243\alpha_3}{16}m^2\sqrt{\tau}\left(1-\frac{1}{8\tau}-\frac{1}{128\tau^2}+O(\tau^{-3})\right)e^{-5\tau/3} \,, \nn\\
a & = &  {\alpha_3}{\sqrt{\tau}}\left(1-\frac{13}{8\tau}-\frac{25}{128\tau^2}+O(\tau^{-3})\right) \,, \label{Rmode3}
\end{eqnarray}    
    
    \item $\alpha_6$ mode
\begin{eqnarray}
\phi_3 & = &  \frac{\alpha_6}{16\sqrt{6}}\frac{1}{\sqrt{\tau}}\left(128+\frac{112}{\tau}+\frac{111}{\tau^2}+O(\tau^{-3})\right) \,,\nn\\
C_2^+ & = & 0 \,, \nn\\
C_2^- & = & \alpha_6(1 + O(\tau^{-4}))e^{\tau} \,, \nn\\
B_2 & = & \alpha_6(1 + O(\tau^{-4})) e^{\tau} \,, \nn\\
C & = & 8 \alpha_6 + O(\tau^{-4})\,, \nn\\
B & = & 0 \,, \nn\\
a & = &  \frac{3\alpha_6}{2\sqrt{6}}\frac{1}{\sqrt{\tau}}\left(8+\frac{1}{\tau}+O(\tau^{-2})\right)\,,
\end{eqnarray}
    
    \item $\alpha_8$ mode
 \begin{eqnarray}
\phi_3 & = & {2\alpha_8}{\sqrt{\frac{2}{3}}}{\sqrt{\tau}}\left(1+\frac{13}{128\tau^2}+O(\tau^{-3})\right)  \,,\nn\\
C_2^+ & = & {\alpha_8}{{\tau}}\left(1-\frac{13}{8\tau}+O(\tau^{-6})\right) \,, \nn\\
C_2^- & = &  0\,, \nn\\
B_2 & = & {2\alpha_8}{{\tau}}\left(-1+\frac{2}{\tau}+O(\tau^{-6})\right)e^{-\tau} \,, \nn\\
C & = & 2{\alpha_8}\left(1+O(\tau^{-6})\right) \,, \nn\\
B & = & {2\sqrt{6}\alpha_8}{\sqrt{\tau}}\left(1-\frac{1}{8\tau}-\frac{1}{128\tau^2}+O(\tau^{-3})\right)e^{-\tau} \,, \nn\\
a & = &  \frac{5\alpha_8}{4}\sqrt{\frac{3}{2}}\frac{1}{\sqrt{\tau}}\left(-1-\frac{1}{8\tau}-\frac{3}{128\tau^2}+O(\tau^{-3})\right) \,,
\end{eqnarray}
Note that this mode has linear $\tau$ behavior of $C_2^+$ and is ``locked'' with the constant $C$ mode.   
    
    \item $\alpha_9$ mode
\begin{eqnarray}
\phi_3 & = & \frac{2\alpha_9}{\sqrt{\tau}}\sqrt{\frac{2}{3}}\left(1+ \frac{7}{8\tau}+\frac{111}{128\tau^2}+O(\tau^{-3})\right)  \,,\nn\\
C_2^+ & = & \alpha_9 + \frac{27\alpha_9}{8}m^2\tau\left(1+\frac{17}{4\tau}+O(\tau^{-5})\right)e^{-2\tau/3} \,, \nn\\
C_2^- & = &  \frac{81\alpha_9}{8}m^2\tau\left(1+\frac{39}{16\tau}+O(\tau^{-4})\right)e^{-5\tau/3}\,,\nn \\
B_2 & = & \frac{7533\alpha_9}{128}m^2\left(-1-\frac{35}{380928\tau^3}+O(\tau^{-5})\right)e^{-5\tau/3} \,, \nn\\
C & = & \frac{27\alpha_9}{2}m^2\left(-1+O(\tau^{-5})\right)e^{-2\tau/3} \,, \nn\\
B & = & \frac{27\alpha_9}{2}\sqrt{\frac{3}{2}}m^2\tau^{3/2}\left(-1-\frac{33}{8\tau}+\frac{69}{128\tau^2}+O(\tau^{-3})\right)e^{-5\tau/3} \,, \nn\\
a & = &  \frac{2\alpha_9}{\sqrt{\tau}}\sqrt{\frac{3}{2}}\left(1+\frac{1}{8\tau}+\frac{3}{128\tau^2}+O(\tau^{-3})\right) \,, \label{Rmode9}
\end{eqnarray}
This is a subtle mode, because the expansion of $C_2^+$ occurs in the subleading exponential order, while the leading constant solution drops from most of the equations, which only depend on the derivative.
    
    \item $\alpha_{11}$ mode
\begin{eqnarray}
\phi_3 & = & {4\alpha_{11}}\sqrt{\frac23}{{\tau}^{3/2}}\left(1+ \frac{1}{12\tau}-\frac{111}{128\tau^2}+O(\tau^{-3})\right) e^{-4\tau/3} \,,\nn\\
C_2^+ & = & 2\alpha_{11}\tau^2\left(1+\frac{127}{96\tau^2}+O(\tau^{-5})\right)e^{-4\tau/3} \,, \nn\\
C_2^- & = &  \alpha_{11}\tau\left(1+O(\tau^{-5})\right)e^{-\tau/3}\,, \nn\\
B_2 & = & \frac{11\alpha_{11}}{12}\left(1+O(\tau^{-5})\right)e^{-\tau/3} \,, \nn\\
C & = & \frac{10\alpha_{11}}{9}\tau \left(1+O(\tau^{-5})\right)e^{-4\tau/3}\,, \nn\\
B & = & {2\alpha_{11}}\sqrt{\frac23}{{\tau}^{3/2}}\left(1+ \frac{1}{24\tau}-\frac{11}{384\tau^2}+O(\tau^{-3})\right) e^{-\tau/3} \,, \nn\\
a & = &  \frac{\alpha_{11}}{4}\sqrt{\frac23}{\sqrt{\tau}}\left(1-\frac{227}{8\tau}-\frac{453}{128\tau^2}+O(\tau^{-3})\right) e^{-4\tau/3}\,,
\end{eqnarray}

    \item $\alpha_{13}$ mode
\begin{eqnarray}
\phi_3 & = & \frac{16\alpha_{13}}{3}\sqrt{\frac23}{\sqrt{\tau}}\left(1+ \frac{1}{\tau}+\frac{125}{128\tau^2}+O(\tau^{-3})\right) e^{-4\tau/3} \,,\nn\\
C_2^+ & = & 0 \,, \nn\\
C_2^- & = &  \alpha_{13}\left(1+O(\tau^{-5})\right)e^{-\tau/3}\,, \nn\\
B_2 & = & \frac{\alpha_{13}}{3}\left(-1+O(\tau^{-5})\right)e^{-\tau/3} \,, \nn\\
C & = & \frac{16\alpha_{13}}{9}\tau \left(1-\frac{13}{4\tau}+O(\tau^{-6})\right)e^{-4\tau/3}\,, \nn\\
B & = & \frac{4\alpha_{13}}{3}\sqrt{\frac23}{\sqrt{\tau}}\left(1- \frac{1}{8\tau}-\frac{1}{128\tau^2}+O(\tau^{-3})\right) e^{-\tau/3} \,, \nn\\
a & = &  {4\alpha_{13}}\sqrt{\frac23}{\sqrt{\tau}}\left(1+\frac{5}{8\tau}+\frac{11}{128\tau^2}+O(\tau^{-3})\right) e^{-4\tau/3}\,,
\end{eqnarray}

\end{itemize}{}

\subsection{Regular UV solutions}
\label{sec:UVreg}

The following modes realize the UV regular modes of the system (\ref{eqf3})-(\ref{eqa}):

\begin{itemize}
\item $\alpha_1$ mode 
\begin{eqnarray}
\phi_3 & = & {\alpha_1}\sqrt{\tau}\left(-1-\frac{17}{40\tau}-\frac{303}{640\tau^2}+O(\tau^{-3})\right)e^{-2\tau}  \,,\nn\\
C_2^+ & = & \frac{135\alpha_1}{256}\sqrt{\frac{3}{2}}m^2\tau\left(1-\frac{1}{4\tau}+O(\tau^{-5})\right)e^{-8\tau/3} \,, \nn\\
C_2^- & = &  \frac{27\alpha_1}{64}\sqrt{\frac{3}{2}}m^2\tau^2\left(-1-\frac{863}{280\tau}+\frac{2469}{2450\tau^2}+O(\tau^{-5})\right)e^{-11\tau/3} \,, \nn\\
B_2 & = & \frac{27\alpha_1}{64}\sqrt{\frac{3}{2}}m^2\tau^2\left(1-\frac{653}{280\tau}+\frac{6593}{9800\tau^2}+O(\tau^{-5})\right)e^{-11\tau/3} \,, \nn\\
C & = & \frac{135\alpha_1}{128}\sqrt{\frac{3}{2}}m^2\left(1+O(\tau^{-5})\right)e^{-8\tau/3} \,, \nn\\
B & = & \frac{729\alpha_1}{320}m^2\tau^{3/2}\left(-1+\frac{17}{30\tau}-\frac{91}{1920\tau^2}+O(\tau^{-3})\right)e^{-11\tau/3} \,, \nn\\
a & = & {\alpha_1}\sqrt{\tau}\left(1+\frac{7}{40\tau}+\frac{19}{640\tau^2}+O(\tau^{-3})\right)e^{-2\tau} \,, \label{Rmode1}
\end{eqnarray}

 \item  $\alpha_4$ mode
\begin{eqnarray}
\phi_3 & = & \frac{68\alpha_{4}}{35}\sqrt{\frac{2}{3}}\sqrt{\tau}\left(1+\frac{2441}{4760\tau}+\frac{5987}{10880\tau^2}+O(\tau^{-3})\right)e^{-10\tau/3}  \,,\nn\\
C_2^+ & = & \frac{28\alpha_{4}}{5}\tau\left(-1- \frac{1}{35\tau}+O(\tau^{-5})\right)e^{-10\tau/3} \,, \nn\\
C_2^- & = &  {\alpha_{4}}\tau\left(1+\frac{4}{5\tau}+O(\tau^{-5})\right)e^{-7\tau/3} \,, \nn\\
B_2 & = & {\alpha_{4}}\tau\left(1-\frac{17}{10\tau}+O(\tau^{-5})\right)e^{-7\tau/3} \,, \nn\\
C & = & \frac{16\alpha_{4}}{3}\tau\left(1- \frac{7}{10\tau}+O(\tau^{-5})\right)e^{-10\tau/3} \,, \nn\\
B & = & {5\alpha_{4}}\sqrt{\frac{2}{3}}\sqrt{\tau}\left(-1+\frac{1}{8\tau}+\frac{1}{128\tau^2}+O(\tau^{-3})\right)e^{-7\tau/3} \,, \nn\\
a & = & \frac{3\alpha_{4}}{35}\sqrt{\frac{2}{3}}\sqrt{\tau}\left(-1-\frac{1033}{280\tau}-\frac{2101}{4480\tau^2}+O(\tau^{-3})\right)e^{-10\tau/3} \,, \label{Rmode4}
\end{eqnarray}

    \item  $\alpha_5$ mode (in fact this modes appears to be a linear combination of modes $\alpha_1$ and $\alpha_5$ in solutions~(\ref{UVas1})-(\ref{UVas7}))
\begin{eqnarray}
\phi_3 & = & {4\alpha_{5}}\sqrt{\frac{2}{3}}\sqrt{\tau}\left(1+\frac{5}{8\tau}+\frac{83}{128\tau^2}+O(\tau^{-3})\right)e^{-2\tau}  \,,\nn\\
C_2^+ & = & 2\alpha_{5}\left(-1+O(\tau^{-5})\right)e^{-2\tau} \,, \nn\\
C_2^- & = &  {\alpha_{5}}\left(1+O(\tau^{-5})\right)e^{-\tau} \,, \nn\\
B_2 & = & {\alpha_{5}}\left(-1+O(\tau^{-5})\right)e^{-\tau} \,, \nn\\
C & = & 0 \,, \nn\\
B & = & 0 \,, \nn\\
a & = & 0 \label{Rmode5} \,,
\end{eqnarray}

\item  $\alpha_7$ mode
\begin{eqnarray}
\phi_3 & = & \frac{\alpha_{7}}{\sqrt{\tau}}\sqrt{\frac{2}{3}}\left(-1-\frac{7}{8\tau}-\frac{111}{128\tau^2}+O(\tau^{-3})\right)e^{-4\tau/3}  \,,\nn\\
C_2^+ & = & \alpha_{7}e^{-4\tau/3} \,, \nn\\
C_2^- & = &  \frac{13\alpha_{7}}{5}\left(-1+O(\tau^{-5})\right)e^{-7\tau/3} \,, \nn\\
B_2 & = & \frac{7\alpha_{7}}{5}\left(1+O(\tau^{-5})\right)e^{-7\tau/3} \,, \nn\\
C & = & \frac{8\alpha_{7}}{3}\left(-1+O(\tau^{-5})\right)e^{-4\tau/3} \,, \nn\\
B & = & {4\alpha_{7}}\sqrt{\frac{2}{3}}\sqrt{\tau}\left(1-\frac{1}{8\tau}-\frac{1}{128\tau^2}+O(\tau^{-3})\right)e^{-7\tau/3} \,, \nn\\
a & = & \frac{\alpha_{7}}{\sqrt{\tau}}\sqrt{\frac{3}{2}}\left(-1-\frac{1}{8\tau}-\frac{3}{128\tau^2}+O(\tau^{-3})\right)e^{-4\tau/3} \,, \label{Rmode7}
\end{eqnarray}

\item  $\alpha_{10}$ mode
\begin{eqnarray}
\phi_3 & = & \frac{\alpha_{10}}{\sqrt{\tau}}\sqrt{\frac{2}{3}}\left(-1-\frac{27}{8\tau}-\frac{391}{128\tau^2}+O(\tau^{-3})\right)e^{-4\tau/3}  \,,\nn\\
C_2^+ & = & {\alpha_{10}}\tau\left(1+\frac{5}{8\tau}\right)e^{-4\tau/3} \,, \nn\\
C_2^- & = &  \frac{4\alpha_{10}}{5}\tau^2\left(1+\frac{207}{160\tau^2}+O(\tau^{-5})\right)e^{-7\tau/3} \,, \nn\\
B_2 & = & \frac{4\alpha_{10}}{5}\tau^2\left(1-\frac{553}{160\tau^2}+O(\tau^{-5})\right)e^{-7\tau/3} \,, \nn\\
C & = & \frac{4\alpha_{10}}{3}\tau\left(-1+O(\tau^{-5})\right)e^{-4\tau/3} \,, \nn\\
B & = & {4\alpha_{10}}\sqrt{\frac{2}{3}}{\tau}^{3/2}\left(-1-\frac{33}{20\tau}+\frac{147}{640\tau^2}+O(\tau^{-3})\right)e^{-7\tau/3} \,, \nn\\
a & = & {\alpha_{10}}\sqrt{\frac{3}{2}}\sqrt{\tau}\left(-1-\frac{21}{8\tau}-\frac{43}{128\tau^2}+O(\tau^{-3})\right)e^{-4\tau/3} \,, \label{Rmode13}
\end{eqnarray}

    \item  $\alpha_{12}$-mode (this mode also appears to be mixed with $\alpha_1$ and $\alpha_5$ of solutions~(\ref{UVas1})-(\ref{UVas7}))
\begin{eqnarray}
\phi_3 & = & \frac{16\alpha_{12}}{5}\sqrt{\frac{2}{3}}{\tau}^{3/2}\left(1+\frac{47}{32\tau}+\frac{1689}{1280\tau^2}+O(\tau^{-3})\right)e^{-2\tau}  \,,\nn\\
C_2^+ & = & 0 \,, \nn\\
C_2^- & = &  {\alpha_{12}}{{\tau}}\left(1+O(\tau^{-5})\right)e^{-\tau} \,, \nn\\
B_2 & = & {\alpha_{12}}{{\tau}}\left(-1+\frac{1}{\tau}+O(\tau^{-5})\right)e^{-\tau} \,, \nn\\
C & = & 0 \,, \nn\\
B & = & {\alpha_{12}}{\sqrt{6\tau}}\left(1-\frac{1}{8\tau}-\frac{1}{128\tau^2}+O(\tau^{-3})\right)e^{-\tau} \,, \nn\\
a & = & \frac{4\alpha_{12}}{5}\sqrt{\frac{2}{3}}{\tau}^{3/2}\left(1+\frac{537}{640\tau^2}+O(\tau^{-3})\right)e^{-2\tau} \,, \label{Rmode12}
\end{eqnarray}

\item $\alpha_{14}$ mode
\begin{eqnarray}
\phi_3 & = & \frac{2\alpha_{14}}{3\sqrt{\tau}}\left(1+\frac{7}{8\tau}+\frac{111}{128\tau^2}+O(\tau^{-3})\right)e^{-2\tau/3}  \,,\nn\\
C_2^+ & = & 0 \,, \nn\\
C_2^- & = &  \frac{81\alpha_{14}}{320}\sqrt{\frac{3}{2}}m^2\left(1+O(\tau^{-5})\right)e^{-7\tau/3} \,, \nn\\
B_2 & = & \frac{189\alpha_{14}}{320}\sqrt{\frac{3}{2}}m^2\left(-1+O(\tau^{-5})\right)e^{-7\tau/3} \,, \nn\\
C & = & 0 \,, \nn\\
B & = & \frac{27\alpha_{14}}{16}m^2\sqrt{\tau}\left(-1+\frac{1}{8\tau}+\frac{1}{128\tau^2}+O(\tau^{-3})\right)e^{-7\tau/3} \,, \nn\\
a & = & \frac{\alpha_{14}}{\sqrt{\tau}}\left(1+\frac{1}{8\tau}+\frac{3}{128\tau^2}+O(\tau^{-3})\right)e^{-2\tau/3} \,, \label{Rmode14}
\end{eqnarray}

\end{itemize}{}

\subsection{Analysis of the UV modes}
\label{sec:UVcomp}

Now we would like to check the compatibility of the regular UV solutions with equation~(\ref{eqA}). The leading exponential asymptotic of this equation reads
\begin{multline}
\label{UVeq8}
   \frac{3 m^2 (4 \tau -1) }{4 \sqrt[3]{2}}e^{-\frac{4 \tau }{3}}a'  \ -\ \frac{3 m^2 }{2 \sqrt[3]{2}}e^{-\frac{4 \tau }{3}}a \ + \ \frac{8}{3} 2^{2/3} e^{-\frac{5 \tau }{3}} B' \ + \  \frac{8\ 2^{2/3}  (20 \tau -23) }{9 (1-4 \tau )}e^{-\frac{5 \tau }{3}}B \\ - \ \frac{64 \sqrt[6]{2}  (\tau -1)}{3 \sqrt{12 \tau -3}}e^{-\frac{5 \tau }{3}}B_2' \ -\ \frac{64 \sqrt[6]{2} \tau}{3 \sqrt{12 \tau -3}}e^{-\frac{5 \tau }{3}}{B_2}  \ + \ \frac{16 \sqrt[6]{2} }{3 \sqrt{12 \tau -3}}e^{-\frac{2 \tau }{3}} C \\ + \ \frac{64 \sqrt[6]{2}  (\tau -2) }{3 \sqrt{12 \tau -3}}e^{-\frac{5 \tau }{3}}{C_2^-}' \ + \ \frac{64 \sqrt[6]{2}  (\tau -1)}{3 \sqrt{12 \tau -3}}e^{-\frac{5 \tau }{3}}{C_2^-} \ - \ \frac{32 \sqrt[6]{2} }{3 \sqrt{12 \tau -3}}e^{-\frac{2 \tau }{3}} {C_2^+}' \\ + \ \frac{12\ 2^{2/3} m^2  (1-\tau )^2 }{1-4 \tau }e^{-\frac{4 \tau }{3}}\phi_3 \ = \ 0\,.
\end{multline}{}

Substituting the modes found above we checked that the modes $\alpha_{1}$~(\ref{Rmode1}), $\alpha_4$~(\ref{Rmode4}), $\alpha_5$~(\ref{Rmode5}), $\alpha_7$~(\ref{Rmode7}) and $\alpha_{12}$~(\ref{Rmode12}) satisfy equation~(\ref{UVeq8}) up to five orders in the leading exponential expansion. Although the modes $\alpha_{10}$~(\ref{Rmode13}) and $\alpha_{14}$~(\ref{Rmode14}) do not satisfy equation~(\ref{UVeq8}), their linear combination does satisfy the equation if one imposes the condition
    \beq
    \alpha_{10} \ = \ \frac{27}{16} \sqrt{\frac{3}{2}}m^2 \alpha_{14}\,.
    \eeq
As expected the physical system contains only six independent modes.

\begin{figure}
    \centering
    \includegraphics[width=0.47\linewidth]{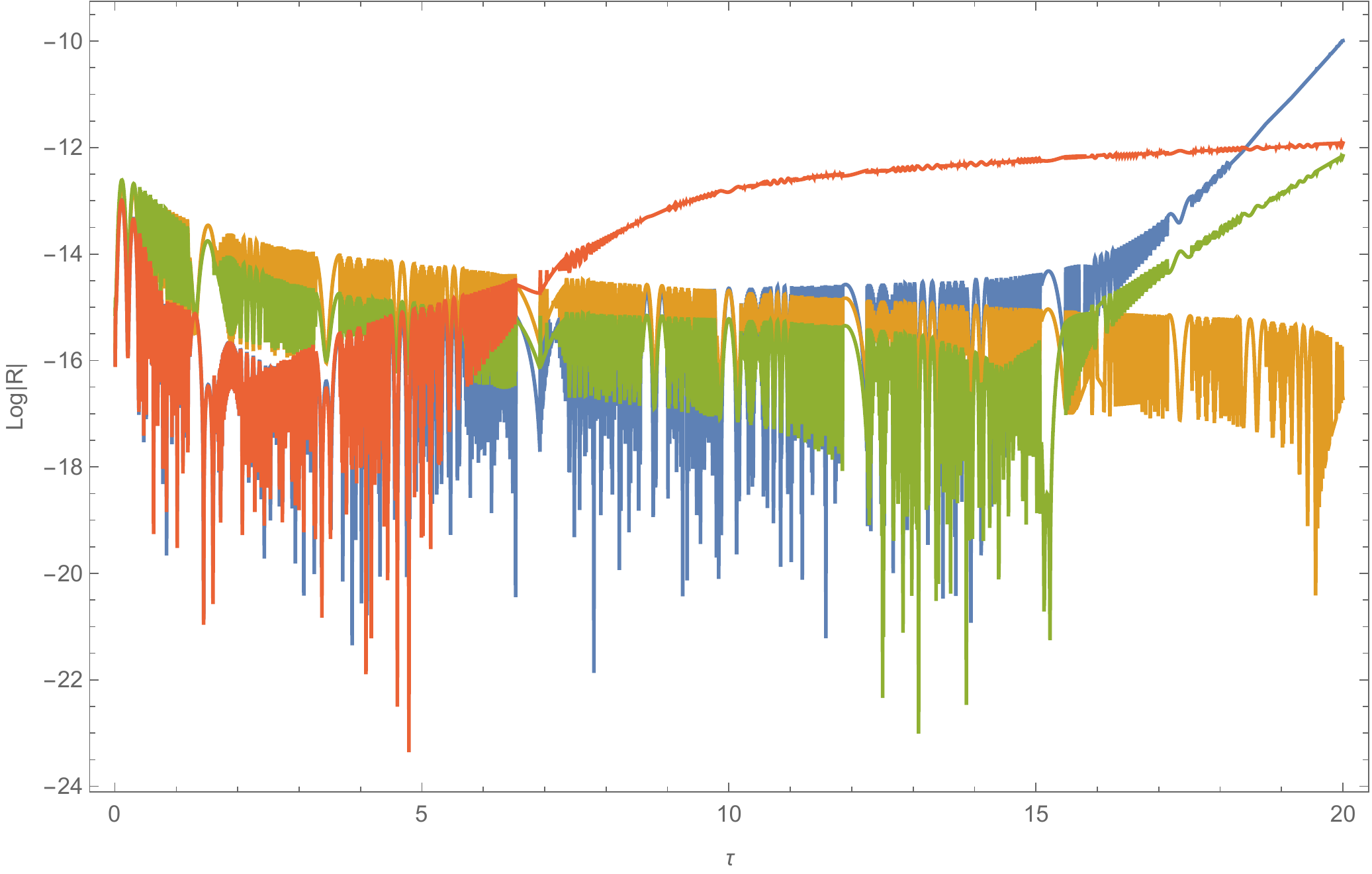}
    \hfill
    \includegraphics[width=0.47\linewidth]{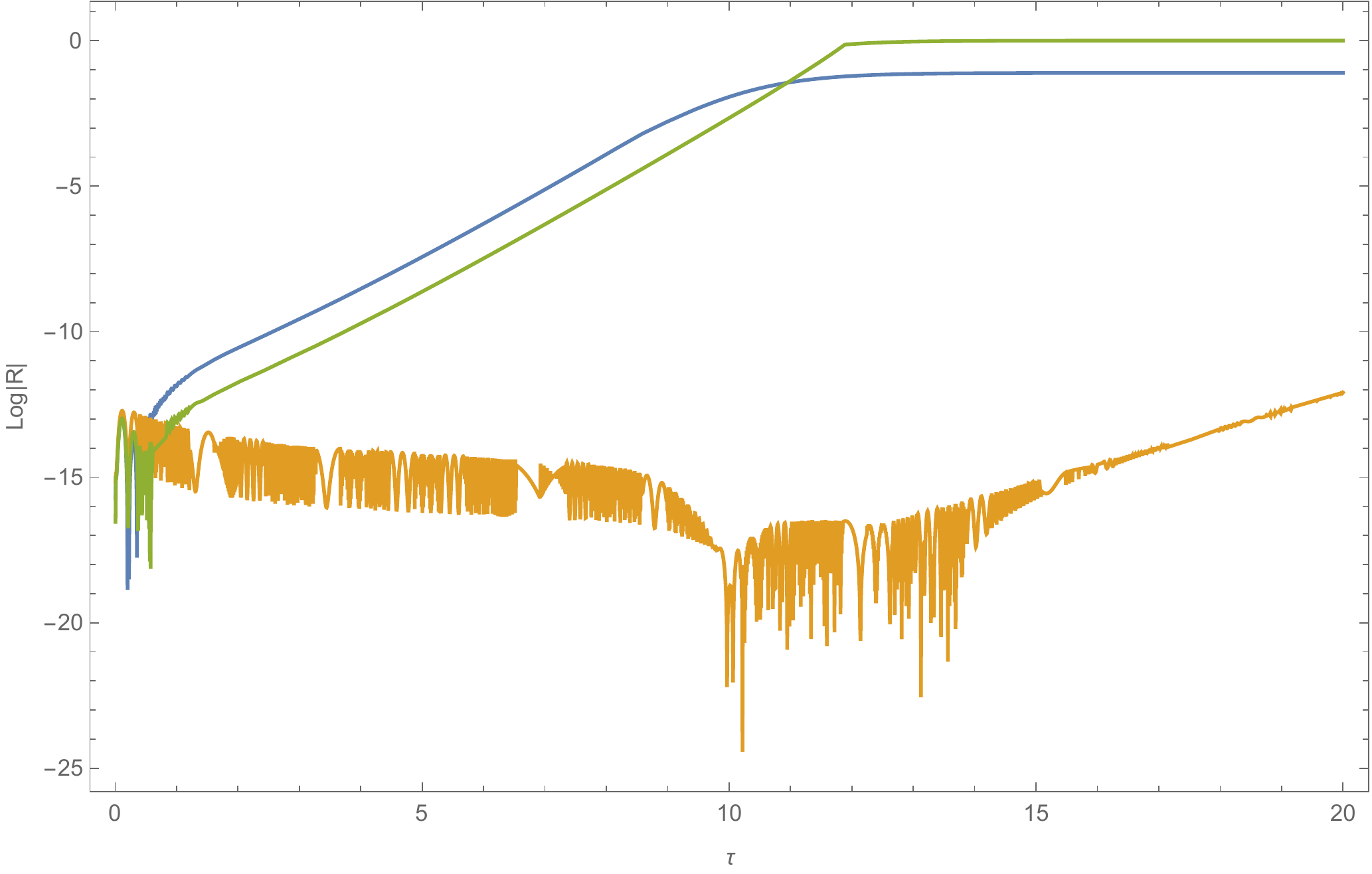}
    \caption{The logarithm of the ratio of the left hand side of equation~(\ref{eqA}) over the sum of absolute values of the terms of the same equation for seven regular UV solutions of the linearized equations. It can be seen that five of the seven regular modes satisfy equation~(\ref{eqA}) with the accuracy of the numerical analysis.}
    \label{fig:uvtoir}
\end{figure}

One can make some numerical tests of the exact constraint~(\ref{eqA}). For each regular $\alpha_i$ we plotted the (logarithm of the) ratio of the left hand side of equation~(\ref{eqA}) (which is supposed to vanish) over the sum of absolute values of the terms of the same equation. The results are shown on figure~\ref{fig:uvtoir}. As can be seen from that figure, all but two modes ($\alpha_{10}$ and $\alpha_{14}$) satisfy equation~(\ref{eqA}) with good accuracy. In fact, in the limit $\tau\to 0$, the numerical solution evolves $\alpha_{10}$ and $\alpha_{14}$ to modes that satisfy the constraint in the IR. Similarly, one can show that the above linear combination of $\alpha_{10}$ and $\alpha_{14}$ always satisfies the constraint.

To conclude the analysis of the UV asymptotic we need to comment on the regular versus singular classification of the modes. For this we  evaluate the action functional for the modes substituting the fields in the type IIB action. This gives:
\begin{multline}
    \mathcal{G}_{ab}\partial_N\phi^a\partial^M\phi^b\ \approx\ \frac{ e^{\frac{2 \tau }{3}} (4 \tau)  \left((\pt_\mu{B_2})^2\right)+ e^{\frac{4 \tau }{3}}  \left(({B_2}')^2+({B_2})^2\right)}{4 \tau}+\frac{e^{4\tau/3}C\ {C_2^+}'}{4\tau}+\\+\frac{e^{4\tau/3}C^2}{4\tau}+\frac{ 4 \tau e^{\frac{2 \tau }{3}} \left[(\pt_\mu{C_2^-})^2+(\pt_\mu{C_2^+})^2\right)]+ e^{\frac{4 \tau }{3}}  \left[({C_2^-}')^2+{C_2^-}+({C_2^+}')^2\right]}{4 \tau}+\\+e^{\frac{2 \tau }{3}}  \left[ e^{\frac{2 \tau }{3}}  (C')^2+ 4 \tau(\pt_\mu C)^2\right]+m^2 16 \tau^2 e^{-\frac{4 \tau }{3}}  (\phi_3)^2+\tau e^{-\frac{2 \tau }{3}}\left( \pt_\mu a+  \pt_\mu{\phi_3}' -\frac{\pt_\mu \phi_3}{2 \tau} \right)^2+\\+\tau^2\frac{e^{-\frac{4 \tau }{3}} \left(-12 \tilde{m}^2   {\phi_3}'+16\tilde{m}^2  \phi_3\right)^2}{72 \sqrt[3]{2}}+\tau^2 e^{-\frac{2 \tau }{3}} \left(\frac{\pt_\mu \phi_3}{2 \tau} - {\pt_\mu \phi_3}' \right)^2+\frac{8 \tau}{3} e^{-\frac{2 \tau }{3}}   \left(\pt_\mu \phi_3\right)^2+\\+\frac{e^{-\frac{4 \tau }{3}}}{3\  \sqrt{3} (4 \tau )^{5/2} } \bigg\{ 48\tau  e^{\frac{2 \tau }{3}}  B \big[(48 \tau^2  \pt_\mu \pt_\mu B-2  e^{\frac{2 \tau }{3}} \left(16 \tau {B'}+4 \tau  {B''}\right)\big]+8  e^{\frac{4 \tau }{3}} (80 \tau^2  B^2+\\+12 \tau   \pt_\mu a \bigg[96  \tau ^2 \pt_\mu {a'}-144 \tau^2 \pt_\mu a''+256\tau^2 \pt_\mu a \bigg]\bigg\}.
\label{UVaction}
\end{multline}
For this expression to be finite, the following conditions must be satisfied:
\ber
  \phi_3 & < & \tau^{-3/2}e^{\tau/3} \,, \\
  C_2^+ & < & e^{-2\tau/3}\,, \\
  C_2^- & < & e^{-2\tau/3}\,, \\
  B_2 & < &  e^{-2\tau/3} \,,\\
  C & < & e^{-2\tau/3}\,, \\
  B & < &  \tau^{-1/4} \,,\\
  a & < &  \tau^{-1}e^{\tau/3}\,.
\eer

We conclude that the above separation of the modes into singular and regular lists was correct.

\section{IR asymptotics}
\label{sec:IRasympt}

In this section we will summarize the asymptotic solution of linearized equations~(\ref{eqf3})-(\ref{eqA}) in the regime $\tau\to 0$ (IR).

First, we summarize the form of the equations when only the leading $\tau\to0$ terms in the coefficients are taken into account. We also neglect the terms which are subleading inhomogeneous (e.g. $a''\sim a/\tau^2 \gg a/\tau$).
\begin{multline}
\phi_3''+\frac{8 \phi_3'}{\tau}+\frac{10 \phi_3}{\tau^2}-\frac{2 a'}{\tau^2}-\frac{6 a}{\tau^3}+\frac{2\ 6^{2/3} \sqrt{I_0} {B_2}}{\tau^4}+\frac{2\ 6^{2/3} \sqrt{I_0} {C_2^-}}{\tau^5}+\\+\frac{2\ 6^{2/3} \sqrt{I_0} {C_2^+}}{\tau^5}=0
\label{f3eqIR}
\end{multline}
\begin{multline}
 {C_2^-}''-\frac{2 {C_2^+}'}{\tau}+\frac{C'}{6}-\frac{6^{1/3}B'}{6I_0^{1/2}\tau}-\frac{2{C_2^-}'}{\tau}-\frac{2\times6^{2/3}B_2}{9I_0\tau}-\frac{2^{1/3}\tilde{m}^2\tau^3\phi_3}{8\times 3^{1/3}I_0^{1/2}}+\\+\frac{6^{2/3}\tau C}{54I_0}+\frac{6^{1/3}B}{2I_0^{1/2}\tau^2}+\frac{3\tilde{m}^2I_0^{1/2}a}{2^{7/3}}=0
    \label{C2meqIR}
\end{multline}
\begin{multline}
{C_2^+}''-\frac{2{C_2^-}'}{\tau}-\frac{C'}{6}+\frac{6^{1/3}B'}{6I_0^{1/2}\tau}+\frac{2C_2^-}{\tau^2}+\frac{2\times 6^{2/3}B_2}{9I_0\tau}+\frac{2^{1/3}\tilde{m}^2\tau^3\phi_3}{8\times 3^{1/3}I_0^{1/2}}-\\-\frac{6^{2/3}\tau C}{54I_0}-\frac{6^{1/3}B}{2I_0\tau^2}-\frac{3 \tilde{m}^2I_0^{1/2}a}{2^{7/3}}=0
       \label{C2peqIR}
\end{multline}
\begin{multline}
B_2'' - \frac{2}{\tau^2}B_2 - \frac{\tau}{6}C' + \frac{6^{1/3}}{3I_0^{1/2}}B' - \frac{2^{2/3}\tau}{3^{4/3}I_0}C_2^- + \frac{3^{1/3}\tilde{m}^2\tau^4}{12\times 6^{2/3}I_0^{1/2}}\phi_3
\\ - \frac{6^{2/3}\tau^2}{54I_0}C - \frac{6^{1/3}}{6I_0^{1/2}}B - \frac{\tilde{m}^2I_0^{1/2}\tau}{2^{7/3}}a = 0\,,
    \label{B2eqIR}
\end{multline}
\begin{multline}
    C^{\prime\prime} + \frac{2}{\tau}C^\prime + \frac{6^{2/3}}{I_0\tau^2}C_2^{+\prime} + \frac{6^{2/3}}{I_0\tau^2}C_2^{-\prime}+\frac{2^{2/3}}{3^{1/3} I_0\tau}B_2^\prime - \frac{2^{2/3}}{3^{1/3} I_0\tau}C_2^-  \\+\frac{2^{2/3}}{3^{1/3}I_0\tau^2}B_2 + \frac{4}{3I_0^{3/2}}B = 0\,,
    \label{CeqIR}
\end{multline}
\begin{multline}
    B^{\prime\prime}-\frac{2\times 6^{1/3}}{3I_0^{1/2}\tau}C_2^{-\prime} - \frac{2\times 6^{1/3}}{3I_0^{1/2}\tau}C_2^{+\prime} - \frac{4\times 6^{1/3}}{3I_0^{1/2}}B_2^\prime  \\ - \frac{2\times 6^{1/3}}{3I_0^{1/2}\tau}B_2 + \frac{4\times 6^{1/3}\tau}{9I_0^{1/2}}C + \frac{3^{2/3}\tilde{m}^2I_0}{2^{2/3}}a = 0\,,
    \label{BeqIR}
\end{multline}
\begin{multline}
a^{\prime\prime} + \frac{2a^{\prime}}{\tau} - \frac{2a}{\tau^2} - \frac{6^{1/3}\tau^4}{27I_0^2}\phi_3^\prime -\frac{2}{\tau^2}B \\ + \frac{2\times6^{1/3}}{3I_0^{1/2}\tau^2}C_2^-  + \frac{2\times6^{1/3}}{3I_0^{1/2}\tau^2}C_2^+   + \frac{2\times6^{1/3}}{3I_0^{1/2}\tau}B_2  = 0\,,
\label{aeqIR}
\end{multline}
\begin{multline}
\label{AeqIR}
{C_2^+}'+{C_2^-}' + \tau {B_2}' -\tau C_2^- + \frac{2\tau^2}{3}C - \frac{3I_0^{1/2}}{6^{1/3}}B' \\ - \frac{3^{4/3}}{4}\tilde{m}^2I_0^{3/2}\tau^2a' + \frac{3^{1/3}\tilde{m}^2\tau^6}{6^{5/3}I_0^{1/2}}\phi_3 = 0\,.
\end{multline}
Note that we diagonalized the pair of the equations containing the second derivatives of $C_2^\pm$. The pair of the resulting equations imply a very simple equations for the sum $P=C_2^++C_2^-$:
\beq
P'' - \frac{2}{\tau}P' \ = \ 0\,.
\eeq

We have not attempted to completely solve the reduced system, but instead considered a simplified system~(\ref{eqf3IRs})-(\ref{eqaIRs}), which gave the same spectrum of exponents $\beta_i$, according to the notation in~(\ref{IRexpansion}). Below we list the expansion of the solutions to the full system. 

In the IR limit the expressions are typically more bulky, so we restrict ourselves to either the leading order (LO) or NLO. The general principle of organization of the expressions below is that only those orders are listed that can be fit in a single line. Indeed, more complex form of the coefficients typically mean that they correspond to higher orders and are likely to be neglected. In the numerical calculations we use fuller expressions if necessary, to control the precision. The expressions here are listed in order for the results to be reproducible. 

We label the modes in accordance to the classification of the simplified system~(\ref{Bmode12})-(\ref{Bmode1314}). We will first list the modes with singular behavior at $\tau\to 0$ in section~\ref{sec:IRsingular} and then the modes with regular behavior in section~\ref{sec:IRregular}. We will justify the classification in section~\ref{sec:IRaction}.

\subsection{Singular IR solutions}
\label{sec:IRsingular}

The system of seven second order differential equations has seven modes with singular behavior at $\tau\to 0$. These modes are analogous to $\beta_i$ with odd $i$ of the simplified system~(\ref{Bmode12})-(\ref{Bmode1314}). Here we list the modes following their simplified classification

\begin{itemize}
    \item the mode similar to $\beta_1$ in~(\ref{Bmode12}) is
{\small
\ber
\phi_3 & = & \frac{\beta_1}{\tau^5}\bigg(1+\left(\frac{1}{2} \left(\frac{3}{2}\right)^{2/3} I_0 m^2-\frac{1}{18 \sqrt[3]{6} I_0}+\frac{7}{15}\right)\tau^2+ O(\tau^4)\bigg)\,,  \nn \\
C_2^+ & = &  \beta_1\left(\frac{\tilde{m}^2}{8\ 2^{2/3} \sqrt[3]{3} \sqrt{I_0}}-\frac{1}{27 I_0^{3/2}}\right)\left(1+\frac{\tilde{m}^2 \tau^2 \left(-405\ 2^{2/3} I_0^2 \tilde{m}^2+126 \sqrt[3]{6} I_0+100\right)}{540 \sqrt[3]{6} I_0 \tilde{m}^2-320\ 3^{2/3}}\right)\,, \nn \\
C_2^- & = &  \frac{\beta_1\tilde{m}^2}{8\ 2^{2/3} \sqrt[3]{3} \sqrt{I_0}}\left(-1 +\tau^2 \left(\frac{1}{2} \left(\frac{3}{2}\right)^{2/3} I_0 \tilde{m}^2-\frac{5}{27 \sqrt[3]{6} I_0}-\frac{23}{45}\right)+ O(\tau^4)\right)\,, \nn\\
B_2 & = &  \frac{\beta_1\tilde{m}^2 \tau}{24\ 2^{2/3} \sqrt[3]{3} \sqrt{I_0}}\left(1+\tau^2 \left(\frac{4}{15}-\frac{1}{2} \left(\frac{3}{2}\right)^{2/3} I_0 m^2-\frac{5}{3 \sqrt[3]{6} I_0}\right)+O(\tau^4)\right)\,, \nn \\
C & = & \frac{\beta_1\tilde{m}^2 \tau^3 \left(4860 \sqrt[3]{2} I_0^2 \tilde{m}^2+3 \sqrt[3]{3} I_0 \left(55 \sqrt[3]{3} \tilde{m}^2-1152\right)-176\ 2^{2/3}\right)}{1080 I_0^{3/2} \left(648 I_0+11\ 6^{2/3}\right)} + O(\tau^5)\,, \nn \\
B & = &  \frac{7\beta_1 \tilde{m}^2 \tau^2}{108 \sqrt[3]{2} I_0}\bigg(1-\frac{\tau^2\left(-30456 I_0^2+7043\ 6^{2/3} I_0+770 \sqrt[3]{6}\right) }{420 I_0 \left(648 I_0+11\ 6^{2/3}\right)} \nn \\
&& - \frac{15 \sqrt[3]{3} I_0\tilde{m}^2 \tau^2 \left(108 \sqrt[3]{6} I_0+11\right) }{9072 I_0+154\ 6^{2/3}}+O(\tau^4)\bigg)\,, \nn \\
a & = &  \frac{2\beta_1\sqrt[3]{2}}{27\ 3^{2/3} I_0^2}\left(1+\tau^2\left(\frac{1}{6 \sqrt[3]{6} I_0}+\frac{3}{5}\right)+O(\tau^4)\right)\,,
\label{beta1}
\eer
}

  \item $\beta_3$ mode
{\small
\ber
\phi_3 & = & -\frac{4\beta_3 \left(45\ 6^{2/3} I_0^2 \tilde{m}^2-18 \sqrt[3]{2} I_0-5\ 3^{2/3}\right)}{135 I_0^{3/2} \tilde{m}^2 \tau^3} + O(\tau^{-1})  \,,\nn\\
C_2^+ & = & \frac{\beta_{3}}{\tau^2}\left(-1 + \frac{\tau^2 \left(-135\ 3^{2/3} I_0^3 \tilde{m}^4+3 \sqrt[3]{6} I_0 \left(35 \sqrt[3]{3} \tilde{m}^2+16\right)-80\right)}{270\ 2^{2/3} I_0^2 \tilde{m}^2} + O(\tau^4)\right) \,, \nn\\
C_2^- & = & \frac{\beta_3}{\tau^2}\left(1+ \tau^2 \left(\frac{1}{2} \left(\frac{3}{2}\right)^{2/3} I_0 \tilde{m}^2-\frac{7}{6 \sqrt[3]{6} I_0}+\frac{1}{6}\right) + O(\tau^4)\right) \,, \nn\\
B_2 & = & \frac{\beta_3\tau}{6 \sqrt[3]{6} I_0}\left(-1+ \tau^2 \left(\left(\frac{3}{2}\right)^{2/3} I_0 \tilde{m}^2-\frac{5}{3 \sqrt[3]{6} I_0}+\frac{14}{15}\right) + O(\tau^4)\right) \,, \nn\\
C & = & -\frac{\beta_3\tau^3 \left(68040 \sqrt[3]{3} I_0^3 \tilde{m}^2-9 I_0^2 \left(5904\ 6^{2/3}-385\ 2^{2/3} \tilde{m}^2\right)+115548 \sqrt[3]{6} I_0+12320\right)}{9450 I_0^2 \left(648 I_0+11\ 6^{2/3}\right)}\,, \nn\\
B & = & \frac{\beta_3\tau^2 \left(-9\ 2^{2/3} I_0^2 \tilde{m}^2-24 \sqrt[3]{6} I_0+28\right)}{108 I_0^{3/2}} + O(\tau^4) \,, \nn\\
a & = &  \frac{4\beta_3}{3 \sqrt[3]{3} I_0^{3/2} \tilde{m}^2 \tau^2}\left(-1+ \tau^2 \left(\sqrt[3]{\frac{2}{3}} (-I_0) \tilde{m}^2+\frac{5}{18 \sqrt[3]{6} I_0}+\frac{4}{15}\right) + O(\tau^4)\right)\,,
\eer
}

  \item $\beta_5$ mode
{\small
\ber
\phi_3 & = &  \frac{\beta_56^{2/3} \sqrt{I_0}}{\tau^3}\left(-1+ \tau^2 \left(-\frac{1}{2} \left(\frac{3}{2}\right)^{2/3} I_0 \tilde{m}^2-\frac{1}{6 \sqrt[3]{6} I_0}+\frac{3}{10}\right) + O(\tau^4)\right) \,,\nn\\
C_2^+ & = &  \beta_5\left(1-\frac{\tilde{m}^2 \tau^2}{8\ 3^{2/3}}-\frac{\tilde{m}^2 \tau^4 \left(270\ 2^{2/3} \sqrt[3]{3} I_0^2 \tilde{m}^2+45 \sqrt[3]{2} I_0 \left(8\ 3^{2/3}-5 \tilde{m}^2\right)+322 \sqrt[3]{3}\right)}{8640}\right), \nn\\
C_2^- & = & \frac{\left(6\beta_5 \sqrt[3]{6} I_0+1\right) \tilde{m}^2 \tau^2}{8\ 3^{2/3}} + O(\tau^4) \,, \nn\\
B_2 & = &  -\frac{\beta_5}{8} \sqrt[3]{3} \tilde{m}^2 \tau^3 + O(\tau^5)\,, \nn\\
C & = & \frac{3\beta_5}{2} \sqrt[3]{3} \tilde{m}^2 \tau \left(-1 + \tau^2 \left(\frac{I_0 \tilde{m}^2}{2\ 2^{2/3} \sqrt[3]{3}}+\frac{17}{90}\right) + O(\tau^4)\right)\,, \nn\\
B & = & \frac{\beta_5\sqrt{I_0} \tilde{m}^2 \tau ^2}{2 \sqrt[3]{2}}\left(1 + \frac{\tau^2 \left(110\ 6^{2/3}-7290\ 2^{2/3} I_0^3 \tilde{m}^2+9 \sqrt[3]{6} I_0^2 \left(55 \sqrt[3]{3} \tilde{m}^2-1332\right)-798 I_0\right)}{360 I_0 \left(54 \sqrt[3]{6} I_0-11\right)}\right) , \nn\\
a & = &  \frac{\beta_5\sqrt[3]{2}}{3^{2/3} \sqrt{I_0}}\left(1 + \tau^2\left(\frac{1}{6 \sqrt[3]{6} I_0}-\frac{3}{10}\right)  + O(\tau^4)\right) \,,
\eer
}

  \item $\beta_7$ mode (this mode in fact is a combination of $\beta_7$ and $\beta_{13}$, while the proper $\beta_{13}$ has logarithmic coefficients, as one can see below)
\ber
\phi_3 & = &  \frac{2\beta_7 \left(45\ 6^{2/3} I_0^2 \tilde{m}^2+36 \sqrt[3]{2} I_0+10\ 3^{2/3}\right)}{135 I_0^{3/2} \tilde{m}^2 \tau ^3} + O(\tau^{-1}) \,, \nn\\
C_2^+ & = &  \beta_7\left(-\frac{4 \sqrt[3]{2}}{27 I_0^2 \tilde{m}^2}+\frac{4 \left(\frac{2}{3}\right)^{2/3}}{15 I_0 \tilde{m}^2}+\frac{7}{6 \sqrt[3]{6} I_0}-\frac{1}{3}\right) + O(\tau^2)\,, \nn\\
C_2^- & = & -\frac{7\beta_7}{6 \sqrt[3]{6} I_0} + O(\tau^2)\,, \nn\\
B_2 & = & \frac{\beta_7}{\tau}\left(1 + \tau ^2 \left(\frac{1}{2} \left(\frac{3}{2}\right)^{2/3} I_0 \tilde{m}^2-\frac{1}{6 \sqrt[3]{6} I_0}-\frac{1}{6}\right) + O(\tau^4)\right) \,, \nn\\
C & = &  -\frac{\beta_7\tau^3 \left(68040 \sqrt[3]{3} I_0^3 \tilde{m}^2-9 I_0^2 \left(5904\ 6^{2/3}-385\ 2^{2/3} \tilde{m}^2\right)+115548 \sqrt[3]{6} I_0+12320\right)}{9450 I_0^2 \left(648 I_0+11\ 6^{2/3}\right)}, \nn\\
B & = &  \frac{\beta_7\tau^2 \left(45\ 2^{2/3} I_0^2 \tilde{m}^2-24 \sqrt[3]{6} I_0+28\right)}{108 I_0^{3/2}} + O(\tau^4)\,, \nn\\
a & = & \frac{4\beta_7}{3^{4/3} I_0^{3/2} \tilde{m}^2 \tau ^2}\left(-1 + \tau ^2 \left(\frac{I_0 \tilde{m}^2}{2^{2/3} \sqrt[3]{3}}+\frac{5}{18 \sqrt[3]{6} I_0}+\frac{4}{15}\right) + O(\tau^4)\right),
\label{beta3}
\eer

  \item $\beta_9$ mode
\ber
\phi_3 & = &  \frac{\beta_9}{\tau^3}\left(\frac{4 \left(\frac{2}{3}\right)^{2/3} \sqrt{I_0}}{15 \tilde{m}^2}-I_0^{3/2}\right) + O(\tau^{-1})\,,\nn\\
C_2^+ & = & \beta_9\left(\frac{1}{12}-\frac{I_0}{6^{2/3}}+\frac{4}{45 \sqrt[3]{3} \tilde{m}^2}\right) + \frac{\beta_9\tau ^2 \left(41-15 \sqrt[3]{2} 3^{2/3} I_0 \tilde{m}^2\right)}{1080} + O(\tau^4) \,, \nn\\
C_2^- & = & \frac{\beta_9}{12}\left(-1 + \frac{\tau ^2}{90}  \left(94-135\ 2^{2/3} I_0^2 \tilde{m}^2+3 \sqrt[3]{6} I_0 \left(5 \sqrt[3]{3} \tilde{m}^2+8\right)\right) + O(\tau^4)\right) , \nn\\
B_2 & = & \frac{\beta_9\tau}{12}\left(1 + \frac{2\tau^2}{15} + O(\tau^4)\right) \,, \nn\\
C & = & \frac{\beta_9}{\tau}\left(1 + O(\tau^4)\right)\,, \nn\\
B & = &  \frac{\beta_9}{540} \tau^2 \left(4\ 6^{2/3} \sqrt{I_0}-45 \sqrt[3]{3} I_0^{3/2} \tilde{m}^2\right)\,, \nn\\
a & = & \beta_9 \frac{8 \sqrt[3]{2}-15\ 6^{2/3} I_0 \tilde{m}^2}{270 \sqrt{I_0} \tilde{m}^2} + O(\tau^2)\,,
\eer

\item $\beta_{11}$ mode
{\small
\ber
\phi_3 & = &  0 \,,\nn\\
C_2^+ & = & -\frac{\beta_{11}\sqrt[3]{3}}{2\ 2^{2/3} \sqrt{I_0}} \,, \nn\\
C_2^- & = & \frac{\beta_{11}\sqrt[3]{3}}{2\ 2^{2/3} \sqrt{I_0}}\left(1-\frac{\tau ^2}{6}+\frac{7 \tau ^4}{360} + O(\tau^6)\right) \,, \nn\\
B_2 & = & \frac{\beta_{11}\tau}{2\ 6^{2/3} \sqrt{I_0}}\left(-1+\frac{7 \tau^2}{30}-\frac{31 \tau^4}{840}+O(\tau^6)\right) \,, \nn\\
C & = & 0\,, \nn\\
B & = & \beta_{11}\left(1 -\tau^2\left(\frac{1}{6 \sqrt[3]{6} I_0}+\frac{1}{10}\right)  +\frac{\tau^4\left(83592 I_0^3+61899\ 6^{2/3} I_0^2-12740 \sqrt[3]{6} I_0-1925\right) }{12600 I_0^2 \left(648 I_0+11\ 6^{2/3}\right)}\right) , \nn\\
a & = & \beta_{11}\left(-1 + \tau^2\left(\frac{1}{6 \sqrt[3]{6} I_0}-\frac{1}{5}\right)  + O(\tau^4)\right) \,,
\eer
}

\item $\beta_{13}$ mode (this mode mixes with $\beta_7$ and hence has logarithmic coefficients)
{\small
\ber
\phi_3 & = &  \beta_{13}\left(\frac{1}{4} \left(\frac{3}{2}\right)^{2/3} I_0 \tilde{m}^2-1+\frac{1}{2} \left(\frac{3}{2}\right)^{2/3} 3 I_0 \tilde{m}^2 \log \tau + O(\tau^2)\right) \,,\nn\\
C_2^+ & = & \frac{\beta_{13}\sqrt{I_0} \tilde{m}^2 (6 \log\tau +1)}{8 \sqrt[3]{2}} + O(\tau^2)\,, \nn\\
C_2^- & = &  \frac{\beta_{13}\tilde{m}^2 \tau ^2 \left(480\ 3^{2/3} I_0^2 \tilde{m}^2-60\ 2^{2/3} \sqrt[3]{3} I_0 \tilde{m}^2 \log\tau + 2^{2/3} I_0 \left(35 \sqrt[3]{3} \tilde{m}^2-108\right)-60 \sqrt[3]{2} 3^{2/3}\right)}{3840 \sqrt{I_0}} \,, \nn\\
B_2 & = & \frac{\beta_{13}\sqrt{I_0} \tilde{m}^2 \tau  (3-2 \log\tau)}{8 \sqrt[3]{2}} + O(\tau^3) \,, \nn\\
C & = & -\frac{\beta_{13}\tilde{m}^2 \tau  \left(27 \sqrt[3]{2} I_0 \tilde{m}^2 \log \tau  - 9 \sqrt[3]{2} I_0 \tilde{m}^2+2 \sqrt[3]{3}\right)}{8\ 6^{2/3} \sqrt{I_0}} + O(\tau^3) \,, \nn\\
B & = & \beta_{13}\left(\frac{3}{2}\right)^{2/3} I_0 \tilde{m}^2 \log\tau + O(\tau^2) \,, \nn\\
a & = &  \frac{\beta_{13}}{\tau^2}\left(1+O(\tau^2)\right)\,,
\eer
}
\end{itemize}

\subsection{Regular IR solutions}
\label{sec:IRregular}

In this section we list the seven regular asymptotic solutions of system~(\ref{eqf3})-(\ref{eqA}) in the IR limit. In terms of the modes~(\ref{Bmode12})-(\ref{Bmode1314}) these contain

\begin{itemize}
    \item $\beta_{2}$ mode
{\small
\ber
\phi_3 & = & \frac{\beta_2}{\tau^2}\left(1+\tau^2 \left(-\frac{1}{10} \left(\frac{3}{2}\right)^{2/3} I_0 m^2-\frac{13}{30 \sqrt[3]{6} I_0}+\frac{2}{3}\right)+ O(\tau^4)\right)  \,, \nn\\
C_2^+ & = &  \frac{\beta_2\tilde{m}^2 \tau^3}{40\ 2^{2/3} \sqrt[3]{3} \sqrt{I_0}}\left(-1+O(\tau^2)\right)\,, \nn \\
C_2^- & = & \frac{\beta_2\tilde{m}^2 \tau^3}{40\ 2^{2/3} \sqrt[3]{3} \sqrt{I_0}}\left(1 + \frac{\tau^2 \left(135\ 2^{2/3} I_0^2 \tilde{m}^2+60 \sqrt[3]{6} I_0+304\right)}{1260 \sqrt[3]{6} I_0}+ O(\tau^4)\right)\,, \nn\\
B_2 & = &  -\frac{\beta_2\tilde{m}^2 \tau^4}{120\ 2^{2/3} \sqrt[3]{3} \sqrt{I_0}} + O(\tau^6) \,, \nn\\
C & = & \frac{\beta_2\left(15 \sqrt[3]{3} I_0+2^{2/3}\right) \tilde{m}^2 \tau^4}{120 I_0^{3/2} \left(45 I_0+6^{2/3}\right)} + O(\tau^6)\,, \nn\\
B & = & -\frac{\beta_2\left(45\ 2^{2/3} I_0+2 \sqrt[3]{2} 3^{2/3}\right) \tilde{m}^2 \tau^5}{300 I_0 \left(45 I_0+6^{2/3}\right) + O(\tau^7)} \,, \nn\\
a & = &  \frac{\beta_2\tau^3}{45\ 6^{2/3} I_0^2}+ O(\tau^5)\,,
\eer
}

\item $\beta_{4}$ mode
{\small
\ber
\phi_3 & = & \frac{\beta_42^{5/3} \sqrt{I_0}}{5 \sqrt[3]{3}}\left(1+\frac{\tau^2 \left(171 \sqrt[3]{6} I_0-675 \sqrt[3]{3} I_0^3 \tilde{m}^2+45 I_0^2 \left(41\ 6^{2/3}-2^{2/3} \tilde{m}^2\right)-10\right)}{70\ 6^{2/3} I_0 \left(45 I_0+6^{2/3}\right)}\right)  \,, \nn\\
C_2^+ & = & \beta_4\tau\left(-1+ \tau^2 \left(\frac{1}{10} \left( \frac{3}{2}\right)^{2/3} I_0 \tilde{m}^2+\frac{4}{15}\right)+O(\tau^4)\right) \,,\nn \\
C_2^- & = &  \beta_4\tau\left(1+ \tau^2 \left(\frac{17}{30}-\frac{1}{10} \left(\frac{3}{2}\right)^{2/3} I_0 \tilde{m}^2\right)+O(\tau^4)\right)\,,\nn \\
B_2 & = &  \frac{\beta_4\tau^6 \left(90 \sqrt[3]{3} I_0^2 \tilde{m}^2+6\ 2^{2/3} I_0 \tilde{m}^2+285\ 6^{2/3} I_0+38 \sqrt[3]{6}\right)}{1050 I_0 \left(45 I_0+6^{2/3}\right)} + O(\tau^8)\,,\nn \\
C & = &\frac{\beta_4 2^{2/3} \tau^2}{\sqrt[3]{3} I_0}\left(-1+\frac{\tau^2 \left(45 \left(\frac{3}{2}\right)^{2/3} I_0^2 \tilde{m}^2+3 I_0 \left(\sqrt[3]{3} \tilde{m}^2+25\right)+\frac{5\ 2^{2/3}}{\sqrt[3]{3}}\right)}{10 \left(45 I_0+6^{2/3}\right)}+O(\tau^4)\right)\,, \nn\\
B & = &  \frac{\beta_4\tau^5 \left(-45\ 2^{2/3} I_0^2 \tilde{m}^2-2 \sqrt[3]{6} I_0 \left(\sqrt[3]{3} \tilde{m}^2+180\right)-48\right)}{150 \sqrt{I_0} \left(45 I_0+6^{2/3}\right)} + O(\tau^7)\,,\nn \\
a & = & \frac{2\beta_4 \sqrt[3]{2} \tau^3}{15\ 3^{2/3} \sqrt{I_0}} + O(\tau^5) \,,
\eer
}

\item $\beta_{6}$ mode
{\small
\ber
\phi_3 & = & -\frac{3\beta_6 \sqrt[3]{6} I_0}{5}  + O(\tau^2) \,, \nn\\
C_2^+ & = &  \frac{\beta_6\left(4\ 6^{2/3} I_0-\sqrt[3]{6}\right) \tau^3}{20 \sqrt{I_0}} + O(\tau^5) \,, \nn\\
C_2^- & = &\frac{\beta_6\left(6\ 6^{2/3} I_0+\sqrt[3]{6}\right) \tau^3}{20 \sqrt{I_0}} \left(1 - \frac{\tau^2 \left(360\ 2^{2/3} I_0^2 \tilde{m}^2+5 \sqrt[3]{6} I_0 \left(7 \sqrt[3]{3} \tilde{m}^2+36\right)+1016\right)}{420 \left(6 \sqrt[3]{6} I_0+1\right)} \right)\,, \nn\\
B_2 & = & \frac{\beta_6 2^{4/3} \tau^4}{5\ 3^{2/3} \sqrt{I_0}} \left(-1 + \frac{\tau^2 \left(2025\ 2^{2/3} I_0^2 \tilde{m}^2+30 \sqrt[3]{6} I_0 \left(3 \sqrt[3]{3} \tilde{m}^2+122\right)+488\right)}{840 \left(15 \sqrt[3]{6} I_0+2\right)}+O(\tau^4)\right)\,, \nn\\
C & = &  \frac{3\beta_6 \sqrt[3]{3} \tau^2}{2^{2/3} \sqrt{I_0}}\left(-1+\frac{\tau^2 \left(135\ 2^{2/3} I_0^2 \tilde{m}^2+6 \sqrt[3]{6} I_0 \left(\sqrt[3]{3} \tilde{m}^2+25\right)+20\right)}{60 \left(15 \sqrt[3]{6} I_0+2\right)} + O(\tau^4)\right)\,, \nn\\
B & = & \beta_6\tau^3\left(1 - \frac{t^2 \left(675 \sqrt[3]{2} 3^{2/3} I_0^3 \tilde{m}^2+90 I_0^2 \left(\sqrt[3]{3} \tilde{m}^2+39\right)+453\ 6^{2/3} I_0+50 \sqrt[3]{6}\right)}{300 I_0 \left(45 I_0+6^{2/3}\right)} \right) \,, \nn\\
a & = &\frac{\beta_6\left(-270 I_0^2+69\ 6^{2/3} I_0+10 \sqrt[3]{6}\right) \tau^5}{525 I_0 \left(45 I_0+6^{2/3}\right)} + O(\tau^7) \,,
\eer
}

\item $\beta_{8}$ mode
{\small
\ber
\phi_3 & = & \frac{\beta_8 2^{5/3} \sqrt{I_0}}{5 \sqrt[3]{3}} + O(\tau^2) \,,\nn\\
C_2^+ & = &  -\frac{2}{3}\beta_8\tau^3 + O(\tau^5)\,, \nn\\
C_2^- & = &\beta_8\tau^3\left(-1+\frac{\tau^2 \left(150 \sqrt[3]{2} 3^{2/3} I_0^2 \tilde{m}^2+7 \sqrt[3]{3} I_0 \tilde{m}^2+225 I_0+164\ 6^{2/3}\right)}{3150 I_0}+ O(\tau^4)\right) \,, \nn\\
B_2 & = & \beta_8\tau^2\left(1+\tau^2 \left(\frac{1}{30}-\frac{1}{10} \left(\frac{3}{2}\right)^{2/3} I_0 \tilde{m}^2\right)+O(\tau^4)\right) \,, \nn\\
C & = &  \frac{\beta_8 2^{5/3} \tau^2}{\sqrt[3]{3} I_0}\left(1-\frac{\tau^2 \left(45 \left(\frac{3}{2}\right)^{2/3} I_0^2 \tilde{m}^2+3 I_0 \left(\sqrt[3]{3} \tilde{m}^2+25\right)+\frac{5\ 2^{2/3}}{\sqrt[3]{3}}\right)}{10 \left(45 I_0+6^{2/3}\right)}+O(\tau^4)\right) \,, \nn\\
B & = & \frac{\beta_8\tau^5 \left(-45\ 2^{2/3} I_0^2 \tilde{m}^2-2 \sqrt[3]{6} I_0 \left(\sqrt[3]{3} \tilde{m}^2-360\right)+96\right)}{150 \sqrt{I_0} \left(45 I_0+6^{2/3}\right)}+ O(\tau^7) \,, \nn\\
a & = & \frac{\beta_8 2^{4/3} \tau^3}{15\ 3^{2/3} \sqrt{I_0}} + O(\tau^5) \,,
\eer
}

\item $\beta_{10}$ mode
{\small
\ber
\phi_3 & = &  -\frac{\beta_{10} 2^{5/3} \sqrt{I_0} }{15 \sqrt[3]{3}} + O(\tau^2)\,,\nn\\
C_2^+ & = & \frac{\beta_{10}}{360} \tau^3 \left(32-\sqrt[3]{2} 3^{2/3} I_0 \tilde{m}^2\right) + O(\tau^5) \,, \nn\\
C_2^- & = & \frac{\beta_{10}}{360} \tau^3 \left(\sqrt[3]{2} 3^{2/3} I_0 \tilde{m}^2+48\right) + O(\tau^5) \,, \nn\\
B_2 & = & -\frac{\beta_{10}I_0 \tilde{m}^2 \tau^4}{60\ 2^{2/3} \sqrt[3]{3}} + O(\tau^6) \,, \nn\\
C & = & \beta_{10}\left(1-\frac{\tau^2 \left(3 \sqrt[3]{2} 3^{2/3} I_0^2 \tilde{m}^2+2\ 6^{2/3}\right)}{36 I_0}+O(\tau^4)\right)\,, \nn\\
B & = & \frac{\beta_{10}\tau^5 \left(-45\ 2^{2/3} I_0^2 \tilde{m}^2-2 \sqrt[3]{6} I_0 \left(\sqrt[3]{3} \tilde{m}^2+45\right)-12\right)}{225 \sqrt{I_0} \left(45 I_0+6^{2/3}\right)} + O(\tau^5) \,, \nn\\
a & = & -\frac{\beta_{10}2^{4/3}\tau^3}{45\ 3^{2/3} \sqrt{I_0}} + O(\tau^5) \,,
\eer
}

\item $\beta_{12}$ mode (note that this mode contains logarithmic terms in the expansion)
{\small
\ber
\phi_3 & = &  -\frac{2\beta_{12}  \left(5 I_0 \left(9 I_0 \tilde{m}^2+4\ 2^{2/3} \sqrt[3]{3}\right) \log\tau+2 \sqrt[3]{2} \left(29 \sqrt[3]{6} I_0-35\right)\right)}{15\tilde{m}^2\sqrt{I_0} \tau^2}  + O(\tau^0)\,, \nn \\
C_2^+ & = &  - \beta_{12}\tau\log\tau + O(\tau^5)\,, \nn\\
C_2^- & = &  \beta_{12}\tau\log\tau + O(\tau^5)\,, \nn\\
B_2 & = &  \frac{\beta_{12}}{2}\tau^2\log\tau + O(\tau^4)
\,, \nn\\
C & = & \frac{2^{2/3} \beta_{12} \tau^2}{\sqrt[3]{3} I_0} + O(\tau^4)\,, \nn\\
B & = &  -\frac{3\times 3^{2/3}\beta_{12} \sqrt{I_0} \tau}{\sqrt[3]{2}} + O(\tau^3)\,, \nn\\
a & = &  -6^{2/3}\beta_{12} \sqrt{I_0} \tau \log\tau + O(\tau^3)\,,
\eer
}

\item $\beta_{14}$ mode
{\small
\ber
\phi_3 & = &  \frac{\beta_{14}}{20} \left(16-3\ 2^{2/3} I_0^2 \tilde{m}^2\right) + O(\tau^2)\,,\nn\\
C_2^+ & = & \frac{\beta_{14}}{40} \sqrt{I_0} \left(4 \sqrt[3]{3} I_0+2^{2/3}\right) \tilde{m}^2 \tau^3 + O(\tau^5) \,, \nn\\
C_2^- & = & \frac{\beta_{14}}{40} \sqrt{I_0} \left(6 \sqrt[3]{3} I_0-2^{2/3}\right) \tilde{m}^2 \tau^3 + O(\tau^5) \,, \nn\\
B_2 & = & \frac{\beta_{14}\tilde{m}^2 \tau^6 \left(540\ 3^{2/3} I_0^3 \tilde{m}^2+9\ 2^{2/3} I_0^2 \left(4 \sqrt[3]{3} \tilde{m}^2+285\right)-816 \sqrt[3]{2} 3^{2/3} I_0-124 \sqrt[3]{3}\right)}{8400 \sqrt{I_0} \left(45 I_0+6^{2/3}\right)}\,, \nn\\
C & = & \frac{3\beta_{14} \sqrt{I_0} \tilde{m}^2 \tau^2}{4 \sqrt[3]{2}}\left(-1+ \frac{\tau^2 \left(135 \sqrt[3]{2} 3^{2/3} I_0^3 \tilde{m}^2+18 I_0^2 \left(\sqrt[3]{3} \tilde{m}^2+25\right)-20\ 6^{2/3} I_0-4 \sqrt[3]{6}\right)}{60 I_0 \left(45 I_0+6^{2/3}\right)}\right)\,, \nn\\
B & = & -\frac{3\beta_{14} \tilde{m}^2 \tau^5 \left(15\ 2^{2/3} \sqrt[3]{3} I_0^3 \tilde{m}^2+2 \sqrt[3]{2} I_0^2 \left(\tilde{m}^2+30\ 3^{2/3}\right)-52 \sqrt[3]{3} I_0-4\ 2^{2/3}\right)}{200 \left(45 I_0+6^{2/3}\right)} + O(\tau^7)\,, \nn\\
a & = & \beta_{14}\tau\left(1+ \frac{\tau^2 \left(-\sqrt[3]{2} 3^{2/3} I_0^2 \tilde{m}^2+4 I_0+6^{2/3}\right)}{60 I_0} + O(\tau^4)\right) \,.
\eer
}

\end{itemize}

\subsection{Analysis of the IR modes}
\label{sec:IRaction}

\begin{figure}
    \centering
    \includegraphics[width=0.47\linewidth]{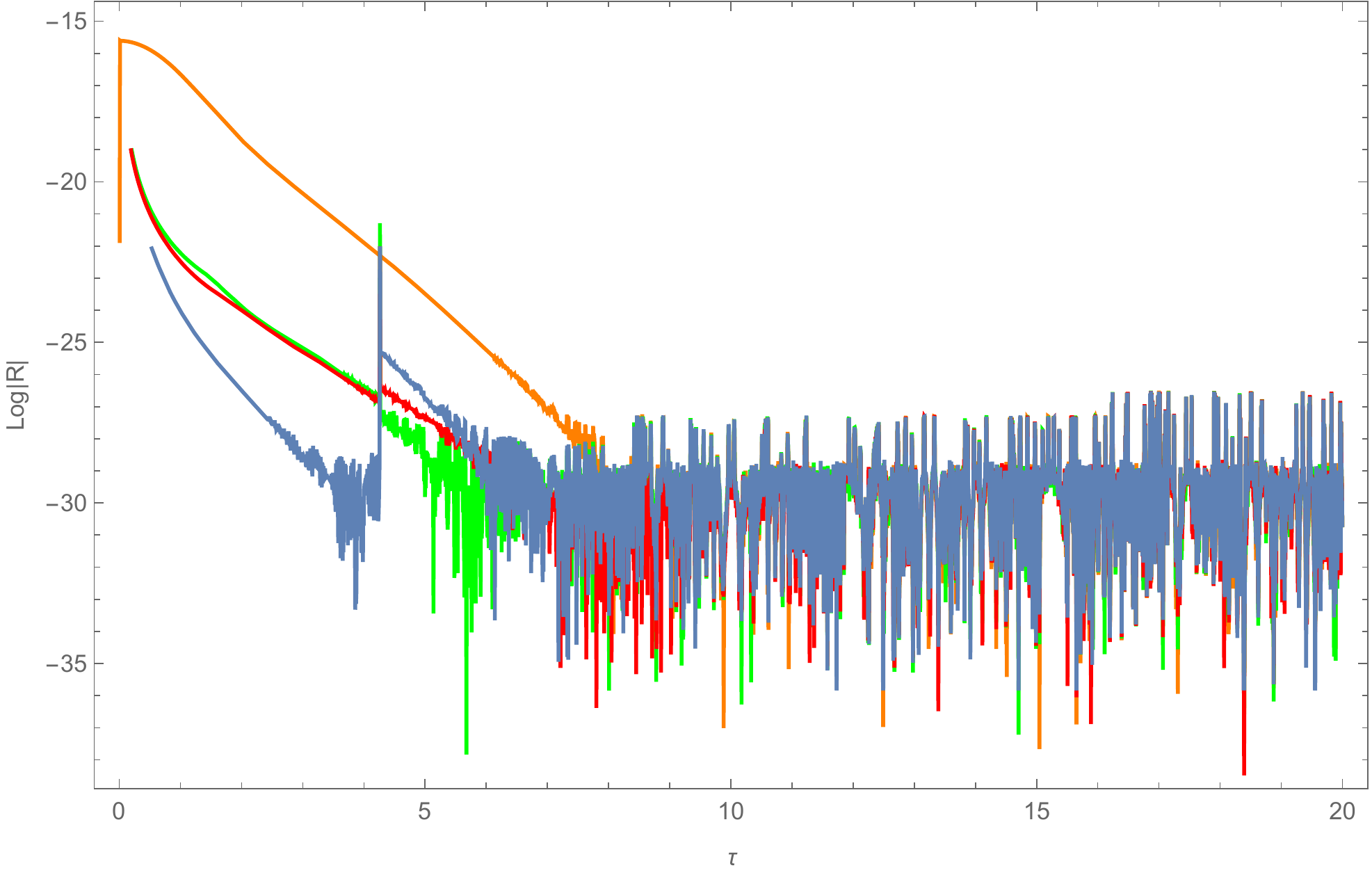}
    \hfill
    \includegraphics[width=0.47\linewidth]{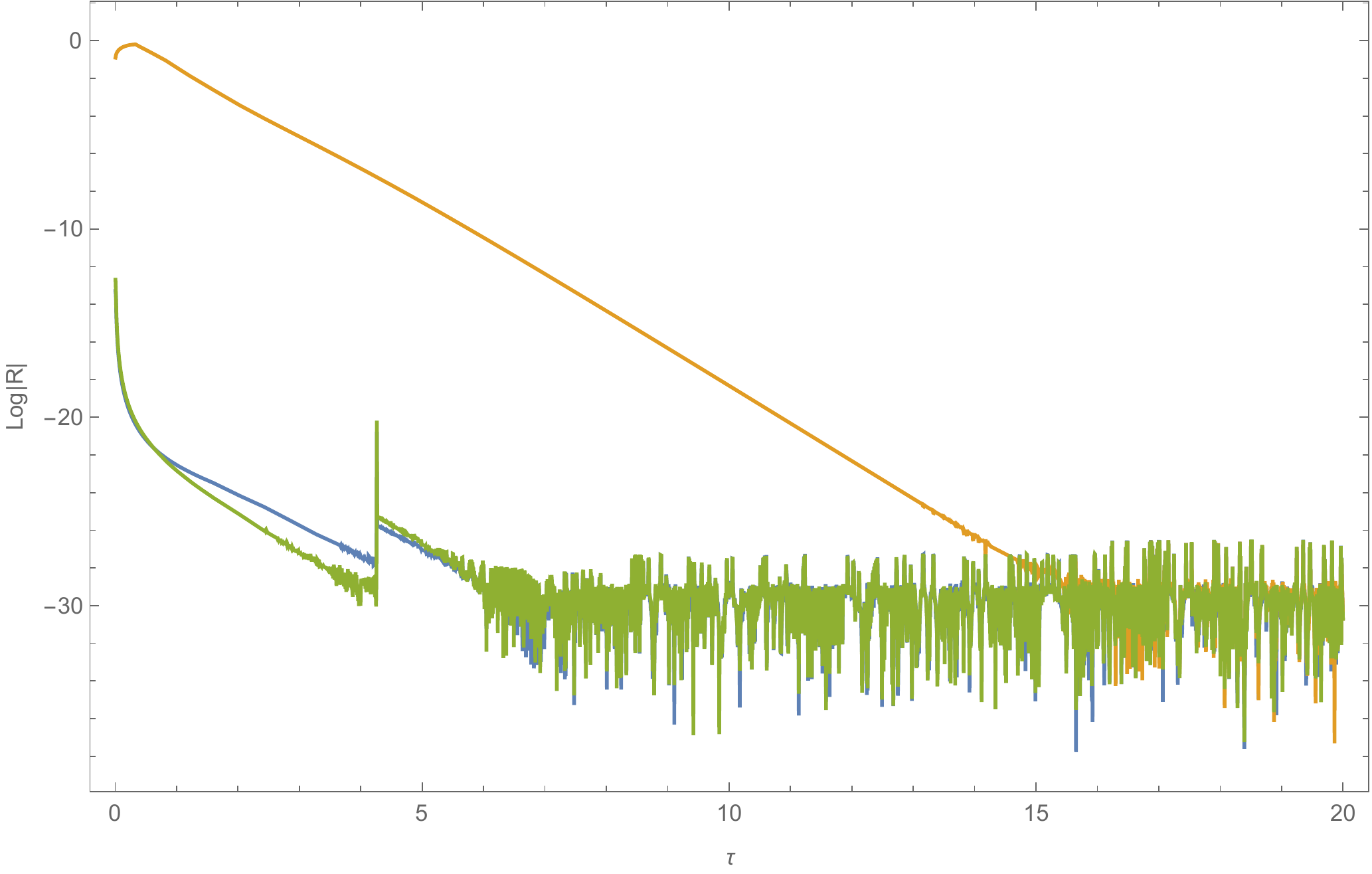}
    \caption{The logarithm of the ratio of the left hand side of equation~(\ref{eqA}) over the sum of absolute values of the terms of the same equation for seven regular IR solutions of the linearized equations. Six of the seven regular modes satisfy equation~(\ref{eqA}) with the accuracy of the numerical analysis.}
    \label{fig:irtouv}
\end{figure}

In the previous sections we separated the IR modes in the singular and the regular ones. This separation follows from the analysis of the asymptotic form of the action functional for these modes, which we deduced from the type IIB action. The relevant terms in the functional  read
\ber
  \mathcal{G}_{ab}\partial_M\phi^a\partial_M\phi^b & \approx & \tau^2\partial_M C\partial_M C+ \partial_M B_2\partial_M B_2+ \partial_M B\partial_M B \nn
  \\ & + &  \frac{1}{\tau^2}\partial_M(C_2^-+C_2^+)\partial_M(C_2^-+C_2^+)  +\tau^4(\partial_\mu a +\tau^2\partial_\mu\phi_3^{\prime})^2  \nn
  \\ & + &   C[ B_2 - \tau C_2^-+ \tau B_2^{\prime} + (C_2^{-\prime}+C_2^{+\prime})]+\tau^8(\partial_\mu \phi_3^\prime)^2+\tau^6(\partial_\mu\phi_3)^2 \nn \\
  & + & (C_2^-)^2 +\frac{B_2^2}{\tau^2} +\tau^2C^2 +\tau^8(\phi_3)^2\,.
  \label{IRaction}
\eer
Here index $\mu$ labels only the Minkowski directions, while index $M$ can also take value $M=\tau$. The $\tau$ integral of this function is convergent for small $\tau$ if
\ber
  \phi_3 & < & \tau^{-7/2} \,, \\
  C_2^+ & < & \tau^{3/2}\,, \\
  C_2^- & < & \tau^{3/2}\,, \\
  B_2 & < &  \tau^{1/2} \,,\\
  C & < & \tau^{-1/2}\,, \\
  B & < &  \tau^{1/2} \,,\\
  a & < &  \tau^{-5/2}\,.
\eer
These, however, are sufficient, but not necessary conditions. For, example, one can substitute the conditions for $C_2^{\pm}$ by a weaker condition 
\beq
P \ = \ C_2^+ + C_2^- \ < \ \tau^{3/2}\,.
\eeq
It is important that the most singular term of $C_2^{\pm}$ in~(\ref{IRaction}) only depends on the sum $P$.

One can check that all the modes $\beta_i$ with even $i$ satisfy the above conditions with a subtlety, that the modes $\beta_4$ and $\beta_{12}$ only satisfy the weak condition.

We should also find out which of the modes are physical, that is satisfy the constraint imposed by equation~(\ref{eqA}), whose leading asymptotic form is equation~(\ref{AeqIR}). We checked that up to the NNLO all the modes $\beta_i$ with even $i$, except $\beta_{12}$ do satisfy equation~(\ref{eqA}), which confirms the fact that the system we are studying contains only six physical modes.

Finally, one can make a numerical check of equation~(\ref{eqA}), similar to the one made in appendix~\ref{sec:IRasympt}. As demonstrated by figure~\ref{fig:irtouv}, six of the seven regular mode satisfy the constraint with the numerical accuracy. Again, one sees that the mode that does not satify the constraint in the IR, tends to do so at the opposite end.


\end{appendix}

\bibliographystyle{hieeetr}
\bibliography{references}

\begin{thebibliography}{10}

\bibitem{Klebanov:2000hb}
I.~R. Klebanov and M.~J. Strassler, ``{Supergravity and a confining gauge
  theory: Duality cascades and chi SB resolution of naked singularities},''
  {\em JHEP}, vol.~08, p.~052, 2000, hep-th/0007191.

\bibitem{Klebanov:1998hh}
I.~R. Klebanov and E.~Witten, ``{Superconformal field theory on three-branes at
  a Calabi-Yau singularity},'' {\em Nucl. Phys. B}, vol.~536, pp.~199--218,
  1998, hep-th/9807080.

\bibitem{Gubser:1998fp}
S.~S. Gubser and I.~R. Klebanov, ``{Baryons and domain walls in an N=1
  superconformal gauge theory},'' {\em Phys. Rev. D}, vol.~58, p.~125025, 1998,
  hep-th/9808075.

\bibitem{Klebanov:1999rd}
I.~R. Klebanov and N.~A. Nekrasov, ``{Gravity duals of fractional branes and
  logarithmic RG flow},'' {\em Nucl. Phys. B}, vol.~574, pp.~263--274, 2000,
  hep-th/9911096.

\bibitem{Klebanov:2000nc}
I.~R. Klebanov and A.~A. Tseytlin, ``{Gravity duals of supersymmetric SU(N) x
  SU(N+M) gauge theories},'' {\em Nucl. Phys.}, vol.~B578, pp.~123--138, 2000,
  hep-th/0002159.

\bibitem{Strassler:2005qs}
M.~J. Strassler, ``{The Duality cascade},'' in {\em {Progress in string theory.
  Proceedings, Summer School, TASI 2003, Boulder, USA, June 2-27, 2003}},
  pp.~419--510, 2005, hep-th/0505153.

\bibitem{Gubser:2004qj}
S.~S. Gubser, C.~P. Herzog, and I.~R. Klebanov, ``{Symmetry breaking and
  axionic strings in the warped deformed conifold},'' {\em JHEP}, vol.~09,
  p.~036, 2004, hep-th/0405282.

\bibitem{Gubser:2004tf}
S.~S. Gubser, C.~P. Herzog, and I.~R. Klebanov, ``{Variations on the warped
  deformed conifold},'' {\em Comptes Rendus Physique}, vol.~5, pp.~1031--1038,
  2004, hep-th/0409186.

\bibitem{Benna:2006ib}
M.~K. Benna, A.~Dymarsky, and I.~R. Klebanov, ``{Baryonic Condensates on the
  Conifold},'' {\em JHEP}, vol.~08, p.~034, 2007, hep-th/0612136.

\bibitem{Ceresole:1999ht}
A.~Ceresole, G.~Dall'Agata, and R.~D'Auria, ``{K K spectroscopy of type IIB
  supergravity on AdS(5) x T**11},'' {\em JHEP}, vol.~11, p.~009, 1999,
  hep-th/9907216.

\bibitem{Morningstar:1999rf}
C.~J. Morningstar and M.~J. Peardon, ``{The Glueball spectrum from an
  anisotropic lattice study},'' {\em Phys. Rev.}, vol.~D60, p.~034509, 1999,
  hep-lat/9901004.

\bibitem{Chen:2005mg}
Y.~Chen {\em et~al.}, ``{Glueball spectrum and matrix elements on anisotropic
  lattices},'' {\em Phys. Rev.}, vol.~D73, p.~014516, 2006, hep-lat/0510074.

\bibitem{Teper:1998kw}
M.~J. Teper, ``{Glueball masses and other physical properties of SU(N) gauge
  theories in D = (3+1): A Review of lattice results for theorists},'' 1998,
  hep-th/9812187.

\bibitem{Lucini:2004my}
B.~Lucini, M.~Teper, and U.~Wenger, ``{Glueballs and k-strings in SU(N) gauge
  theories: Calculations with improved operators},'' {\em JHEP}, vol.~06,
  p.~012, 2004, hep-lat/0404008.

\bibitem{Lucini:2010nv}
B.~Lucini, A.~Rago, and E.~Rinaldi, ``{Glueball masses in the large N limit},''
  {\em JHEP}, vol.~08, p.~119, 2010, 1007.3879.

\bibitem{Holligan:2019lma}
J.~Holligan, E.~Bennett, D.~K. Hong, J.-W. Lee, C.-J.~D. Lin, B.~Lucini,
  M.~Piai, and D.~Vadacchino, ``{$Sp(2N)$ Yang-Mills towards large $N$},'' in
  {\em {37th International Symposium on Lattice Field Theory}}, 2019,
  1912.09788.

\bibitem{Bennett:2020hqd}
Bennett, J.~Holligan, D.~K. Hong, J.-W. Lee, C.-J.~D. Lin, B.~Lucini, M.~Piai,
  and D.~Vadacchino, ``{Color dependence of tensor and scalar glueball masses
  in Yang-Mills theories},'' 4 2020, 2004.11063.

\bibitem{Athenodorou:2020ani}
A.~Athenodorou and M.~Teper, ``{The glueball spectrum of SU(3) gauge theory in
  3+1 dimension},'' 7 2020, 2007.06422.

\bibitem{Gregory:2012hu}
E.~Gregory, A.~Irving, B.~Lucini, C.~McNeile, A.~Rago, C.~Richards, and
  E.~Rinaldi, ``{Towards the glueball spectrum from unquenched lattice QCD},''
  {\em JHEP}, vol.~10, p.~170, 2012, 1208.1858.

\bibitem{Berg:2005pd}
M.~Berg, M.~Haack, and W.~Mueck, ``{Bulk dynamics in confining gauge
  theories},'' {\em Nucl. Phys.}, vol.~B736, pp.~82--132, 2006, hep-th/0507285.

\bibitem{Dymarsky:2006hn}
A.~{\relax Ya}. Dymarsky and D.~G. Melnikov, ``{On the glueball spectrum in the
  Klebanov-Strassler model},'' {\em JETP Lett.}, vol.~84, pp.~368--371, 2006.
\newblock [Pisma Zh. Eksp. Teor. Fiz.84,440(2006)].

\bibitem{Berg:2006xy}
M.~Berg, M.~Haack, and W.~Mueck, ``{Glueballs vs. Gluinoballs: Fluctuation
  Spectra in Non-AdS/Non-CFT},'' {\em Nucl. Phys.}, vol.~B789, pp.~1--44, 2008,
  hep-th/0612224.

\bibitem{Dymarsky:2007zs}
A.~Dymarsky and D.~Melnikov, ``{Gravity Multiplet on KS and BB Backgrounds},''
  {\em JHEP}, vol.~05, p.~035, 2008, 0710.4517.

\bibitem{Benna:2007mb}
M.~K. Benna, A.~Dymarsky, I.~R. Klebanov, and A.~Solovyov, ``{On Normal Modes
  of a Warped Throat},'' {\em JHEP}, vol.~06, p.~070, 2008, 0712.4404.

\bibitem{Dymarsky:2008wd}
A.~Dymarsky, D.~Melnikov, and A.~Solovyov, ``{I-odd sector of the
  Klebanov-Strassler theory},'' {\em JHEP}, vol.~05, p.~105, 2009, 0810.5666.

\bibitem{Gordeli:2009nw}
I.~Gordeli and D.~Melnikov, ``{On I-even Singlet Glueballs in the
  Klebanov-Strassler Theory},'' {\em JHEP}, vol.~08, p.~082, 2011, 0912.5517.

\bibitem{Gordeli:2013jea}
I.~Gordeli and D.~Melnikov, ``{Calculation of glueball spectra in
  supersymmetric theories via holography},'' in {\em {International Workshop on
  Low X Physics (Israel 2013) Eilat, Israel, May 30-June 04, 2013}}, 2013,
  1311.6537.

\bibitem{Krasnitz:2000ir}
M.~Krasnitz, ``{A Two point function in a cascading N=1 gauge theory from
  supergravity},'' 11 2000, hep-th/0011179.

\bibitem{Caceres:2000qe}
E.~Caceres and R.~Hernandez, ``{Glueball masses for the deformed conifold
  theory},'' {\em Phys. Lett. B}, vol.~504, pp.~64--70, 2001, hep-th/0011204.

\bibitem{Bianchi:2003ug}
M.~Bianchi, M.~Prisco, and W.~Mueck, ``{New results on holographic three point
  functions},'' {\em JHEP}, vol.~11, p.~052, 2003, hep-th/0310129.

\bibitem{Amador:2004pz}
X.~Amador and E.~Caceres, ``{Spin two glueball mass and glueball regge
  trajectory from supergravity},'' {\em JHEP}, vol.~11, p.~022, 2004,
  hep-th/0402061.

\bibitem{Caceres:2005yx}
E.~Caceres and C.~Nunez, ``{Glueballs of super Yang-Mills from wrapped
  branes},'' {\em JHEP}, vol.~09, p.~027, 2005, hep-th/0506051.

\bibitem{Elander:2009bm}
D.~Elander, ``{Glueball Spectra of SQCD-like Theories},'' {\em JHEP}, vol.~03,
  p.~114, 2010, 0912.1600.

\bibitem{Bianchi:2010cy}
M.~Bianchi and W.~de~Paula, ``{On Exact Symmetries and Massless Vectors in
  Holographic Flows and other Flux Vacua},'' {\em JHEP}, vol.~04, p.~113, 2010,
  1003.2536.

\bibitem{Pufu:2010ie}
S.~S. Pufu, I.~R. Klebanov, T.~Klose, and J.~Lin, ``{Green's Functions and
  Non-Singlet Glueballs on Deformed Conifolds},'' {\em J. Phys. A}, vol.~44,
  p.~055404, 2011, 1009.2763.

\bibitem{Elander:2014ola}
D.~Elander, ``{Light scalar from deformations of the Klebanov-Strassler
  background},'' {\em Phys. Rev.}, vol.~D91, no.~12, p.~126012, 2015,
  1401.3412.

\bibitem{Elander:2017cle}
D.~Elander and M.~Piai, ``{Calculable mass hierarchies and a light dilaton from
  gravity duals},'' {\em Phys. Lett. B}, vol.~772, pp.~110--114, 2017,
  1703.09205.

\bibitem{Elander:2017hyr}
D.~Elander and M.~Piai, ``{Glueballs on the Baryonic Branch of
  Klebanov-Strassler: dimensional deconstruction and a light scalar
  particle},'' {\em JHEP}, vol.~06, p.~003, 2017, 1703.10158.

\bibitem{inprogress}
A.~Dymarsky, D.~Melnikov, and C.~Rodrigues~Filho. to appear.

\bibitem{Maldacena:1997re}
J.~M. Maldacena, ``{The Large N limit of superconformal field theories and
  supergravity},'' {\em Int. J. Theor. Phys.}, vol.~38, pp.~1113--1133, 1999,
  hep-th/9711200.
\newblock [Adv. Theor. Math. Phys.2,231(1998)].

\bibitem{Gubser:1998bc}
S.~S. Gubser, I.~R. Klebanov, and A.~M. Polyakov, ``{Gauge theory correlators
  from noncritical string theory},'' {\em Phys. Lett.}, vol.~B428,
  pp.~105--114, 1998, hep-th/9802109.

\bibitem{Witten:1998qj}
E.~Witten, ``{Anti-de Sitter space and holography},'' {\em Adv. Theor. Math.
  Phys.}, vol.~2, pp.~253--291, 1998, hep-th/9802150.

\bibitem{Aharony:1999ti}
O.~Aharony, S.~S. Gubser, J.~M. Maldacena, H.~Ooguri, and Y.~Oz, ``{Large N
  field theories, string theory and gravity},'' {\em Phys. Rept.}, vol.~323,
  pp.~183--386, 2000, hep-th/9905111.

\bibitem{Nastase:2015wjb}
H.~Nastase, {\em {Introduction to the ADS/CFT Correspondence}}.
\newblock Cambridge University Press, 9 2015.

\bibitem{Erdmenger:2018xqz}
J.~Erdmenger, ``{Introduction to Gauge/Gravity Duality},'' {\em PoS},
  vol.~TASI2017, p.~001, 2018, 1807.09872.

\bibitem{DeWolfe:2018dkl}
O.~DeWolfe, ``{TASI Lectures on Applications of Gauge/Gravity Duality},'' {\em
  PoS}, vol.~TASI2017, p.~014, 2018, 1802.08267.

\bibitem{Harlow:2018fse}
D.~Harlow, ``{TASI Lectures on the Emergence of Bulk Physics in AdS/CFT},''
  {\em PoS}, vol.~TASI2017, p.~002, 2018, 1802.01040.

\bibitem{Schwarz:1983qr}
J.~H. Schwarz, ``{Covariant Field Equations of Chiral N=2 D=10 Supergravity},''
  {\em Nucl. Phys.}, vol.~B226, p.~269, 1983.
\newblock [,269(1983)].

\bibitem{Candelas:1989js}
P.~Candelas and X.~C. de~la Ossa, ``{Comments on Conifolds},'' {\em Nucl.
  Phys.}, vol.~B342, pp.~246--268, 1990.

\bibitem{Csaki:1998qr}
C.~Csaki, H.~Ooguri, Y.~Oz, and J.~Terning, ``{Glueball mass spectrum from
  supergravity},'' {\em JHEP}, vol.~01, p.~017, 1999, hep-th/9806021.

\bibitem{Brower:2000rp}
R.~C. Brower, S.~D. Mathur, and C.-I. Tan, ``{Glueball spectrum for QCD from
  AdS supergravity duality},'' {\em Nucl. Phys.}, vol.~B587, pp.~249--276,
  2000, hep-th/0003115.

\bibitem{Klebanov:2002gr}
I.~R. Klebanov, P.~Ouyang, and E.~Witten, ``{A Gravity dual of the chiral
  anomaly},'' {\em Phys. Rev.}, vol.~D65, p.~105007, 2002, hep-th/0202056.

\bibitem{Butti:2004pk}
A.~Butti, M.~Grana, R.~Minasian, M.~Petrini, and A.~Zaffaroni, ``{The Baryonic
  branch of Klebanov-Strassler solution: A supersymmetric family of SU(3)
  structure backgrounds},'' {\em JHEP}, vol.~03, p.~069, 2005, hep-th/0412187.

\bibitem{Apreda:2003gc}
R.~Apreda, ``{Nonsupersymmetric regular solutions from wrapped and fractional
  branes},'' 1 2003, hep-th/0301118.

\bibitem{Kim:1985ez}
H.~J. Kim, L.~J. Romans, and P.~van Nieuwenhuizen, ``{The Mass Spectrum of
  Chiral N=2 D=10 Supergravity on S**5},'' {\em Phys. Rev.}, vol.~D32, p.~389,
  1985.

\bibitem{Elander:2010wd}
D.~Elander and M.~Piai, ``{Light scalars from a compact fifth dimension},''
  {\em JHEP}, vol.~01, p.~026, 2011, 1010.1964.

\bibitem{Elander:2012yh}
D.~Elander and M.~Piai, ``{On the glueball spectrum of walking backgrounds from
  wrapped-D5 gravity duals},'' {\em Nucl. Phys. B}, vol.~871, pp.~164--180,
  2013, 1212.2600.

\bibitem{Elander:2018aub}
D.~Elander, M.~Piai, and J.~Roughley, ``{Holographic glueballs from the circle
  reduction of Romans supergravity},'' {\em JHEP}, vol.~02, p.~101, 2019,
  1811.01010.

\bibitem{Romans:1985tw}
L.~Romans, ``{The F(4) Gauged Supergravity in Six-dimensions},'' {\em Nucl.
  Phys. B}, vol.~269, p.~691, 1986.

\end{thebibliography}

\end{document}